\documentclass[letterpaper, 12pt]{article}[2000/05/19]
\usepackage[english]{babel}
\usepackage{amsfonts,amsmath,amssymb,amsthm,latexsym,amscd,mathrsfs}
\usepackage{ifthen,cite}
\usepackage[bookmarksnumbered=true]{hyperref}

\hypersetup{pdfpagetransition={Split}}

\newcommand{\evenhead}{Author \ name}
\newcommand{\oddhead}{Article \ name}
\newcommand{\theArticleName}{Article name}

\newcommand{\FirstPageHeading}[1]{\thispagestyle{empty}%
\noindent\raisebox{0pt}[0pt][0pt]{\makebox[\textwidth]{\protect\footnotesize \sf }}\par}

\newcommand{\ArticleName}[1]{\renewcommand{\theArticleName}{#1}\vspace{-2mm}\par\noindent {\LARGE\bf  #1\par}}
\newcommand{\Author}[1]{\vspace{5mm}\par\noindent {\Large  #1\par} \par\vspace{2mm}\par}
\newcommand{\Address}[1]{\vspace{2mm}\par\noindent {\it #1} \par}
\newcommand{\Email}[1]{\ifthenelse{\equal{#1}{}}{}{\par\noindent {\rm E-mail: }{\it  #1} \par}}
\newcommand{\URLaddress}[1]{\ifthenelse{\equal{#1}{}}{}{\par\noindent {\rm URL: }{\tt  #1} \par}}
\newcommand{\EmailD}[1]{\ifthenelse{\equal{#1}{}}{}{\par\noindent {$\phantom{\dag}$~\rm E-mail: }{\it  #1} \par}}
\newcommand{\URLaddressD}[1]{\ifthenelse{\equal{#1}{}}{}{\par\noindent {$\phantom{\dag}$~\rm URL: }{\tt  #1} \par}}

\newcommand{\Abstract}[1]{\vspace{6mm}\par\noindent\hspace*{10mm}
\parbox{140mm}{\small {\bf Abstract.} #1}\par}
\newcommand{\Keywords}[1]{\vspace{3mm}\par\noindent\hspace*{10mm}
\parbox{140mm}{\small {\bf Key words:} \rm #1}\par}
\newcommand{\Classification}[1]{\vspace{3mm}\par\noindent\hspace*{10mm}
\parbox{140mm}{\small {\it 2000 Mathematics Subject Classification:} \rm #1}\vspace{3mm}\par}
\newcommand{\ShortArticleName}[1]{\renewcommand{\oddhead}{#1}}
\newcommand{\AuthorNameForHeading}[1]{\renewcommand{\evenhead}{#1}}

\long\def\@makecaption#1#2{
  \sbox\@tempboxa{\small \textbf{#1.}\ \ #2}%
  \ifdim \wd\@tempboxa >\hsize
    {\small \textbf{#1.}\ \ #2}\par \else
    \global \@minipagefalse
    \hb@xt@\hsize{\hfil\box\@tempboxa\hfil}%
  \fi \vskip\belowcaptionskip}

\def\numberwithin#1#2{\@ifundefined{c@#1}{\@nocounterr{#1}}{%
  \@ifundefined{c@#2}{\@nocnterr{#2}}{%
  \@addtoreset{#1}{#2}%
  \toks@\@xp\@xp\@xp{\csname the#1\endcsname}%
  \@xp\xdef\csname the#1\endcsname
    {\@xp\@nx\csname the#2\endcsname.\the\toks@}}}}
\def\E^#1{{\buildrel #1 \over\vee}}
\newtheorem{theorem}{Theorem}

{\theoremstyle{definition}

}

\begin{document}

\FirstPageHeading{V.I. Gerasimenko}

\ShortArticleName{Theory of quantum evolution equations}

\AuthorNameForHeading{V.I. Gerasimenko}

\ArticleName{Introduction to the Theory of Evolution\\ Equations of Quantum Many-Particle Systems}

\Author{V.I. Gerasimenko\footnote{E-mail: \emph{gerasym@imath.kiev.ua}}}

\Address{Institute of Mathematics of NAS of Ukraine,\\
         3, Tereshchenkivs'ka Str., 01601 Kyiv-4, Ukraine}

\medskip

\Abstract{
In the paper we review some recent results of the theory of hierarchies of quantum evolution equations.

The evolution of infinite-particle quantum systems is described within the framework of the evolution
of states by the quantum BBGKY hierarchy for marginal density operators or within the framework of an equivalent
approach in terms of the marginal observables by the dual quantum BBGKY hierarchy. The nonperturbative solutions
of the Cauchy problem of the dual quantum BBGKY hierarchy and the quantum BBGKY hierarchy are constructed.

The origin of the microscopic description of non-equilibrium quantum correlations is considered. It
consists in one more approach to the description of the evolution of states of quantum many-particle
systems by means of correlation operators. The correlation operators and the marginal correlation
operators are governed by the von Neumann hierarchy and the quantum nonlinear BBGKY hierarchy respectively.
The nonperturbative solutions of the Cauchy problem of both these hierarchies are constructed.

We develop also an approach to a description of the evolution of states by means of the quantum kinetic
equations. For initial states which are specified in terms of a one-particle density operator the
equivalence of the description of the evolution of quantum many-particle states by the Cauchy problem
of the quantum BBGKY hierarchy and by the Cauchy problem of the generalized quantum kinetic equation
together with a sequence of explicitly defined functionals of a solution of stated kinetic equation is
established. The relationships of the specific quantum kinetic equations with the generalized quantum
kinetic equation are discussed.

In conclusion the mean field asymptotic behavior of stated hierarchy solutions are established. The
constructed asymptotics are governed by the quantum Vlasov hierarchy for limit states, the nonlinear
quantum Vlasov hierarchy for limit correlations and the dual quantum Vlasov hierarchy for limit
observables respectively.}

\medskip

\medskip
\Classification{35Q40; 47H20; 47J35; 82C10; 82C22.}

\makeatletter
\renewcommand{\@evenhead}{
\hspace*{-3pt}\raisebox{-15pt}[\headheight][0pt]{\vbox{\hbox to \textwidth {\thepage \hfil \evenhead}\vskip4pt \hrule}}}
\renewcommand{\@oddhead}{
\hspace*{-3pt}\raisebox{-15pt}[\headheight][0pt]{\vbox{\hbox to \textwidth {\oddhead \hfil \thepage}\vskip4pt\hrule}}}
\renewcommand{\@evenfoot}{}
\renewcommand{\@oddfoot}{}
\makeatother

\newpage
\vphantom{math}
\protect\tableofcontents

\vspace{0.5cm}

\section{Introduction: motivations and results}
Experimental advances in the Bose condensation of atomic gases and in the strong correlated Fermi
systems have stimulated interesting problems in the theory of evolution equations of quantum many-particle systems.
Among them it is a description of collective behavior of interacting particles by quantum kinetic equations,
i.e. the evolution equations for a one-particle marginal density operator \cite{CGP97,CIP}. The
rigorous derivation of quantum kinetic equations is a very trendy subject nowadays, and the number
of publications devoted to it has grown last decade. In particular the considerable progress in the
rigorous derivation of the nonlinear Schr\"{o}dinger equation and the Gross-Pitaevskii equation for
the Bose condensate \cite{AGT,AA,BGGM2,FL,ESchY2,EShY10,LSSY,M10,PP09,CP,GMM,S-R} in mean field scaling
limit as well as the quantum Boltzmann equation \cite{BCEP3,ESY} is observed. Owing to the intrinsic
complexity and richness of this problem, first of all it is necessary to develop an adequate mathematical
theory of evolution equations of quantum many-particle systems underlying of kinetic equations.

In this review we expound the foundation of the theory of quantum evolution equations
of statistical mechanics. We construct nonperturbative solutions of the Cauchy problem of the
quantum BBGKY hierarchy for marginal density operators and the dual quantum BBGKY hierarchy for
marginal observables on appropriate Banach spaces. On basis of obtained results the origin of the
microscopic description of non-equilibrium quantum correlations is considered. In particular, the
nonperturbative solutions of the Cauchy problem of the von Neumann hierarchy for correlation operators
and the nonlinear quantum BBGKY hierarchy for marginal correlation operators are constructed. We develop
also one more approach to the description of the evolution of quantum many-particle systems by means of
the quantum kinetic equations. For initial states which are specified in terms of a one-particle density
operator the equivalence of the description of the evolution of quantum infinite-particle states by the
Cauchy problem of the quantum BBGKY hierarchy and by the Cauchy problem of the generalized quantum kinetic
equation is established. We also describe the mean field scaling asymptotic behavior of
nonperturbative solutions of hierarchies of quantum evolution equations under consideration.

We outline the structure of the paper and the main results.
In introductory section 2 we set forth the traditional approach to the description of the evolution of
quantum many-particle systems. It is well known that a description of quantum many-particle systems is
formulated in terms of two sets of objects: observables and states. The functional of the mean value
of observables defines a duality between observables and states and as a consequence there exist two
equivalent approaches to the description of the evolution, namely in terms of the evolution equations
for observables (the Heisenberg equation) and for states (the von Neumann equation). We adduce some
preliminary facts about dynamics of finitely many quantum particles described by these evolution
equations within the framework of nonequilibrium grand canonical ensemble.

In section 3 one more an equivalent approach to the description of the evolution of states of quantum
many-particle systems is given by means of operators which are interpreted as correlation operators.
To justify the von Neumann hierarchy for correlation operators, at first we consider in detail the
motivation of the introduction of correlation operators to the description of states or in other
words, the origin of the microscopic description of correlations in quantum many-particle systems is
considered.

In section 4 for the description of the evolution of infinite-particle quantum systems we introduce the
hierarchies of evolution equations for the marginal observables, density operators and correlation
operators. They are derived as the evolution equations for one more but an equivalent method of the
description of states and observables of finitely many particles. The nonperturbative solutions of
the Cauchy problem of the dual quantum BBGKY hierarchy for the marginal observables, the quantum BBGKY
hierarchy for marginal density operators and the nonlinear quantum BBGKY hierarchy for marginal correlation
operators are constructed as an expansion over particle clusters which evolution is governed by the
corresponding-order cumulant (semi-invariant) of the groups of operators of the Heisenberg equations,
the von Neumann equations and the von Neumann hierarchy of finitely many particles respectively.

In section 5 we develop an approach to a description of the evolution of states by means of the quantum
kinetic equations. For initial states which are specified in terms of a one-particle density operator
the equivalence of the description of the evolution of quantum many-particle states by the Cauchy problem
of the quantum BBGKY hierarchy and by the Cauchy problem of the generalized quantum kinetic equation
together with a sequence of explicitly defined correlation functionals of a solution of stated quantum
kinetic equation is established. The relationship of the generalized quantum kinetic equation
with the specific quantum kinetic equations is discussed.

In section 6 the mean field (self-consistent field) asymptotic behavior of constructed above solutions
of the quantum hierarchies and the generalized quantum kinetic equation is established. In particular,
the quantum Vlasov hierarchy for limit states, the nonlinear quantum Vlasov hierarchy for limit correlations
and the dual quantum Vlasov hierarchy for limit observables are rigorously derived.

Finally in section 7 we conclude with some observations and perspectives for future research.


\section{On evolution equations of quantum many-particle systems}
It is well known that a description of quantum many-particle systems is formulated in terms of two sets
of objects: observables and states. The functional of the mean value of observables defines a duality
between observables and states and as a consequence there exist two approaches to the description of the
evolution, namely in terms of the evolution equations for observables (the Heisenberg equation) and for
states (the von Neumann equation). In this section we adduce some preliminary facts about dynamics of
finitely many quantum particles described within the framework of nonequilibrium grand canonical ensemble.

\subsection{Preliminaries}
We consider a quantum system of a non-fixed, i.e. arbitrary but finite, number of identical (spinless)
particles (nonequilibrium grand canonical ensemble) obeying the Maxwell-Boltzmann statistics in the space
$\mathbb{R}^{\nu}$. We will use units where $h={2\pi\hbar}=1$ is a Planck constant, and $m=1$ is the
mass of particles.

Let $\mathcal{H}$ be a one-particle Hilbert space, then the $n$-particle space $\mathcal{H}_{n}$, is the
tensor product of $n$ Hilbert spaces $\mathcal{H}$. We adopt the usual convention that $\mathcal{H}^{\otimes 0}
=\mathbb{C}$. We denote by $\mathcal{F}_{\mathcal{H}}={\bigoplus\limits}_{n=0}^{\infty}\mathcal{H}_{n}$
the Fock space over the Hilbert space $\mathcal{H}$.

Let a sequence $g=(g_{0},g_{1},\ldots,g_{n},\ldots)$ be an infinite sequence of self-adjoint
bounded operators defined on the Fock space $\mathcal{F}_{\mathcal{H}}$ and $g_{0}\in\mathbb{C}$.
The operator $g_{n}$ defined on the $n$-particle Hilbert space $\mathcal{H}_{n}=\mathcal{H}^{\otimes n}$
will be denoted by $g_{n}(1,\ldots,n)$. Let the space $\mathfrak{L}(\mathcal{F}_\mathcal{H})$
be the space of sequences $g=(g_{0},g_{1},\ldots,g_{n},\ldots)$ of bounded operators $g_{n}$
defined on the Hilbert space $\mathcal{H}_n$ that satisfy symmetry condition:
$g_{n}(1,\ldots,n)=g_{n}(i_1,\ldots,i_n)$, for arbitrary $(i_{1},\ldots,i_{n})\in(1,\ldots,n)$,
equipped with the operator norm $\|.\|_{\mathfrak{L}(\mathcal{H}_{n})}$ \cite{DauL}. We will
also consider a more general space $\mathfrak{L}_{\gamma}(\mathcal{F}_\mathcal{H})$ with a norm
\begin{eqnarray*}
   &&\big\|g\big\|_{\mathfrak{L}_{\gamma} (\mathcal{F}_\mathcal{H})}\doteq
       \max\limits_{n\geq 0}\,\frac{\gamma^{n}}{n!}\,
       \big\|g_{n}\big\|_{\mathfrak{L}(\mathcal{H}_{n})},
\end{eqnarray*}
where $0<\gamma<1$. We denote by $\mathfrak{L}_{\gamma,0}(\mathcal{F}_\mathcal{H})
\subset\mathfrak{L}_{\gamma}(\mathcal{F}_\mathcal{H})$ the everywhere dense set of finite
sequences of degenerate operators with infinitely differentiable kernels with compact
supports. Observables of finitely many quantum particles are sequences of self-adjoint
operators from the space $\mathfrak{L}_{\gamma}(\mathcal{F}_\mathcal{H})$ \cite{DauL}.

Let $\mathfrak{L}^{1}(\mathcal{F}_\mathcal{H})={\bigoplus\limits}_{n=0}^{\infty}\mathfrak{L}^{1}
(\mathcal{H}_{n})$ be the space of sequences $f=(f_{0},f_{1},\ldots,f_{n},\ldots)$ of trace class
operators $f_{n}=f_{n}(1,\ldots,n)\in\mathfrak{L}^{1}(\mathcal{H}_{n})$ and $f_{0}\in\mathbb{C}$,
satisfying the mentioned above symmetry condition, equipped with the trace norm
\begin{eqnarray*}
    &&\big\|f\big\|_{\mathfrak{L}^{1}(\mathcal{F}_\mathcal{H})}=
       \sum\limits_{n=0}^{\infty}\,\big \|f_{n}\big\|_{\mathfrak{L}^{1}(\mathcal{H}_{n})}
       \doteq\sum\limits_{n=0}^{\infty}\,\mathrm{Tr}_{1,\ldots,n}|f_{n}(1,\ldots,n)|,
\end{eqnarray*}
where $\mathrm{Tr}_{1,\ldots,n}$ is the partial trace over $1,\ldots,n$ particles. The everywhere
dense set of finite sequences of degenerate operators with infinitely differentiable kernels with
compact supports from the space $\mathfrak{L}^{1}(\mathcal{F}_\mathcal{H})$ we denote by
$\mathfrak{L}^{1}_0(\mathcal{F}_\mathcal{H})$. The sequences of self-adjoint operators
$f_{n}\in\mathfrak{L}^{1}(\mathcal{H}_{n}),\,n\geq1$, which kernels are known as density matrices,
defined on the $n$-particle Hilbert space $\mathcal{H}_{n}=L^{2}(\mathbb{R}^{\nu n})$, describe
states of a quantum system of non-fixed number of particles.

The space $\mathfrak{L}(\mathcal{F}_\mathcal{H})$ is dual to the space
$\mathfrak{L}^{1}(\mathcal{F}_\mathcal{H})$ with respect to the bilinear form
\begin{eqnarray}\label{average}
    &&(g,f)\doteq\sum\limits_{n=0}^{\infty}\frac{1}{n!}
         \,\mathrm{Tr}_{1,\ldots,n}\,g_{n}f_{n},
\end{eqnarray}
where $g_{n}\in\mathfrak{L}(\mathcal{H}_{n})$ and $f_{n}\in\mathfrak{L}^{1}(\mathcal{H}_{n})$. The
range of positive continuous linear functional \eqref{average} on the space of observables
$\mathfrak{L}(\mathcal{F}_\mathcal{H})$ is interpreted as the mean values (averages) of observables.

The Bose-Einstein and Fermi-Dirac statistics endow observables and states with additional symmetry
properties \cite{BR} in comparison with the Maxwell-Boltzmann statistics. Let $\mathcal{H}$ be a
one-particle Hilbert space, then the $n$-particle spaces $\mathcal{H}^{\pm}_n$, are correspondingly
symmetric and antisymmetric tensor products of $n$ Hilbert spaces $\mathcal{H}$ that are associated
with $n$-particle systems of bosons and fermions \cite{BR}. We denote by $\mathcal{F}^{\pm}_{\mathcal{H}}={\bigoplus\limits}_{n=0}^{\infty}\mathcal{H}_{n}^{\pm}$
the Bose and Fermi Fock spaces over the Hilbert space $\mathcal{H}$ respectively.

The symmetrization operator $\mathcal{S}^{+}_{n}$ and the anti-symmetrization operator
$\mathcal{S}^{-}_{n}$ on $\mathcal{H}^{\otimes n}$ are defined by the formula
\begin{eqnarray}\label{Sn}
    &&\mathcal{S}^{\pm}_{n}\doteq\frac {1}{n!}
       \sum\limits_{\pi\epsilon \mathfrak{S}_{n}}(\pm1)^{|\pi|}p_{\pi},
\end{eqnarray}
where the operator $p_{\pi}$ is a transposition operator of the permutation $\pi$ from the
permutation group $\mathfrak{S}_{n}$ of the set $(1,\ldots,n)$ and $|\pi|$ denotes the
number of transpositions in the permutation $\pi$. The operators $\mathcal{S}^{\pm}_{n}$
are orthogonal projectors, i.e. $({\mathcal{S}^{\pm}_{n}})^{2}=\mathcal{S}^{\pm}_{n}$,
ranges of which are correspondingly the symmetric tensor product $\mathcal{H}_{n}^{+}$ and
the antisymmetric tensor product $\mathcal{H}_{n}^{-}$ of $n$ Hilbert spaces $\mathcal{H}$.

\subsection{The Heisenberg equation: the evolution of observables}
The Hamiltonian $H_{n}$ of $n$-particle system is a self-adjoint operator with the domain
$\mathcal{D}(H_{n})\subset\mathcal{H}_{n}$
\begin{eqnarray}\label{H}
    &&H_{n}=\sum\limits_{i=1}^{n}K(i)+\sum\limits_{i_{1}<i_{2}=1}^{n}\Phi(i_{1},i_{2}),
\end{eqnarray}
where $K(i)$ is the operator of a kinetic energy of the $i$ particle and $\Phi(i_{1},i_{2})$ is
the operator of a two-body interaction potential. The operator $K(i)$ acts on functions $\psi_n$
that belong to the subspace $L^{2}_{0}(\mathbb{R}^{\nu n})\subset\mathcal{D}(H_n)
\subset L^{2}(\mathbb{R}^{\nu n})$ of infinitely differentiable functions with compact supports
according to the formula: $K(i)\psi_n=-\frac{1}{2}\Delta_{q_i}\psi_n$. Correspondingly we have:
$\Phi(i_{1},i_{2})\psi_{n}=\Phi(q_{i_{1}},q_{i_{2}})\psi_{n}$, and we assume that the function
$\Phi(q_{i_{1}},q_{i_{2}})$ is symmetric with respect to permutations of its arguments,
translation-invariant and bounded function.

The evolution of observables $A(t)=(A_0,A_{1}(t,1),\ldots,A_{n}(t,1,\ldots,n),\ldots)$ is described
by the initial-value problem of a sequence of the Heisenberg equations \cite{Pe95,DauL}
\begin{eqnarray}
  \label{H-N1}
     &&\frac{d}{d t}A(t)=\mathcal{N}A(t),\\
  \label{H-N12}
     &&A(t)|_{t=0}=A(0),
\end{eqnarray}
where $A(0)=(A_0,A_{1}^{0}(1),\ldots,A_{n}^{0}(1,\ldots,n),\ldots)\in \mathfrak{L}(\mathcal{F}_\mathcal{H})$,
if $g\in \mathcal{D}(\mathcal{N})\subset \mathfrak{L}(\mathcal{F}_\mathcal{H})$, the generator $\mathcal{N}={\bigoplus\limits}_{n=0}^{\infty}\mathcal{N}_{n}$ is defined by the formula
\begin{eqnarray}\label{dkomyt}
     &&\mathcal{N}_{n}\,g_{n}\doteq-i(g_{n}H_{n}-H_{n}g_{n}),
\end{eqnarray}
and $H_{n}$ is the Hamiltonian \eqref{H}.

To determine a solution of the Cauchy problem \eqref{H-N1}-\eqref{H-N12} we introduce on the space
$\mathfrak{L}(\mathcal{F}_\mathcal{H})$ the following one-parameter family of operators
$\mathcal{G}(t)=\oplus^{\infty}_{n=0}\mathcal{G}_{n}(t)$:
\begin{eqnarray}\label{grG}
     &&\mathcal{G}_{n}(t)g_{n}\doteq e^{itH_{n}}g_{n}e^{-itH_{n}}.
\end{eqnarray}
On the space $\mathfrak{L}_{\gamma}(\mathcal{F}_\mathcal{H})$ the one-parameter mapping
$\mathbb{R}^1\ni t\mapsto\mathcal{G}(t)g$ defines an isometric $\ast$-weak continuous group of
operators, i.e. it is a $C_{0}^{\ast}$-group \cite{DauL,BR}. The infinitesimal generator $\mathcal{N}={\bigoplus\limits}_{n=0}^{\infty}~\mathcal{N}_{n}$ of this group of operators is a
closed operator for the $\ast$-weak topology and on its domain of definition $\mathcal{D}(\mathcal{N}_{n})
\subset\mathfrak{L}(\mathcal{H}_{n})$ which is everywhere dense for the $\ast$-weak topology,
$\mathcal{N}_{n}$ is defined as follows in the sense of the $\ast$-weak convergence of the space
$\mathfrak{L}(\mathcal{H}_{n})$
\begin{eqnarray}\label{infOper1}
    &&\mathrm{w^{\ast}-}\lim\limits_{t\rightarrow 0}\frac{1}{t}(\mathcal{G}_{n}(t)g_{n}-g_{n})
       =-i(g_{n}H_{n}-H_{n}g_{n}),
\end{eqnarray}
where the operator $\mathcal{N}_{n}g_{n}=-i(g_{n}H_{n}-H_{n}g_{n})$ is defined
on the domain $\mathcal{D}(H_{n})\subset\mathcal{H}_{n}$.

The group of operators \eqref{grG} preserves the self-adjointness of operators and the canonical
commutation relations which are determined the physical meaning of operators characterizing
particles and the structure of generator \eqref{dkomyt} of the Heisenberg equation \eqref{H-N1}.

On the space $\mathfrak{L}_{\gamma}(\mathcal{F}_\mathcal{H})$ for initial-value problem
\eqref{H-N1}-\eqref{H-N12} the following statement holds.

\begin{theorem}
A unique solution of initial-value problem \eqref{H-N1}-\eqref{H-N12} of the Heisenberg
equation is determined by the formula
\begin{eqnarray}\label{sH}
    &&A(t)=\mathcal{G}(t)A(0),
\end{eqnarray}
where the one-parameter family $\{\mathcal{G}(t)\}_{t\in \mathbb{R}}$ of operators is defined
by expression \eqref{grG}. For $A(0)\in\mathcal{D}(\mathcal{N})\subset\mathfrak{L}_{\gamma}
(\mathcal{F}_\mathcal{H})$, operator \eqref{sH} is a classical solution and for arbitrary initial
data $A(0)\in\mathfrak{L}_{\gamma}(\mathcal{F}_\mathcal{H})$, it is a generalized solution.
\end{theorem}

The average values of observables (mean values of observables) are defined by the positive
continuous linear functional on the space $\mathfrak{L}(\mathcal{F}_\mathcal{H})$
\begin{eqnarray}\label{averageD}
     &&\langle A\rangle(t)=(A(t),D(0))\doteq(I,D(0))^{-1}\sum\limits_{n=0}^{\infty}\frac{1}{n!}
         \,\mathrm{Tr}_{1,\ldots,n}\,A_{n}(t)\,D_{n}^0,
\end{eqnarray}
where $\mathrm{Tr}_{1,\ldots,n}$ are the partial traces over $1,\ldots,n$ particles,
$D(0)=(1,D_{1}^0,\ldots,D_{n}^0,\ldots)$ is a sequence of self-adjoint positive density operators
defined on the Fock space $\mathcal{F}_{\mathcal{H}}$ that describes the states of a quantum system
of a non-fixed number of particles and $(I,D(0))={\sum\limits}_{n=0}^{\infty}\frac{1}{n!}
\mathrm{Tr}_{1,\ldots,n}D_{n}^0$ is a normalizing factor (the grand canonical partition function). For
$D(0)\in\mathfrak{L}^{1}(\mathcal{F}_\mathcal{H})$ and $A(t)\in\mathfrak{L}(\mathcal{F}_\mathcal{H})$
average value functional \eqref{averageD} exists and determines a duality between observables and states.

We note that in case of a system of fixed number $N$ of particles the observables and states are
one-component sequences $A^{(N)}(t)=(0,\ldots,0,A_{N}(t),0,\ldots)$ and $D^{(N)}(0)=(0,\ldots,0,D_{N}^0,0,\ldots)$,
respectively, and therefore, average value formula \eqref{averageD} reduces to the functional
\begin{eqnarray*}
   &&\langle A^{(N)}\rangle(t)=(\mathrm{Tr}_{1,\ldots,N}D_{N}^0)^{-1}
     \mathrm{Tr}_{1,\ldots,N}A_{N}(t)D_{N}^0.
\end{eqnarray*}
It is usually assumed that the normalizing condition: $\mathrm{Tr}_{1,\ldots,N}D_{N}^0=1$, take place.

On the spaces $\mathfrak{L}(\mathcal{H}_{n}^{\pm})$ mapping \eqref{grG} is defined and have the similar
properties as in case of the Maxwell-Boltzmann statistics stated above.

Symmetrization and anti-symmetrization operators \eqref{Sn} are integrals of motion of the Heisenberg
equation \eqref{H-N1}, and as a consequence it holds
\begin{eqnarray*}
   &&\mathcal{G}_{n}(t)\mathcal{S}^{\pm}_{n}=\mathcal{S}^{\pm}_{n}\mathcal{G}_{n}(t),
\end{eqnarray*}
and
\begin{eqnarray*}
   &&\mathcal{N}_{n}\mathcal{S}^{\pm}_{n}=\mathcal{S}^{\pm}_{n}\mathcal{N}_{n},
\end{eqnarray*}
where the operator $\mathcal{N}_{n}$ is defined by \eqref{dkomyt}. Thus, the symmetry of
observables is preserved in process of the evolution.

\subsection{The von Neumann equation: the evolution of states}
As a consequence of the validity for functional \eqref{averageD} of the following equality
\begin{eqnarray*}
     &&(A(t),D(0))=(I,D(0))^{-1}\sum\limits_{n=0}^{\infty}\frac{1}{n!}
         \,\mathrm{Tr}_{1,\ldots,n}\,\mathcal{G}_{n}(t)A_{n}^0\,D_{n}^0=\\
     &&=(I,\mathcal{G}(-t)D(0))^{-1}\sum\limits_{n=0}^{\infty}\frac{1}{n!}\,
         \mathrm{Tr}_{1,\ldots,n}\,A_{n}^0\,\mathcal{G}_{n}(-t)D_{n}^0\equiv(I,D(t))^{-1}(A(0),D(t)),
\end{eqnarray*}
where $D(0)=(1,D_{1}^{0}(1),\ldots, D_{n}^{0}(1,\ldots,n),\ldots)\in\mathfrak{L}^{1}(\mathcal{F}_\mathcal{H})$,
it is possible to describe the evolution within the framework of the evolution of states.
Indeed, the evolution of all possible states, i.e. the sequence $D(t)=(1,D_{1}(t,1),\ldots, D_{n}(t,1,\ldots,n),\ldots)\in\mathfrak{L}^{1}(\mathcal{F}_\mathcal{H})$ of the density
operators $D_{n}(t),\, n\geq1$, is described by the initial-value problem of a sequence
of the von Neumann equations (the quantum Liouville equations) \cite{BQ,Pe95}
\begin{eqnarray}
  \label{vonNeumannEqn}
     &&\frac{d}{d t}D(t)=-\mathcal{N}D(t),\\
  \label{F-N12}
     &&D(t)|_{t=0}=D(0).
\end{eqnarray}
The generator $(-\mathcal{N})=\oplus^{\infty}_{n=0}(-\mathcal{N}_{n})$ of the von Neumann
equation \eqref{vonNeumannEqn} is the adjoint operator to generator \eqref{dkomyt} of the
Heisenberg equation \eqref{H-N1} in the sense of functional \eqref{averageD}, and for
$f\in \mathfrak{L}^{1}_{0}(\mathcal{F}_\mathcal{H})\subset\mathcal{D}(\mathcal{N})\subset
\mathfrak{L}^{1}(\mathcal{F}_\mathcal{H})$ it is defined by the formula
\begin{eqnarray}\label{infOper}
   &&(-\mathcal{N}_{n})f_{n}\doteq-i(H_{n}f_{n}-f_{n}H_{n}),
\end{eqnarray}
where the operator $H_{n}$ is the Hamiltonian \eqref{H}.

On the space of sequences of trace class operators $\mathfrak{L}^{1}(\mathcal{F}_\mathcal{H})$
for initial-value problem \eqref{vonNeumannEqn}-\eqref{F-N12} the following statement is true.

\begin{theorem}
A unique solution of initial-value problem \eqref{vonNeumannEqn}-\eqref{F-N12} of the von Neumann
equation is determined by the formula
\begin{eqnarray}\label{rozv_fon-N}
    &&D(t)=\mathcal{G}(-t)D(0),
\end{eqnarray}
where the one-parameter family of operators $\mathcal{G}(-t)=\oplus_{n=0}^{\infty}\mathcal{G}_{n}(-t)$,
is defined by
\begin{eqnarray}\label{groupG}
    &&\mathcal{G}_{n}(-t)f_{n}\doteq e^{-i t H_{n}}f_{n}e^{i t H_{n}}.
\end{eqnarray}
If $D(0)\in\mathfrak{L}^{1}_{0}(\mathcal{F}_\mathcal{H})\subset\mathfrak{L}^{1}(\mathcal{F}_\mathcal{H})$,
operator \eqref{rozv_fon-N} is a strong (classical) solution and for arbitrary
$D(0)\in\mathfrak{L}^{1}(\mathcal{F}_\mathcal{H})$ it is a weak (generalized) solution.
\end{theorem}

On the space $\mathfrak{L}^{1}(\mathcal{H}_{n})$ the mapping $t\rightarrow\mathcal{G}_{n}(-t)f_{n}$
is an isometric strongly continuous group that preserves positivity and self-adjointness of operators.
If $f_{n}\in\mathfrak{L}_{0}^{1}(\mathcal{H}_{n})\subset\mathcal{D}(-\mathcal{N}_{n})$, in the sense of
the norm convergence of the space $\mathfrak{L}^{1}(\mathcal{H}_{n})$ there exists the limit by which
the infinitesimal generator of the group of evolution operators \eqref{groupG} is determined by
\begin{eqnarray*}\label{infOperl}
    &&\lim\limits_{t\rightarrow 0}\frac{1}{t}(\mathcal{G}_{n}(-t)f_{n}-f_{n})=-i(H_{n}f_{n}-f_{n}H_{n}),
\end{eqnarray*}
where $H_{n}$ is the Hamiltonian \eqref{H} and operator \eqref{infOper} is defined on the domain
$\mathcal{D}(H_{n})\subset\mathcal{H}_{n}$.

On the spaces $\mathfrak{L}^{1}(\mathcal{H}_{n}^{\pm})$ mapping \eqref{groupG} is defined and have
the similar properties as in case of the Maxwell-Boltzmann statistics stated above.

Symmetrization and anti-symmetrization operators \eqref{Sn} are integrals of motion of
the von Neumann equation \eqref{vonNeumannEqn}, and as a consequence it holds
\begin{eqnarray*}
   &&\mathcal{G}_{n}(-t)\mathcal{S}^{\pm}_{n}=\mathcal{S}^{\pm}_{n}\mathcal{G}_{n}(-t),
\end{eqnarray*}
and
\begin{eqnarray*}
   &&(-\mathcal{N}_{n})\mathcal{S}^{\pm}_{n}=\mathcal{S}^{\pm}_{n}(-\mathcal{N}_{n}),
\end{eqnarray*}
where the operator $(-\mathcal{N}_{n})$ is defined by formula \eqref{infOper}. Thus, the symmetry
of states is preserved in process of the evolution.


\section{The evolution equations for quantum correlations}
One more an equivalent approach to the description of the evolution of states of quantum
many-particle systems is given by means of operators which are interpreted as correlation
operators. The correlation operators are governed by the von Neumann \cite{GerS},\cite{GP}.
To justify such hierarchy of quantum evolution equations, at first we consider in detail the
motivation of the description of states within the framework of correlation
operators or in other words, we consider the origin of the microscopic description of correlations
in quantum many-particle systems.

\subsection{Correlation operators}
We introduce a sequence of operators $g(t)=(0,g_{1}(t,1),\ldots,g_{s}(t,1,\ldots,s),\ldots)
\in\mathfrak{L}^{1}(\mathcal{F}_\mathcal{H})$ defined by the cluster expansions of the density
operators $D(t)=(1,D_{1}(t,1),\ldots,D_{s}(t,1,\ldots,s),$ $\ldots)\in\mathfrak{L}^{1}(\mathcal{F}_\mathcal{H})$
\begin{eqnarray}\label{D_(g)N}
    &&D_{s}(t,Y)= g_{s}(t,Y)+\sum\limits_{\mbox{\scriptsize $\begin{array}{c}\mathrm{P}:Y=
        \bigcup_{i}X_{i},\\|\mathrm{P}|>1 \end{array}$}}
        \prod_{X_i\subset \mathrm{P}}g_{|X_i|}(t,X_i),\quad s\geq1,
\end{eqnarray}
where ${\sum\limits}_{\mathrm{P}:Y=\bigcup_{i} X_{i},\,|\mathrm{P}|>1}$ is the sum over all possible
partitions $\mathrm{P}$ of the set $Y\equiv(1,\ldots,s)$ into $|\mathrm{P}|>1$ nonempty mutually
disjoint subsets $X_i\subset Y$. The operators defined by recursion relations \eqref{D_(g)N} are
interpreted as the correlation operators of quantum many-particle systems, i.e. operators that
characterize the correlations of particle states.

In order to construct a solution of recursion relations \eqref{D_(g)N}, i.e. to express the correlation
operators in terms of the density operators, we introduce some notions. On sequences of operators
$f,\widetilde{f}\in \mathfrak{L}^{1}(\mathcal{F}_\mathcal{H})$ we define the $\ast$-product
\begin{eqnarray}\label{Product}
    &&(f\ast\widetilde{f})_{|Y|}(Y)=\sum\limits_{Z\subset Y}\,f_{|Z|}(Z)
        \,\widetilde{f}_{|Y\backslash Z|}(Y\backslash Z),
\end{eqnarray}
where $\sum_{Z\subset Y}$ is the sum over all subsets $Z$ of the set $Y\equiv(1,\ldots,s)$.
By means of definition \eqref{Product} of the $\ast$-product we introduce the mapping
${\mathbb E}\mathrm{xp}_{\ast}$ and the inverse mapping ${\mathbb L}\mathrm{n}_{\ast}$
on sequences $h=(0,h_1(1),\ldots,h_n(1,\ldots,n),\ldots)$ of operators $h_n\in
\mathfrak{L}^{1}(\mathcal{H}_{n})$ by the expansions
\begin{eqnarray}\label{circledExp}
   &&({\mathbb E}\mathrm{xp}_{\ast}\,h )_{|Y|}(Y)=\big(\mathbb{I}+
      \sum\limits_{n=1}^{\infty} \frac{h^{\ast n}}{n!}\big)_{|Y|}(Y)=\\
   &&=I\delta_{|Y|,0}+\sum\limits_{\mathrm{P}:\,Y=\bigcup_{i}X_{i}}\,
      \prod_{X_i\subset \mathrm{P}}h_{|X_i|}(X_i),\nonumber
\end{eqnarray}
where we use the notations accepted in \eqref{D_(g)N}, $\delta_{|Y|,0}$ is a Kronecker symbol,
$\mathbb{I}=(1,0,\ldots,0,\ldots)$, and respectively,
\begin{eqnarray}\label{circledLn}
   &&({\mathbb L}\mathrm{n}_{\ast}(\mathbb{I}+h))_{|Y|}(Y)=
       \big(\sum\limits_{n=1}^{\infty} (-1)^{n-1}\,\frac{h^{\ast n}}{n}\big)_{|Y|}(Y)=\\
   &&=\sum\limits_{\mathrm{P}:\,Y=\bigcup_{i}X_{i}}(-1)^{|\mathrm{P}|-1}(|\mathrm{P}|-1)!\,
       \prod_{X_i\subset \mathrm{P}}h_{|X_i|}(X_i).\nonumber
\end{eqnarray}
Hence in terms of sequences of operators recursion relations \eqref{D_(g)N} are rewritten
in the form
\begin{eqnarray}\label{DtoGcircledStar}
    &&D(t)={\mathbb E}\mathrm{xp}_{\ast}\,\,g(t),
\end{eqnarray}
where $D(t)=\mathbb{I}+(0,D_1(t,1),\ldots,D_n(t,1,\ldots,n),\ldots)$. From this equation we obtain
\begin{eqnarray*}
    &&g(t)={\mathbb L}\mathrm{n}_{\ast}\,\,D(t).
\end{eqnarray*}

Thus, according to definition \eqref{Product} of the $\ast$-product and mapping \eqref{circledLn},
in the component-wise form solutions of recursion relations \eqref{D_(g)N} are represented by the
expansions
\begin{eqnarray}\label{gfromDFB}
   &&g_{s}(t,Y)=D_{s}(t,Y)+\sum\limits_{\mbox{\scriptsize $\begin{array}{c}\mathrm{P}:Y=
       \bigcup_{i}X_{i},\\|\mathrm{P}|>1\end{array}$}}(-1)^{|\mathrm{P}|-1}(|\mathrm{P}|-1)!\,
       \prod_{X_i\subset \mathrm{P}}D_{|X_i|}(t,X_i), \quad s\geq1.
\end{eqnarray}
The structure of expansions \eqref{gfromDFB} means that the correlation operators have a sense of
cumulants (semi-invariants) of the density operators governed by the Cauchy problem of the von Neumann
equations \eqref{vonNeumannEqn}-\eqref{F-N12}. Therefore correlation operators \eqref{gfromDFB} give
an alternative approach to the description of the state evolution of quantum many-particle systems,
namely within the framework of dynamics of correlations.

We emphasize that the possibility of the description of states within the framework of
correlations arises naturally as a result of dividing of the series in expression \eqref{averageD}
by the normalizing factor series, i.e. in consequence of redefining of functional \eqref{averageD}.

To justify this proposition, i.e. to construct the mean-value functional in terms of correlation
operators \eqref{gfromDFB}, we introduce necessary notions and formulate some equalities. For
arbitrary $f=(f_{0},f_{1},\ldots,f_{n},\ldots)\in\mathfrak{L}^{1}(\mathcal{F}_\mathcal{H})$ and
$Y\equiv(1,\ldots,s)$ we define the linear mapping $\mathfrak{d}_{Y}:f\rightarrow \mathfrak{d}_{Y}f$,
by the formula
\begin{eqnarray}\label{oper_d}
    &&(\mathfrak{d}_{Y} f)_{n}\doteq f_{|Y|+n}(Y,|Y|+1,\ldots,|Y|+n),\quad n\geq0.
\end{eqnarray}
For the set $\{Y\}$ consisting of one element $Y=(1,\ldots,s)$ we have respectively
\begin{eqnarray}\label{oper_c}
    &&(\mathfrak{d}_{\{Y\}} f)_{n}\doteq f_{1+n}(\{Y\},s+1,\ldots,s+n),\quad n\geq0.
\end{eqnarray}
On sequences $\mathfrak{d}_{Y}f$ and $\mathfrak{d}_{Y'}\widetilde{f}$ we introduce the $\ast$-product
\begin{eqnarray}\label{roundStarProduct}
    &&(\mathfrak{d}_{Y}f\ast\mathfrak{d}_{Y'}\widetilde{f})_{|X|}(X)\doteq
       \sum\limits_{Z\subset X}f_{|Z|+|Y|}(Y,Z)\,\widetilde{f}_{|X\backslash Z|+|Y'|}(Y',X\backslash Z),
\end{eqnarray}
where $X,Y,Y'$ are the sets, which elements characterize clusters of particles, and $\sum_{Z\subset X}$
is the sum over all subsets $Z$ of the set $X$. In particular case ($Y=\emptyset$, $Y'=\emptyset$)
definition \eqref{roundStarProduct} reduces to \eqref{Product}.

For $f=(0,f_{1},\ldots,f_{n},\ldots)$,  $f_{n}\in\mathfrak{L}^{1}(\mathcal{H}_{n})$, according
to definitions of mappings \eqref{circledExp} and \eqref{oper_c}, the following equality holds
\begin{eqnarray}\label{d_gamma}
    &&\mathfrak{d}_{\{Y\}}\mathbb{E}\mathrm{xp}_{\ast}f=
      \mathbb{E}\mathrm{xp}_{\ast}f\ast\mathfrak{d}_{\{Y\}}f,
\end{eqnarray}
and for mapping \eqref{oper_d} correspondingly,
\begin{eqnarray}\label{oper_dex}
    &&\mathfrak{d}_{Y}\mathbb{E}\mathrm{xp}_{\ast}f=
       \mathbb{E}\mathrm{xp}_{\ast} f\ast\sum\limits_{\mathrm{P}:\,Y=\bigcup_i X_{i}}
       \mathfrak{d}_{X_1}f\ast\ldots\ast \mathfrak{d}_{X_{|\mathrm{P}|}}f,
\end{eqnarray}
where ${\sum\limits}_{\mathrm{P}:\,Y=\bigcup_i X_{i}}$ is the sum over all possible partitions
$\mathrm{P}$ of the set $Y\equiv(1,\ldots,s)$ into $|\mathrm{P}|$ nonempty mutually disjoint
subsets $X_i\subset Y$.

According to the definition
\begin{eqnarray*}
    &&(I,f)\doteq\sum\limits_{n=0}^{\infty}\frac{1}{n!}\mathrm{Tr}_{1,\ldots,{n}}f_{n},
\end{eqnarray*}
for sequences $f,\widetilde{f}\in\mathfrak{L}^{1}(\mathcal{F}_\mathcal{H})$, the equality holds
\begin{eqnarray}\label{efg}
    &&(I,f\ast \widetilde{f})=(I,f)(I,\widetilde{f}).
\end{eqnarray}

In terms of mappings \eqref{oper_d} and \eqref{oper_c} the generalized cluster expansions of solution
\eqref{rozv_fon-N} of the von Neumann equation
\begin{eqnarray}\label{DClusters}
    &&D_{s+n}(t,Y,\,X\setminus Y)=\sum\limits_{\mathrm{P}:(\{Y\},\,X\setminus Y)=\bigcup_i X_i}
      \prod_{X_i\subset \mathrm{P}}g(t,X_i),\quad s\geq1,
\end{eqnarray}
where $X\setminus Y\equiv(s+1,\ldots,s+n)$, take the form
\begin{eqnarray}\label{gcea}
    &&\mathfrak{d}_{Y}D(t)=\mathfrak{d}_{\{Y\}}{\mathbb E}\mathrm{xp}_{\ast}\,\,g(t).
\end{eqnarray}

We now deduce from the definition of functional \eqref{averageD} the functional for average
values of observables in terms of correlation operators \eqref{gfromDFB}, for example,
for $s$-ary observables $A^{(s)}=(0,\ldots,0,a_{s}(1,\ldots,s),
\ldots,\sum_{i_{1}<\ldots<i_{s}=1}^{n}a_s(i_{1},\ldots,$ $i_{s}),\ldots)$, i.e.
\begin{eqnarray*}
    &&\langle A^{(s)}\rangle(t)=\frac{1}{s!}\,(I,D(t))^{-1}\,
        \mathrm{Tr}_{Y}\,a_{s}(Y)(I,\mathfrak{d}_{Y}D(t)),
\end{eqnarray*}
where $\mathrm{Tr}_{Y}\equiv\mathrm{Tr}_{1,\ldots,s}$. Using generalized cluster expansions
\eqref{gcea} and as a consequence of equalities \eqref{d_gamma} and \eqref{efg}, we find
\begin{eqnarray*}
    &&(I,D(t))^{-1}(I,\mathfrak{d}_{Y}D(t))=(I,D(t))^{-1}
       (I,\mathfrak{d}_{\{Y\}}{\mathbb E}\mathrm{xp}_{\ast}\,\,g(t))=\\
    &&=(I,D(t))^{-1}(I,\mathbb{E}\mathrm{xp}_{\ast}g(t)\ast\mathfrak{d}_{\{Y\}}g(t))
       =(I,D(t))^{-1}(I,\mathbb{E}\mathrm{xp}_{\ast}g(t))(I,\mathfrak{d}_{\{Y\}}g(t)).
\end{eqnarray*}
According to generalized cluster expansions \eqref{gcea}, as a final result we derive the following
representation of the mean-value functional in case of $s$-ary observables
\begin{eqnarray*}
    &&\langle A^{(s)}\rangle(t)=\frac{1}{s!}\,\mathrm{Tr}_{Y}\,a_{s}(Y)(I,\mathfrak{d}_{\{Y\}}g(t)),
\end{eqnarray*}
or in the componentwise form
\begin{eqnarray}\label{averagegs}
    &&\langle A^{(s)}\rangle(t)=\frac{1}{s!}\sum\limits_{n=0}^{\infty}\frac{1}{n!}
        \,\mathrm{Tr}_{1,\ldots,s+n}\,a_{s}(Y)g_{1+n}(t,\{Y\},s+1,\ldots,s+n),
\end{eqnarray}
where the correlation operators of particle clusters $g_{1+n}(t)$ are defined as solutions
of generalized cluster expansions \eqref{gcea}, namely
\begin{eqnarray}\label{gClusters}
    &&g_{1+n}(t,\{Y\},X\setminus Y)=\sum\limits_{\mathrm{P}:(\{Y\},\,X\setminus Y)=\bigcup_i X_i}
      (-1)^{|\mathrm{P}|-1}(|\mathrm{P}| -1)!\,\prod_{X_i\subset \mathrm{P}}D(t,X_i),\quad s\geq1,
\end{eqnarray}
where the density operator $D(t,X_i)$ is solution \eqref{rozv_fon-N} of the von Neumann equation
\eqref{vonNeumannEqn}. For $A^{(s)}\in\mathfrak{L}(\mathcal{F}_\mathcal{H})$ and $g^{(s)}\in
\mathfrak{L}^{1}(\mathcal{F}_\mathcal{H})$ functional \eqref{averagegs} exists.

It should be emphasized that correlation operators that belong to the spaces
$\mathfrak{L}^{1}(\mathcal{F}_\mathcal{H})$ describe only finitely many particles, i.e. systems
with finite average number of particles.

\subsection{The von Neumann hierarchy}
We consider quantum systems of particles obeying the Maxwell-Boltzmann
statistics with the Hamiltonian
\begin{eqnarray}\label{Hn}
    &&H_{n}=\sum\limits_{i=1}^{n}K(i)
        +\sum\limits_{k=1}^{n}\sum\limits_{i_{1}<\ldots<i_{k}=1}^{n}
        \Phi^{(k)}(i_{1},\ldots,i_{k}),
\end{eqnarray}
where $\Phi^{(k)}$ is a $k$-body interaction potential. The evolution of all possible states is described
in terms of the correlation operators
$g(t)=(0,g_{1}(t,1),$ $\ldots,g_{s}(t,1,\ldots,s),\ldots)\in\mathfrak{L}^{1}(\mathcal{F}_\mathcal{H})$
governed by the following Cauchy problem
\begin{eqnarray}
 \label{vNh}
   &&\frac{d}{dt}g_{s}(t,Y)=-\mathcal{N}_{s}(Y)g_{s}(t,Y)+\\
   &&+\sum\limits_{\mbox{\scriptsize $\begin{array}{c}\mathrm{P}:Y=\bigcup_{i}X_{i},\\
     |\mathrm{P}|>1\end{array}$}}\hskip-2mm
     \sum\limits_{\mbox{\scriptsize$\begin{array}{c}{Z_{1}\subset X_{1}},\\Z_{1}\neq\emptyset\end{array}$}}
     \ldots\sum\limits_{\mbox{\scriptsize $\begin{array}{c} {Z_{|\mathrm{P}|}\subset X_{|\mathrm{P}|}},\\
     Z_{|\mathrm{P}|}\neq\emptyset \end{array}$}}
     \hskip-1mm \big(-\mathcal{N}_{\mathrm{int}}^{(\sum\limits_{r=1}^{|\mathrm{P}|}|Z_{{r}}|)}
     (Z_{{1}},\ldots,Z_{{|\mathrm{P}|}})\big)\prod_{X_{i}\subset \mathrm{P}}g_{|X_{i}|}(t,X_{i}),\nonumber\\
     \nonumber\\
 \label{vNhi}
   &&g_{s}(t,Y)\big|_{t=0}=g_{s}^0(Y),\quad s\geq1,
\end{eqnarray}
where the operator $(-\mathcal{N}^{(n)}_{\mathrm{int}})$ is defined by
\begin{eqnarray}\label{Nintn}
   &&(-\mathcal{N}^{(n)}_{\mathrm{int}})f_{n}\doteq
      -i(\Phi^{(n)}f_{n}-f_{n}\Phi^{(n)}),
\end{eqnarray}
${\sum\limits}_{\mbox{\scriptsize $\begin{array}{c}\mathrm{P}:Y=\bigcup_{i} X_{i},\,|\mathrm{P}|>1\end{array}$}}$
is the sum over all possible partitions $\mathrm{P}$ of the set $Y\equiv(1,\ldots,s)$ into $|\mathrm{P}|>1$
nonempty mutually disjoint subsets $X_i\subset Y$ and $\sum_{Z_{j}\subset X_{j},\,Z_{j}\neq\emptyset}$
is a sum over nonempty subsets $Z_{j}\subset X_{j}$. We refer to the hierarchy of equations \eqref{vNh}
as the von Neumann hierarchy \cite{GerS}.

It should be noted that the von Neumann hierarchy \eqref{vNh} is the evolution recurrence equations set.
We cite an instance of the typical equations of hierarchy \eqref{vNh}
\begin{eqnarray*}
    &&\frac{d}{dt}g_{1}(t,1)=-\mathcal{N}(1)g_{1}(t,1),\\
    &&\frac{d}{dt}g_{2}(t,1,2)=-\mathcal{N}_{2}(1,2)g_{2}(t,1,2)-
       \mathcal{N}_{\mathrm{\mathrm{int}}}^{(2)}(1,2)g_{1}(t,1)g_{1}(t,2),
\end{eqnarray*}
\begin{eqnarray*}
    &&\frac{d}{dt}g_{3}(t,1,2,3)=-\mathcal{N}_{3}(1,2,3)g_{3}(t,1,2,3)-\\
    &&\quad-\big(\mathcal{N}_{\mathrm{int}}^{(2)}(1,2)+\mathcal{N}_{\mathrm{int}}^{(2)}(1,3)
       +\mathcal{N}_{\mathrm{int}}^{(3)}(1,2,3)\big)g_{1}(t,1)g_{2}(t,2,3)-\\
    &&\quad-\big(\mathcal{N}_{\mathrm{int}}^{(2)}(1,2)+\mathcal{N}_{\mathrm{int}}^{(2)}(2,3)+
       \mathcal{N}_{\mathrm{int}}^{(3)}(1,2,3)\big)g_{1}(t,2)g_{2}(t,1,3)-\\
    &&\quad-\big(\mathcal{N}_{\mathrm{int}}^{(2)}(1,3)+\mathcal{N}_{\mathrm{int}}^{(2)}(2,3)+
       \mathcal{N}_{\mathrm{int}}^{(3)}(1,2,3)\big)g_{1}(t,3)g_{2}(t,1,2)-\\
    &&\quad-\mathcal{N}_{\mathrm{int}}^{(3)}(1,2,3)(1,2,3)g_{1}(t,1)g_{1}(t,2)g_{1}(t,3).
\end{eqnarray*}
In case of a two-body interaction potential \eqref{H} the von Neumann hierarchy \eqref{vNh}
reduces to the following form
\begin{eqnarray}\label{vonNeumannTwoBody}
   &&\frac{d}{dt}g_{s}(t,Y)=-\mathcal{N}_{s}(Y)g_s(t,Y)+\\
   &&+\sum\limits_{\mathrm{P}:\,Y=X_{1}\bigcup X_2}\,\sum\limits_{i_{1}\in X_{1}}\sum\limits_{i_{2}\in X_{2}}
       \big(-\mathcal{N}_{\mathrm{int}}(i_{1},i_{2})\big)
       g_{|X_{1}|}(t,X_{1})g_{|X_{2}|}(t,X_{2}),\quad s\geq1,\nonumber
\end{eqnarray}
where ${\sum\limits}_{\mathrm{P}:\,Y=X_{1}\bigcup X_2}$ is the sum over all possible partitions $\mathrm{P}$
of the set $Y\equiv(1,\ldots,s)$ into two nonempty mutually disjoint subsets $X_1\subset Y$ and $X_2\subset Y$.
In terms of kernels of correlation operators the first two equations of the von Neumann hierarchy
\eqref{vonNeumannTwoBody} have the form
\begin{eqnarray*}
   &&i\frac{\partial}{\partial t}g_1(t,q_1;q'_1)=-\frac{1}{2}(\Delta_{q_{1}}-\Delta_{q^{'}_{1}})g_1(t,q_1;q'_1),\\
   &&i\frac{\partial}{\partial t}g_2(t,q_1,q_2;q'_1,q'_2)= \\
   &&=\big(-\frac{1}{2}\sum_{i=1}^2(\Delta_{q_{i}}-
      \Delta_{q^{'}_{i}})+(\Phi(q_1-q_2)-\Phi(q'_1-q'_2))\big)g_2(t,q_1,q_2;q'_1,q'_2)+\\
   &&+\big(\Phi(q_1-q_2)-\Phi(q'_1-q'_2)\big)g_1(t,q_1;q'_1)g_1(t,q_2;q'_2).
\end{eqnarray*}
We remark that the von Neumann hierarchy \eqref{vNh} (or \eqref{vonNeumannTwoBody}) is rigorously
derived on basis of solutions \eqref{gfromDFB} of cluster expansions \eqref{D_(g)N} of density operators
governed by the von Neumann equation \eqref{vonNeumannEqn}.

To construct a solution of the Cauchy problem \eqref{vNh}-\eqref{vNhi} we first consider its structure
for physically motivated example of initial data, namely initial data satisfying a chaos property.
A chaos property means the absence of state correlations in a system at the initial time. In this case
the sequence of initial correlation operators is the following one-component sequence
\begin{eqnarray}\label{posl_g(0)}
    &&g(0)=(0,g_{1}^0(1),0,\ldots).
\end{eqnarray}
In fact, in terms of the sequence of the density operators this condition means that $D(0)=
(1,D_{1}^0(1),D_{1}^0(1)D_{1}^0(2),\ldots,\prod^n_{i=1} D_{1}^0(i),\ldots)$ in case of
the Maxwell-Boltzmann statistics, then from \eqref{gfromDFB} we derive \eqref{posl_g(0)}.

On basis of representation \eqref{gfromDFB} of the correlation operators
in terms of the density operators and formula \eqref{rozv_fon-N} we have
\begin{eqnarray*}
   &&g_{s}(t,Y)=\sum\limits_{\mbox{\scriptsize $\begin{array}{c}\mathrm{P}:Y=
       \bigcup_{i}X_{i}\end{array}$}}\hskip-1mm
       (-1)^{|\mathrm{P}|-1}(|\mathrm{P}|-1)!\,\prod_{X_i\subset \mathrm{P}}\mathcal{G}_{|X_{i}|}(-t,X_{i})
       \prod_{i=1}^{s}D_{1}^0(i), \quad s\geq1.
\end{eqnarray*}
Taking into account the equality: $D_{1}^0(i)=g_{1}^0(i),\,1\leq i\leq s$, we derive the formula
of a solution of the Cauchy problem \eqref{vNh}-\eqref{vNhi} for initial data \eqref{posl_g(0)}
\begin{eqnarray}\label{rozvChaosN}
    &&g_{s}(t,Y)=\mathfrak{A}_{s}(-t,Y)\,\prod_{i=1}^{s}g_{1}^0(i), \quad s\geq1,
\end{eqnarray}
where $\mathfrak{A}_{s}(-t,Y)$ is the $sth$-order cumulant of the groups of operators \eqref{groupG} of
the von Neumann equations \eqref{vonNeumannEqn}
\begin{eqnarray}\label{cumulants}
    &&\mathfrak{A}_{s}(-t,Y)\doteq\sum\limits_{\mathrm{P}:\,Y=
       \bigcup_i X_i}(-1)^{|\mathrm{P}|-1}({|\mathrm{P}|-1})!
       \prod\limits_{X_i\subset\mathrm{P}}\mathcal{G}_{|X_{i}|}(-t,X_{i}),
\end{eqnarray}
and we use accepted above notations. It should be emphasized that the structure of expansions
\eqref{cumulants} of the evolution operators $\mathfrak{A}_{s}(t,Y)$ means that they are determined by
the cluster expansions of the groups of operators \eqref{groupG} of the von Neumann equation \eqref{vonNeumannEqn}.

Thus, the cumulant nature of correlation operators induces the cumulant structure of a one-parametric
mapping generated by solution \eqref{rozvChaosN}. From \eqref{rozvChaosN} it is clear that in case of
absence of correlations in a system at initial instant the correlations generated by the dynamics of
a system are completely governed by cumulants \eqref{cumulants} of groups \eqref{groupG}.

Let us indicate some properties of a cumulant (semi-invariant) \eqref{cumulants} of groups of operators
\eqref{groupG}. The corresponding-order cumulant is a solution of the cluster expansions of groups of
operators $\mathcal{G}_s(-t)$ \eqref{groupG}
\begin{eqnarray}\label{groupKlast}
   &&\mathcal{G}_{s}(-t,Y)=\sum\limits_{\mathrm{P}\,:\,Y ={\bigcup_i} X_i}\,
       \prod\limits_{X_i\subset \mathrm{P}}\mathfrak{A}_{|X_i|}(-t,X_i), \quad s\geq1,
\end{eqnarray}
where ${\sum}_\mathrm{P}$ is the sum over all possible partitions $\mathrm{P}$ of the set
$Y\equiv(1,\ldots,s)$ into $|\mathrm{P}|$ nonempty mutually disjoint subsets $X_i\subset Y$.
The simplest examples of equations from recurrence relations \eqref{groupKlast} are given by
the expressions
\begin{eqnarray*}
    &&\mathcal{G}_{1}(-t,1)=\mathfrak{A}_{1}(-t,1),\\
    &&\mathcal{G}_{2}(-t,1,2)=\mathfrak{A}_{2}(-t,1,2)+\mathfrak{A}_{1}(-t,1)\mathfrak{A}_{1}(-t,2).
\end{eqnarray*}

In case of a quantum system of non-interacting particles we have: $\mathfrak{A}_{s}(-t)=0,\,s\geq 2$.
Indeed, for non-interacting quantum particles it holds
\begin{eqnarray*}
   &&\mathcal{G}_{s}(-t,1,\ldots,s)=\prod_{i=1}^{s}\mathcal{G}_{1}(-t,i),
\end{eqnarray*}
and hence,
\begin{eqnarray*}
    &&\mathfrak{A}_{s}(-t,Y)=\sum\limits_{\mathrm{P}:\,Y={\bigcup}_i X_i}
       (-1)^{|\mathrm{P}|-1}(|\mathrm{P}|-1)!\prod\limits_{X_{i}\subset
       \mathrm{P}}\prod_{l_i=1}^{|X_{i}|}\mathcal{G}_{1}(-t,l_i)=\\
    &&=\sum\limits_{k=1}^{s}(-1)^{k-1}\mathrm{s}(s,k)(k-1)!\prod_{i=1}^{s}\mathcal{G}_{1}(-t,i)=0.
\end{eqnarray*}
Here $\mathrm{s}(s,k)$ are the Stirling numbers of the second kind and the following equality is used
\begin{eqnarray}\label{Stirl}
    &&\sum\limits_{\mathrm{P}:\,Y={\bigcup}_i X_i}
        (-1)^{| \mathrm{P}|-1}(|\mathrm{P}|-1)!=
        \sum\limits_{k=1}^{s}(-1)^{k-1}\mathrm{s}(s,k)(k-1)!=\delta_{s,1},
\end{eqnarray}
where $\delta_{s,1}$ is a Kroneker symbol.

If $s=1$, for $f_{1}\in\mathfrak{L}_0^{1}(\mathcal{H})\subset\mathfrak{L}^{1}(\mathcal{H})$
in the sense of the norm convergence of the space $\mathfrak{L}^{1}(\mathcal{H})$ the generator
of first-order cumulant \eqref{cumulants} is given by operator \eqref{infOper}
\begin{eqnarray*}
    &&\lim\limits_{t\rightarrow 0}\frac{1}{t}(\mathfrak{A}_{1}(-t,1)-I)f_{1}=
       -\mathcal{N}_{1}f_{1}.
\end{eqnarray*}
For $s\geq2$ we obtain the following equality in the same sense
\begin{eqnarray*}
   &&\lim\limits_{t\rightarrow 0}\frac{1}{t}\,\mathfrak{A}_{s}(-t,1,\ldots,s)f_{s}=
      -\mathcal{N}_{\mathrm{int}}^{(s)}(1,\ldots,s)f_{s},
\end{eqnarray*}
where the operator $(-\mathcal{N}_{\mathrm{int}}^{(s)})$ is defined by formula \eqref{Nintn}.

If $f_{s}\in\mathfrak{L}^{1}(\mathcal{H}_{s})$, then for $sth$-order cumulant \eqref{cumulants}
of groups of operators \eqref{groupG} the estimate is valid
\begin{eqnarray}\label{est}
   &&\big\|\mathfrak{A}_{s}(-t)f_{s}\big\|_{\mathfrak{L}^{1}(\mathcal{H}_{s})}
      \leq\sum\limits_{\mathrm{P}:\,Y={\bigcup}_i X_i}
      (|\mathrm{P}|-1)!\big\|f_{s}\big\|_{\mathfrak{L}^{1}(\mathcal{H}_{s})}\leq \\
   &&\leq\sum\limits_{k=1}^{s}\mathrm{s}(s,k)(k-1)!\big\|f_{s}\big\|_{\mathfrak{L}^{1}(\mathcal{H}_{s})}
      \leq s!e^{s}\big\|f_{s}\big\|_{\mathfrak{L}^{1}(\mathcal{H}_{s})},\nonumber
\end{eqnarray}
where $\mathrm{s}(s,k)$ are the Stirling numbers of the second kind.

Hereafter we use the following notations: $(\{X_1\},\ldots,\{X_{|\mathrm{P}|}\})$ is a set, elements of
which are $|\mathrm{P}|$ mutually disjoint subsets $X_i\subset Y\equiv(1,\ldots,s)$ of the partition
$\mathrm{P}:Y=\cup_{i=1}^{|\mathrm{P}|}X_i$, i.e. $|(\{X_1\},\ldots,\{X_{|\mathrm{P}|}\})|=|\mathrm{P}|$.
In view of these notations we state that $\{Y\}$ is the set consisting of one element $Y=(1,\ldots,s)$
of the partition $\mathrm{P}$ $(|\mathrm{P}|=1)$ and $|\{Y\}|=1$. We define the declasterization mapping
$\theta: (\{X_1\},\ldots,\{X_{|\mathrm{P}|}\})\rightarrow Y$, by the formula
\begin{eqnarray}\label{Theta}
   &&\theta(\{X_1\},\ldots,\{X_{|\mathrm{P}|}\})=Y.
\end{eqnarray}

For arbitrary initial data a solution of the Cauchy problem \eqref{vNh}-\eqref{vNhi} of the von Neumann
hierarchy is given by the expansion \cite{GerS}
\begin{eqnarray}\label{rozvNF-N_F}
    &&g_{s}(t,Y)=\sum\limits_{\mathrm{P}:\,Y=\bigcup_i X_i}
        \mathfrak{A}_{|\mathrm{P}|}(-t,\{X_1\},\ldots,\{X_{|\mathrm{P}|}\})
        \prod_{X_i\subset \mathrm{P}}g_{|X_i|}^0(X_i),\quad s\geq1.
\end{eqnarray}
Here $\mathfrak{A}_{|\mathrm{P}|}(-t)$ is the $|\mathrm{P}|th$-order cumulant of groups of operators
\eqref{groupG} defined by the formula
\begin{eqnarray} \label{cumulantP}
   &&\mathfrak{A}_{|\mathrm{P}|}(-t,\{X_1\},\ldots,\{X_{|\mathrm{P}|}\})\doteq\\
   &&\doteq\sum\limits_{\mathrm{P}^{'}:\,(\{X_1\},\ldots,\{X_{|\mathrm{P}|}\})=
      \bigcup_k Z_k}(-1)^{|\mathrm{P}^{'}|-1}({|\mathrm{P}^{'}|-1})!
      \prod\limits_{Z_k\subset\mathrm{P}^{'}}\mathcal{G}_{|\theta(Z_{k})|}(-t,\theta(Z_{k})),\nonumber
\end{eqnarray}
where $\sum_{\mathrm{P}^{'}:\,(\{X_1\},\ldots,\{X_{|\mathrm{P}|}\})=\bigcup_k Z_k}$
is the sum over all possible partitions $\mathrm{P}^{'}$ of the set $(\{X_1\},\ldots,$ $\{X_{|\mathrm{P}|}\})$
into $|\mathrm{P}^{'}|$ nonempty mutually disjoint subsets $Z_k\subset (\{X_1\},\ldots,$ $\{X_{|\mathrm{P}|}\})$.
The simplest examples of correlation operators \eqref{rozvNF-N_F} are given by the expressions
\begin{eqnarray*}
    &&g_{1}(t,1)=\mathfrak{A}_{1}(-t,1)g_{1}^0(1),\\
    &&g_{2}(t,1,2)=\mathfrak{A}_{1}(-t,\{1,2\})g_{2}^0(1,2)+
       \mathfrak{A}_{2}(-t,1,2)g_{1}^0(1)g_{1}^0(2),\\
    &&g_{3}(t,1,2,3)=\mathfrak{A}_{1}(-t,\{1,2,3\})g_{3}^0(1,2,3)+
       \mathfrak{A}_{2}(-t,\{2,3\},1)g_{1}^0(1)g_{2}^0(2,3)+\\
    &&\qquad+\mathfrak{A}_{2}(-t,\{1,3\},2)g_{1}^0(2)g_{2}^0(1,3)+
       \mathfrak{A}_{2}(-t,\{1,2\},3)g_{1}^0(3)g_{2}^0(1,2)+\\
    &&\qquad+\mathfrak{A}_{3}(-t,1,2,3)g_{1}^0(1)g_{1}^0(2)g_{1}^0(3).
\end{eqnarray*}

The validity of solution expansion \eqref{rozvNF-N_F} can be verified by straightforward differentiation
by time variable and also in the following way. Taking into account the fact that the von Neumann hierarchy
\eqref{vNh} is the evolution recurrence equations set, we can construct a solution of initial-value problem
\eqref{vNh}-\eqref{vNhi} by integrating each equation of the hierarchy as the inhomogeneous von Neumann equation.
For example, using formula \eqref{groupG}, as a result of the integration of the first two equations of hierarchy
\eqref{vNh} we obtain the following equalities
\begin{eqnarray*}
    &&g_{1}(t,1)=\mathcal{G}_{1}(-t,1)g_{1}^0(1),\\
    &&g_{2}(t,1,2)=\mathcal{G}_{2}(-t,1,2)g_{2}^0(1,2)+\\
    &&+\int\limits_{0}^{t}dt_{1}\mathcal{G}_{2}(-t+t_{1},1,2)(-\mathcal{N}_{\mathrm{int}}(1,2))
       \mathcal{G}_{1}(-t_{1},1)\mathcal{G}_{1}(-t_{1},2)g_{1}^0(1)g_{1}^0(2).
\end{eqnarray*}
Then for the second term on the right-hand side of this equation an analog of the Duhamel equation holds
\begin{eqnarray}\label{iter2kum}
    &&\int\limits_{0}^{t}dt_{1}\mathcal{G}_{2}(-t+t_{1},1,2)
       (-\mathcal{N}_{\mathrm{int}}(1,2))\mathcal{G}_{1}(-t_{1},1)\mathcal{G}_{1}(-t_{1},2)=\\
    &&=-\mathcal{G}_{2}(-t,1,2)\int\limits_{0}^{t}dt_{1}\frac{d}{dt_{1}}\big(\mathcal{G}_{2}(t_{1},1,2)
       \mathcal{G}_{1}(-t_{1},1)\mathcal{G}_{1}(-t_{1},2)\big)=\nonumber\\
    &&=\mathcal{G}_{2}(-t,1,2)-\mathcal{G}_{1}(-t,1)\mathcal{G}_{1}(-t,2)=\mathfrak{A}_{2}(-t,1,2),\nonumber
\end{eqnarray}
where $\mathfrak{A}_{2}(-t)$ is the second-order cumulant of groups of operators \eqref{groupG} defined by
formula \eqref{cumulants}. For $s>2$ a solution of the Cauchy problem \eqref{vNh}-\eqref{vNhi}
constructed by iterations is represented by expansions \eqref{rozvNF-N_F} as a consequence of
transformations similar to an analog of the Duhamel equation \eqref{iter2kum}.

We note, that in case of initial data \eqref{posl_g(0)} solution \eqref{rozvChaosN} of the Cauchy problem
\eqref{vNh}-\eqref{vNhi} of the von Neumann hierarchy may be rewritten in another representation.
For $n=1$, we have
\begin{eqnarray*}
     &&g_{1}(t,1)=\mathfrak{A}_{1}(-t,1)g_{1}^0(1).
 \end{eqnarray*}
Then, within the context of the definition of the first-order cumulant, $\mathfrak{A}_{1}(-t)$,
and the dual group of operators $\mathfrak{A}_{1}(t)$, we express the correlation operators
$g_{s}(t), \,s\geq 2$, in terms of the one-particle correlation operator $g_{1}(t)$ using formula
\eqref{rozvChaosN}. Hence for $s\geq2$ formula \eqref{rozvChaosN} is represented in the form of
the functional with respect to one-particle correlation operators
\begin{eqnarray*}
     &&g_{s}(t,Y\mid g_{1}(t))=\widehat{\mathfrak{A}}_{s}(t,Y)\,\prod_{i=1}^{s}\,g_{1}(t,i),\quad s\geq 2,
\end{eqnarray*}
where $\widehat{\mathfrak{A}}_{s}(t,Y)$ is $sth$-order cumulant \eqref{cumulants} of the scattering
operators
\begin{eqnarray}\label{so}
     &&\widehat{\mathcal{G}}_{s}(t,Y)\doteq
        \mathcal{G}_{s}(-t,Y)\prod_{i=1}^{s}\mathcal{G}_{1}(t,i),\quad s\geq1.
\end{eqnarray}
The generator of the scattering operator $\widehat{\mathcal{G}}_{t}(Y)$ is determined by the operator
\begin{eqnarray*}
  &&\frac{d}{dt}\widehat{\mathcal{G}}_{s}(t,Y)|_{t=0}=\sum\limits_{k=2}^{s}\,\,
     \sum\limits_{i_{1}<\ldots<i_{k}=1}^{s}\,(-\mathcal{N}_{\mathrm{int}}^{(k)}(i_{1},\ldots,i_{k})),
\end{eqnarray*}
where the operator $(-\mathcal{N}_{\mathrm{int}}^{(k)})$ acts on $\mathfrak{L}^{1}_{0}
(\mathcal{H}_{s})\subset\mathfrak{L}^{1}(\mathcal{H}_{s})$ according to formula \eqref{Nintn}.

On the space $\mathfrak{L}^{1}(\mathcal{F}_\mathcal{H})$ for initial-value problem \eqref{vNh}-\eqref{vNhi}
of the von Neumann hierarchy the following statement is true \cite{GerS},\cite{GP}.

\begin{theorem}
For $t\in\mathbb{R}$ a solution of the Cauchy problem \eqref{vNh}-\eqref{vNhi} is given
by the following expansion
\begin{eqnarray}\label{rozvNh}
    &&g_{s}(t,Y)=\mathcal{G}(t;Y|g(0))\doteq\\
    &&\doteq\sum\limits_{\mathrm{P}:\,Y=\bigcup_i X_i}
      \mathfrak{A}_{|\mathrm{P}|}(-t,\{X_1\},\ldots,\{X_{|\mathrm{P}|}\})
      \prod_{X_i\subset \mathrm{P}}g_{|X_i|}^0(X_i),\quad s\geq1,\nonumber
\end{eqnarray}
where $\mathfrak{A}_{|\mathrm{P}|}(-t)$ is the $|\mathrm{P}|th$-order cumulant \eqref{cumulantP}
of the groups of operators. For $g_{n}^0\in \mathfrak{L}^{1}_{0}(\mathcal{H}_{n})\subset
\mathfrak{L}^{1}(\mathcal{H}_{n}),\,n\geq1$, expansion \eqref{rozvNh} is a strong (classical)
solution and for arbitrary initial data $g_{n}^0\in\mathfrak{L}^{1}(\mathcal{H}_{n}),\,n\geq1$,
it is a weak (generalized) solution.
\end{theorem}

The intrinsic properties of constructed solution \eqref{rozvNh} are generated by the properties
of cumulants \eqref{cumulantP} of groups of operators of the von Neumann equations. We indicate
some properties of the one-parameter mapping: $t\rightarrow \mathcal{G}(t|f)$, generated by
solution \eqref{rozvNh}.

For $f_{s}\in\mathfrak{L}^{1}(\mathcal{H}_{s}),\,s\geq1$, the mapping $\mathcal{G}(t;Y|f)$
is defined and, according to the inequality
\begin{eqnarray*}
   &&\big\|\mathfrak{A}_{|\mathrm{P}|}(-t,\{X_1\},\ldots,
      \{X_{|\mathrm{P}|}\})f_s\big\|_{\mathfrak{L}^{1}(\mathcal{H}_{s})}
      \leq |\mathrm{P}|!\,e^{|\mathrm{P}|}\big\|f_s\big\|_{\mathfrak{L}^{1}(\mathcal{H}_{s})},
\end{eqnarray*}
the following estimate is true
\begin{eqnarray}\label{gEstimate}
   &&\big\|\mathcal{G}(t;Y|f)\big\|_{\mathfrak{L}^{1}(\mathcal{H}_{s})}\leq s!e^{2s}c^{s},
\end{eqnarray}
where $c\equiv e^{3}\max(1,\max_{\mathrm{P}:\,Y=\bigcup_iX_i}\|f_{|X_{i}|}\|_{\mathfrak{L}^{1}(\mathcal{H}_{|X_{i}|})})$.
The mapping $\mathcal{G}(t|f)$ has the group property, i.e. for $f\in\mathfrak{L}^{1}(\mathcal{F}_\mathcal{H})$
the equalities are fulfilled
\begin{eqnarray*}\label{ocinka}
   &&\mathcal{G}(t_1+t_2|f)=\mathcal{G}(t_1|\mathcal{G}(t_2|f))=\mathcal{G}(t_2|\mathcal{G}(t_1|f)).
\end{eqnarray*}
On the subspaces $\mathfrak{L}^{1}_{0}(\mathcal{H}_{s})\subset \mathfrak{L}^{1}(\mathcal{H}_{s})$, $s\geq1$,
the infinitesimal generator $\mathcal{N}(Y|f)$ of the group $\mathcal{G}(t;Y|f)$ is defined by the operator
\begin{eqnarray}\label{Nnl}
   &&\mathcal{N}(Y|f)\doteq -\mathcal{N}_{s}(Y)f_{s}(Y)+\\
   &&+\sum\limits_{\mbox{\scriptsize $\begin{array}{c}\mathrm{P}:Y=\bigcup_{i}X_{i},\\|\mathrm{P}|\neq1\end{array}$}}
        \hskip-2mm\sum\limits_{\mbox{\scriptsize$\begin{array}{c}{Z_{1}\subset X_{1}},\\Z_{1}\neq\emptyset\end{array}$}}
        \hskip-2mm\ldots \sum\limits_{\mbox{\scriptsize$\begin{array}{c}{Z_{|\mathrm{P}|}\subset X_{|\mathrm{P}|}},\\Z_{|\mathrm{P}|}\neq\emptyset\end{array}$}}
        \big(-\mathcal{N}_{\mathrm{int}}^{(\sum\limits_{r=1}^{|\mathrm{P}|}|Z_{{r}}|)}
        (Z_{{1}},\ldots,Z_{{|\mathrm{P}|}})\big)\prod_{X_{i}\subset \mathrm{P}}f_{|X_{i}|}(X_{i}),\nonumber
\end{eqnarray}

We remark that a particular solution of the steady von Neumann hierarchy is a sequence of
the Ursell operators, for example, its first two elements have the form
\begin{eqnarray*}
   &&g_1(t,1)= e^{-\beta K(1)},\\
   &&g_2(t,1,2)=e^{-\beta \sum\limits_{i=1}^{2}K(i)}(e^{-\beta\Phi(1,2)}-I),
\end{eqnarray*}
where $\beta$ is a parameter inversely proportional to temperature.

For the purpose of further application we introduce the notion of correlation operators of clusters
of particles
\begin{eqnarray}\label{rozvNF-Nclusters}
    &&g_{1+n}(t,\{Y\},X\setminus Y)=\mathcal{G}(t;\{Y\},X\setminus Y|g(0))\doteq\\
    &&\doteq\sum\limits_{\mathrm{P}:\,(\{Y\},\,X\setminus Y)=\bigcup_i X_i}
       \mathfrak{A}_{|\mathrm{P}|}\big(-t,\{\theta(X_1)\},\ldots,\{\theta(X_{|\mathrm{P}|})\}\big)
       \prod_{X_i\subset \mathrm{P}}g_{|X_i|}^0(X_i),\nonumber
\end{eqnarray}
where $\mathfrak{A}_{|\mathrm{P}|}(-t)$ is the $|\mathrm{P}|th$-order cumulant defined by formula
\eqref{cumulantP}.

We remark that the relations between correlation operators of particle clusters $\mathfrak{d}_{\{Y\}}g(t)\in\mathfrak{L}^{1}(\oplus_{n=0}^{\infty}\mathcal{H}_{s+n})$
and correlation operators of particles \eqref{rozvNh} are given by the following equalities
\begin{eqnarray}\label{grel}
  &&g_{1+n}(t,\{Y\},X\setminus Y)=\\
  &&=\sum\limits_{\mathrm{P}:(\{Y\},\,X\setminus Y)=\bigcup_i X_i}
      (-1)^{|\mathrm{P}|-1}(|\mathrm{P}|-1)!\,
      \prod_{X_i\subset \mathrm{P}}\,\,\sum\limits_{\mathrm{P'}:\,\theta(X_{i})=\bigcup_{j_i} Z_{j_i}}
      \prod_{Z_{j_i}\subset \mathrm{P'}}g_{|Z_{j_i}|}(t,Z_{j_i}).\nonumber
\end{eqnarray}
In particular case $n=0$, i.e. the correlation operator of a cluster of $|Y|$ particles, these
relations take the form
\begin{eqnarray*}\label{gCluster0}
  &&g_{1+0}(t,\{Y\})=\sum\limits_{\mathrm{P}:\,Y=\bigcup_{i} X_{i}}
      \prod_{X_{i}\subset \mathrm{P}}g_{|X_{i}|}(t,X_{i}).\nonumber
\end{eqnarray*}
Due to cluster expansions \eqref{D_(g)N} it is possible to express many-particle correlation
operators through the two-particle and one-particle correlation operators
\begin{eqnarray*}
  &&g_{1+n}(t,\{Y\},X\setminus Y)=\\
  &&=\sum\limits_{\substack{Z\subset{X \backslash Y},\,Z\neq \emptyset}}\,g_{2}(t,\{Y\},\{Z\})
     \sum\limits_{\mathrm{P}:\,X\backslash(Y\cup Z)={\bigcup\limits}_i X_i}
     (-1)^{|\mathrm{P}|}|\mathrm{P}|!\,\prod_{i=1}^{|\mathrm{P}|}\,g_{1}(t,\{X_{i}\}).
\end{eqnarray*}

We note also that in case of many-particle systems obeying quantum statistics, i.e. many-particle
systems of fermions or bosons, the von Neumann hierarchy has the form
\begin{eqnarray}\label{vNfb}
    &&\frac{d}{dt}g_{s}(t,Y)=\mathcal{N}(Y|g(t)), \quad s\geq1,
\end{eqnarray}
where the hierarchy generator is determined by
\begin{eqnarray*}\label{Nnlfb}
    &&\mathcal{N}(Y|g)\doteq -\mathcal{N}_{s}(Y)g_{s}(Y)+\\
    &&+\sum\limits_{\mbox{\scriptsize $\begin{array}{c}\mathrm{P}:Y=\bigcup_{i}X_{i},\\|\mathrm{P}|\neq1\end{array}$}}
       \hskip-2mm\sum\limits_{\mbox{\scriptsize$\begin{array}{c}{Z_{1}\subset X_{1}},\\Z_{1}\neq\emptyset\end{array}$}}
       \hskip-2mm\ldots \sum\limits_{\mbox{\scriptsize$\begin{array}{c}{Z_{|\mathrm{P}|}\subset X_{|\mathrm{P}|}},\\Z_{|\mathrm{P}|}\neq\emptyset\end{array}$}}
       \big(-\mathcal{N}_{\mathrm{int}}^{(\sum\limits_{r=1}^{|\mathrm{P}|}|Z_{{r}}|)}
       (Z_{{1}},\ldots,Z_{{|\mathrm{P}|}})\big)\mathcal{S}^{\pm}_{s}
       \prod_{X_{i}\subset\mathrm{P}}g_{|X_{i}|}(X_{i}),\nonumber
\end{eqnarray*}
and the operators $\mathcal{S}_s^{\pm}$ are defined by formula \eqref{Sn}.
A solution of the Cauchy problem of hierarchy \eqref{vNfb} is given by the corresponding expansion
\begin{eqnarray*}
    &&g_{s}(t,Y)=\sum\limits_{\mathrm{P}:\,Y=\bigcup_iX_i}
        \mathfrak{A}_{|\mathrm{P}|}(-t,\{X_1\},\ldots,\{X_{|\mathrm{P}|}\})\,
        \mathcal{S}_s^{\pm}\prod_{X_i\subset \mathrm{P}}g_{|X_i|}^0(X_i),\quad s\geq1.
\end{eqnarray*}


\section{The hierarchies of quantum evolution equations}
For the description of the evolution of infinite-particle quantum systems the hierarchies
of evolution equations for marginal observables, marginal density operators and marginal
correlation operators are used \cite{CGP97},\cite{BQ}. They are constructed as the evolution
equations for one more but an equivalent method of the description of states and observables
of finitely many particles.

Usually the evolution of infinite-particle quantum systems is described within the framework
of the evolution of states by the quantum BBGKY hierarchy for marginal density operators.
An alternative method of the description of state evolution is given in terms of the
nonlinear quantum BBGKY hierarchy for marginal correlation operators. An equivalent approach
to the description of the evolution of quantum systems is given within the framework of the
marginal observables governed by the dual quantum BBGKY hierarchy \cite{GerS06},\cite{BG}.

For a system of a finite average number of particles there exists an equivalent possibility to
describe observables and states, namely, by sequences of marginal observables (the so-called
$s$-particle observables) $B(t)=(B_0,B_{1}(t,1),\ldots,B_{s}(t,1,\ldots,s),\ldots)$ and marginal
states ($s$-particle density operators) $F(0)=(1,F_{1}^0(1),\ldots,F_{s}^0(1,\ldots,s),\ldots)$
\cite{CGP97}, \cite{Pe95}. These sequences are correspondingly introduced instead of sequences
of observables $A(t)$ and density operators $D(0)$, in such way that mean value \eqref{averageD}
does not change, i.e.
\begin{eqnarray}\label{avmar}
   &&\big\langle A\big\rangle(t)=(I,D(0))^{-1}(A(t),D(0))
      =(I,D(0))^{-1}\sum\limits_{n=0}^{\infty}\frac{1}{n!}
      \,\mathrm{Tr}_{1,\ldots,n}\,A_{n}(t)\,D_{n}^0=\\
   &&=(B(t),F(0))=\sum\limits_{s=0}^{\infty}\frac{1}{s!}\,
      \mathrm{Tr}_{1,\ldots,s}\,B_{s}(t,1,\ldots,s)\,F_{s}^0(1,\ldots,s),\nonumber
\end{eqnarray}
where $(I,D(0))={\sum\limits}_{n=0}^{\infty}\frac{1}{n!}\mathrm{Tr}_{1,\ldots,n}D_{n}^0$ is
a normalizing factor and $I$ is the identity operator. Thus, the relationship of marginal
observables and observables is determined by the formula
\begin{eqnarray}\label{mo}
      &&B_{s}(t,Y)\doteq\sum_{n=0}^s\,\frac{(-1)^n}{n!}\sum_{j_1\neq\ldots\neq j_{n}=1}^s
            A_{s-n}(t,Y\backslash(j_1,\ldots,j_{n})), \quad s\geq 1,
\end{eqnarray}
where $Y\equiv(1,\ldots,s)$. In terms of the density operators the marginal density operators
are defined respectively
\begin{eqnarray}\label{ms}
      &&F_{s}^0(Y)\doteq(I,D(0))^{-1} \sum\limits_{n=0}^{\infty}\frac{1}{n!}
          \mathrm{Tr}_{1,\ldots,s+n}\,D_{s+n}^0(1,\ldots,s+n),\quad s\geq 1.
\end{eqnarray}
As we can see from \eqref{mo} one component sequences of marginal observables correspond to observables
of certain structure, namely the marginal observable $B^{(1)}=(0,a_{1}(1),0,\ldots)$ corresponds to the
additive-type observable $A^{(1)}=(0,a_{1}(1),\ldots,{\sum\limits}_{i=1}^{n}a_{1}(i),\ldots)$, and in the
general case the $k$-ary-type marginal observable $B^{(k)}=(0,\ldots,0,a_{k}(1,\ldots,k),0,\ldots)$
corresponds to the $k$-ary-type observable $A^{(k)}=(0,\ldots,0,a_{k}(1,\ldots,k),\ldots,
\sum_{i_{1}<\ldots<i_{k}=1}^{n}a_s(i_{1},\ldots,i_{k}),\ldots)$.

We emphasize that the evolution of marginal observables \eqref{mo} of both finitely and infinitely
many quantum particles is described by the initial-value problem of the dual BBGKY hierarchy. For
finitely many particles the dual quantum BBGKY hierarchy is equivalent to the Heisenberg equations
\eqref{H-N1} (the dual equation to the Heisenberg equation \eqref{H-N1} is the von Neumann equation \eqref{vonNeumannEqn}).

\subsection{The dual quantum BBGKY hierarchy for marginal observables}
The evolution of marginal observables is described by the initial-value problem
of the dual quantum BBGKY hierarchy
\begin{eqnarray}
  \label{dh}
   &&\frac{d}{dt}B_{s}(t,Y)=\big(\sum\limits_{j=1}^{s}\mathcal{N}(j)+
      \sum\limits_{j_1<j_{2}=1}^{s}\mathcal{N}_{\mathrm{int}}(j_1,j_{2})\big)B_{s}(t,Y)+\\
   &&+\sum_{j_1\neq j_{2}=1}^s\mathcal{N}_{\mathrm{int}}(j_1,j_{2})
      B_{s-1}(t,Y\backslash (j_1)),\nonumber\\\nonumber\\
  \label{dhi}
   &&B_{s}(t)\mid_{t=0}=B_{s}^0,\quad s\geq1,
\end{eqnarray}
where on $\mathfrak{L}_{0}(\mathcal{H}_n)\subset\mathfrak{L}(\mathcal{H}_n)$ the operators
$\mathcal{N}(j)$ and $\mathcal{N}_{\mathrm{int}}(j_1,j_{2})$ are correspondingly
defined by formulas
\begin{eqnarray}\label{com}
   &&\mathcal{N}(j)g_n\doteq -i(g_n K(j)-K(j)g_n),\\
   &&\mathcal{N}_{\mathrm{int}}(j_1,j_{2})g_n\doteq -
       i(g_n\Phi(j_1,j_{2})-\Phi(j_1,j_{2})g_n),\nonumber
\end{eqnarray}
We refer to recurrence evolution equations \eqref{dh} as the dual quantum BBGKY hierarchy since
it is the adjoint hierarchy of evolution equations to the quantum BBGKY hierarchy for the marginal
density operators \cite{BQ,Pe95} with respect to bilinear form \eqref{averageD}.
In case of the space $\mathcal{H}=L^{2}(\mathbb{R}^\nu)$, evolution equations \eqref{dh} for
kernels of the operators $B_{s}(t),\,\geq 1$, are given by
\begin{eqnarray*}
    &&i\,\frac{\partial}{\partial t}B_{1}(t,q_1;q'_1)=-\frac{1}{2}(-\Delta_{q_1}+
       \Delta_{q'_1})B_{1}(t,q_1;q'_1),\\
    &&i\,\frac{\partial}{\partial t}B_{s}(t,q_1,\ldots,q_s;q'_1,\ldots,q'_s)=
       \big(-\frac{1}{2}\sum\limits_{j=1}^s(-\Delta_{q_j}+\Delta_{q'_j})+\\
    &&+\sum\limits_{1=j_1<j_2}^s(\Phi(q'_{j_1}-q'_{j_2})-\Phi(q_{j_1}-q_{j_2}))\big)
       B_s(t,q_1,\ldots,q_s;q'_1,\ldots,q'_s)+\\
    &&+\sum\limits_{1=j_1\neq j_2}^s(\Phi(q'_{j_1}-q'_{j_2})-\Phi(q_{j_1}-q_{j_2}))
       B_{s-1}(t,q_1,\ldots,q_{{j_2}-1},q_{{j_2}+1},\ldots,q_s;\\
    &&\hskip+57mm q'_1,\ldots,q'_{{j_2}-1},q'_{{j_2}+1},\ldots,q'_s), \quad s\geq2.
\end{eqnarray*}
We note that in case of many-body interaction potentials \eqref{Hn} hierarchy \eqref{dh}
has the form \cite{G}
\begin{eqnarray*}\label{dhn}
   &&\frac{d}{dt}B_{s}(t,Y)=\mathcal{N}_{s}(Y)B_{s}(t,Y)+\\
   &&+\sum\limits_{n=1}^{s}\frac{1}{n!}\sum\limits_{k=n+1}^s\frac{1}{(k-n)!}
       \sum_{j_1\neq\ldots\neq j_{k}=1}^s\mathcal{N}_{\mathrm{int}}^{(k)}(j_1,\ldots,j_{k})
       B_{s-n}(t,Y\backslash\{j_1,\ldots,j_{n}\}),\quad s\geq 1.
\end{eqnarray*}
On space $\mathfrak{L}(\mathcal{H}_k)$ the operator $\mathcal{N}^{(k)}_{\mathrm{int}}$ is defined by
\begin{eqnarray}\label{gennd}
   &&\mathcal{N}^{(k)}_{\mathrm{int}}g_{k}\doteq-i\big(g_{k}\Phi^{(k)}-
         \Phi^{(k)}g_{k}\big).
\end{eqnarray}
We note that for finitely many particles the dual quantum BBGKY hierarchy \eqref{dh} (or \eqref{dhn})
is rigorously derived on basis of the Heisenberg equation \eqref{H-N1} according to definition \eqref{mo}.
Hence the structure of evolution equations for marginal observables is defined by the structure of
expansion \eqref{mo}.

Let us introduce some abridged notations: $Y\equiv(1,\ldots,s), X\equiv (j_1,\ldots,j_{n})\subset Y$
and $\{Y\backslash X\}$ is the set consisting of one element $Y\backslash X=(1,\ldots,s)
\backslash(j_1,\ldots,j_{n})$, i.e. the set $\{Y\backslash X\}$ is a connected subset of the set $Y$.

To construct a solution of the Cauchy problem \eqref{dh}-\eqref{dhi} we introduce the
$(1+n)th$-order cumulant of groups of operators \eqref{grG} as follows \cite{GerRS},\cite{G}
\begin{eqnarray}\label{cumulant}
    &&\mathfrak{A}_{1+n}(t,\{Y\backslash X\},X)\doteq\\
    &&\doteq\sum\limits_{\mathrm{P}:\,(\{Y\backslash X\},\,X)={\bigcup}_i X_i}
       (-1)^{\mathrm{|P|}-1}({\mathrm{|P|}-1})!
       \prod_{X_i\subset \mathrm{P}}\mathcal{G}_{|\theta(X_i)|}(t,\theta(X_i)),\quad n\geq0,\nonumber
\end{eqnarray}
where ${\sum}_\mathrm{P}$ is the sum over all possible partitions $\mathrm{P}$
of the set $(\{Y\backslash X\},j_1,\ldots,j_{n})$ into $|\mathrm{P}|$ nonempty
mutually disjoint subsets $ X_i\subset(\{Y\backslash X\},X)$. For example,
\begin{eqnarray*}
    &&\mathfrak{A}_{1}(t,\{Y\})=\mathcal{G}_{s}(t,Y),\\
    &&\mathfrak{A}_{2}(t,\{Y\backslash (j)\},j)=\mathcal{G}_{s}(t,Y)
       -\mathcal{G}_{s-1}(t,Y\backslash(j))\mathcal{G}_{1}(t,j).\nonumber
\end{eqnarray*}

Let us indicate some properties of cumulants \eqref{cumulant}. If $n=0$, for
$g_{s}\in\mathfrak{L}_0(\mathcal{H}_s)\subset\mathfrak{L}(\mathcal{H}_s)$ in the sense of
the $\ast$-weak convergence of the space $\mathfrak{L}(\mathcal{H}_s)$ the generator of
first-order cumulant \eqref{cumulant} is given by operator \eqref{dkomyt}
\begin{eqnarray*}
    &&\lim\limits_{t\rightarrow 0}\frac{1}{t}(\mathfrak{A}_{1}(t,\{Y\})-I)g_{s}(Y)=
       \mathcal{N}_{s}g_{s}(Y).
\end{eqnarray*}
In case of $n=1$ for $g_{s}\in\mathfrak{L}_0(\mathcal{H}_{s})\subset\mathfrak{L}(\mathcal{H}_{s})$
we obtain the following equality in the sense of the $\ast$-weak convergence of the space $\mathfrak{L}(\mathcal{H}_{s})$
\begin{eqnarray*}
   &&\lim\limits_{t\rightarrow 0}\frac{1}{t}\,\mathfrak{A}_{2}(t,\{Y\backslash(j)\},j) g_{s}(Y)
       =\sum_{i\in(Y\backslash(j))}\mathcal{N}_{\mathrm{int}}(i,j)g_{s}(Y),
\end{eqnarray*}
where the operator $\mathcal{N}_{\mathrm{int}}(i,j)$ is defined by \eqref{com}, and for
$n\geq2$ as a consequence of the fact that we consider a system of particles interacting by a two-body
potential \eqref{H}, it holds
\begin{eqnarray*}
   &&\lim\limits_{t\rightarrow 0}\frac{1}{t}\,\mathfrak{A}_{1+n}(t,\{Y\backslash X\},X) g_{s}(Y)=0.
\end{eqnarray*}
Correspondingly in case of $n$-body interaction potential \eqref{Hn} it holds
\begin{eqnarray*}
   &&\lim\limits_{t\rightarrow 0}\frac{1}{t}\,\mathfrak{A}_{n}(t,1,\ldots,n) g_{n}=
      \mathcal{N}_{\mathrm{int}}^{(n)}g_{n},
\end{eqnarray*}
where the operator $\mathcal{N}_{\mathrm{int}}^{(n)}$ is defined by formula \eqref{gennd}.

If $g_{s}\in\mathfrak{L}(\mathcal{H}_{s})$, then for $(1+n)th$-order cumulant \eqref{cumulant}
of groups of operators \eqref{grG} the estimate is valid
\begin{eqnarray}\label{estd}
   &&\big\|\mathfrak{A}_{1+n}(t)g_{s}\big\|_{\mathfrak{L}(\mathcal{H}_{s})}
       \leq \sum\limits_{\mathrm{P}:\,(\{Y\backslash X\},\,X)={\bigcup}_i X_i}
       (|\mathrm{P}|-1)!\big\|g_{s}\big\|_{\mathfrak{L}(\mathcal{H}_{s})}\leq \\
   &&\leq \sum\limits_{k=1}^{n+1}\mathrm{s}(n+1,k)(k-1)!\big\|g_{s}\big\|_{\mathfrak{L}(\mathcal{H}_{s})}
        \leq n!e^{n+2}\big\|g_{s}\big\|_{\mathfrak{L}(\mathcal{H}_{s})},\nonumber
\end{eqnarray}
where $\mathrm{s}(n+1,k)$ are the Stirling numbers of the second kind.

On the space $\mathfrak{L}_{\gamma}(\mathcal{F}_\mathcal{H})$ for abstract initial-value problem
\eqref{dh}-\eqref{dhi} the following statement is true \cite{BG}.

\begin{theorem}
If $B(0)\in\mathfrak{L}_{\gamma}(\mathcal{F}_\mathcal{H})$ and $\gamma<e^{-1}$, then for
$t\in\mathbb{R}$ a unique solution of the Cauchy problem \eqref{dh}-\eqref{dhi} of the dual
quantum BBGKY hierarchy exists and it is determined by the expansion
\begin{eqnarray}\label{sdh}
   &&B_{s}(t,Y)=\sum_{n=0}^s\,\frac{1}{n!}\sum_{j_1\neq\ldots\neq j_{n}=1}^s
       \mathfrak{A}_{1+n}(t,\{Y\backslash X\},X)\,B_{s-n}^0(Y\backslash X),\quad s\geq 1,
\end{eqnarray}
where the $(1+n)th$-order cumulant $\mathfrak{A}_{1+n}(t,\{Y\backslash X\},X)$ is defined by formula \eqref{cumulant}.
For $B(0)\in\mathfrak{L}_{\gamma}^0(\mathcal{F}_\mathcal{H})\subset\mathfrak{L}_{\gamma}(\mathcal{F}_\mathcal{H})$
it is a classical solution and for arbitrary initial data $B(0)\in\mathfrak{L}_{\gamma}(\mathcal{F}_\mathcal{H})$
it is a generalized solution.
\end{theorem}

The simplest examples of marginal observables \eqref{sdh} are given by the expressions
\begin{eqnarray*}
   &&B_{1}(t,1)=\mathfrak{A}_{1}(t,1)B_{1}^0(1),\\
   &&B_{2}(t,1,2)=\mathfrak{A}_{1}(t,\{1,2\})B_{2}^0(1,2)+\mathfrak{A}_{2}(t,1,2)(B_{1}^0(1)+B_{1}^0(2)).
\end{eqnarray*}

We remark that at the initial time $t=0$ solution \eqref{sdh} satisfies initial condition \eqref{dhi}.
Indeed, according to definition \eqref{grG} and equality \eqref{Stirl} for $n\geq0$, we have
\begin{eqnarray*}
    &&\mathfrak{A}_{1+n}(0,\{Y\backslash X\},X)=\sum\limits_{\mathrm{P}:\,(\{Y\backslash X\},\,X)=
       {\bigcup}_i X_i}(-1)^{\mathrm{|P|}-1}({\mathrm{|P|}-1})!I=I\delta_{n,0}.
\end{eqnarray*}

On the space $\mathfrak{L}_{\gamma}(\mathcal{F}_\mathcal{H})$ a solution of the initial-value problem
of the dual BBGKY hierarchy defines a one-parametric mapping with the following properties \cite{G}.

Owing to estimate \eqref{estd} for cumulants \eqref{cumulant}, under the condition that $\gamma<e^{-1}$
we have
\begin{eqnarray}\label{es}
    &&\big\|B(t)\big\|_{\mathfrak{L}_{\gamma}(\mathcal{F}_\mathcal{H})}
       \leq e^2(1-\gamma e)^{-1}\big\|B(0)\big\|_{\mathfrak{L}_{\gamma}(\mathcal{F}_\mathcal{H})}.
\end{eqnarray}
Then on the space $g\in\mathfrak{L}_{\gamma}(\mathcal{F}_\mathcal{H})$ the one-parametric mapping
\begin{eqnarray}\label{mapdual}
    &&\mathbb{R}^{1}\ni t\mapsto(U(t)g)_{s}(Y)\doteq\\
    &&\doteq\sum_{n=0}^s\,\frac{1}{(s-n)!}\sum_{j_1\neq\ldots\neq j_{s-n}=1}^s
       \mathfrak{A}_{1+n}(t,\{Y\backslash X\},X)\,g_{s-n}(Y\backslash X),\quad s\geq1,\nonumber
\end{eqnarray}
is a $C_{0}^{\ast}$-group. The infinitesimal generator $\mathcal{D}={\bigoplus\limits}_{s=0}^{\infty}
\mathcal{D}_{s}$ of this group of operators is a closed operator for the $\ast$-weak topology and
on the domain of the definition $\mathcal{D}(\mathcal{D})\subset\mathfrak{L}_{\gamma}(\mathcal{F}_\mathcal{H})$
which is the everywhere dense set for the $\ast$-weak topology of the space
$\mathfrak{L}_{\gamma}(\mathcal{F}_\mathcal{H})$ it is defined by the operator
\begin{eqnarray*}\label{d}
    &&(\mathcal{D}g)_{s}(Y)\doteq\mathcal{N}_{s}(Y)g_{s}(Y)+\\
    &&+\sum\limits_{n=1}^{s}\frac{1}{n!}\sum\limits_{k=n+1}^s \frac{1}{(k-n)!}
       \sum_{j_1\neq\ldots\neq j_{k}=1}^s\mathcal{N}_{\mathrm{int}}^{(k)}(j_1,\ldots,j_{k})
       g_{s-n}(Y\backslash\{j_1,\ldots,j_{n}\}), \nonumber
\end{eqnarray*}
where the operator $\mathcal{N}^{(k)}_{\mathrm{int}}$ is given by formula \eqref{gennd}.

In capacity of initial data we consider the additive-type observables, i.e. the one-component
sequences $B^{(1)}(0)=(0,a_{1}(1),0,\ldots)$ (the $k$-ary marginal observable is the sequence
$B^{(k)}(0)=(0,\ldots,0,a_{k}(1,\ldots,k),0,\ldots)$ \cite{BGer}). In this case solution
expansion \eqref{sdh} attains the form
\begin{eqnarray}\label{af}
     &&B_{s}^{(1)}(t,Y)=\mathfrak{A}_{s}(t,1,\ldots,s)\sum_{j=1}^s a_{1}(j), \quad s\geq 1.
\end{eqnarray}
For such additive-type observable as number of particles, i.e. one-component sequence
$N(0)=(0,I,0,\ldots)$, according to definition of cumulants \eqref{cumulant}, solution expansion
\eqref{af} get the following form
\begin{eqnarray*}
    &&(N(t))_{s}(Y)=\mathfrak{A}_{s}(t,1,\ldots,s)\sum_{j=1}^s I= s\,\delta_{s,1}I, \quad s\geq 1,
\end{eqnarray*}
where $I$ is the identity operator and $\delta_{s,1}$ is a Kroneker symbol. Hence we have
\begin{eqnarray*}
    &&\big|(N(t),F(0))\big|=\big|\mathrm{Tr}_{\mathrm{1}}\,F_{1}^0(1)\big|\leq
       \big\|F_{1}^0\big\|_{\mathfrak{L}^{1}(\mathcal{H})}<\infty.
\end{eqnarray*}
Thus, the marginal density operators from the space $\mathfrak{L}_{\alpha}^{1}(\mathcal{F}_\mathcal{H})$
describe quantum systems of finitely many particles.

We note that solution expansion \eqref{sdh}, i.e. nonperturbative solution expansion, can be derived from
solution \eqref{sH} of the initial-value problem of the Heisenberg equation \eqref{H-N1}-\eqref{H-N12} on
the basis of expansions \eqref{mo}. Since hierarchy \eqref{dh} has the structure of recurrence equations,
we also deduce that the solution can be constructed by successive integration of the inhomogeneous Heisenberg
equations. Solution \eqref{sdh} is represented in the form of the perturbation (iteration) series as a result
of applying of analogs of the Duhamel equation to cumulants \eqref{cumulant} of groups of operators \eqref{grG}.

Cluster expansions \eqref{groupKlast} of group of operators can be put at the basis of all possible solution
representations of the dual quantum BBGKY hierarchy \eqref{dh}. In fact, solving recurrence relations
\eqref{groupKlast} with respect to the $1st$-order cumulants for the separation
terms, which are independent from the variable $Y\backslash X\equiv(j_1,\ldots,j_{s-n})$
\begin{eqnarray*}
    &&\mathfrak{A}_{1+n}(t,\{Y\backslash X\},X)=
        \sum\limits_{\substack{Z\subset X}} \mathfrak{A}_{1}(t,\{Y\backslash X\cup Z\})
        \sum\limits_{\mathrm{P}\,:\,X \backslash Z={\bigcup\limits}_i X_i}
        (-1)^{|\mathrm{P}|}\,|\mathrm{P}|!\,\prod_{i=1}^{|\mathrm{P}|}\mathfrak{A}_{1}(t,\{X_{i}\}),
\end{eqnarray*}
where ${\sum\limits}_{\substack{Z\subset X}}$ is a sum over all subsets $Z\subset X$ of the set $X$
and taking into account the identity
\begin{eqnarray}\label{id}
    &&\hskip-10mm\sum\limits_{\mathrm{P}\,:\,X \backslash Z ={\bigcup\limits}_i X_i}
        (-1)^{|\mathrm{P}|}\,|\mathrm{P}|!\,\prod_{i=1}^{|\mathrm{P}|}\mathfrak{A}_{1}
        (t,\{X_{i}\})g_{s-n}(Y\backslash X)=\sum\limits_{\mathrm{P}:\,X \backslash Z
        ={\bigcup\limits}_i X_i}(-1)^{|\mathrm{P}|}\,|\mathrm{P}|!\,g_{s-n}(Y\backslash X),
\end{eqnarray}
and the equality
\begin{eqnarray}\label{e}
    &&\sum\limits_{\mathrm{P}:\,X\backslash Z=
       {\bigcup\limits}_i X_i}(-1)^{|\mathrm{P}|}\,|\mathrm{P}|!=(-1)^{|X \backslash Z|},
\end{eqnarray}
for expansion \eqref{sdh} we derive
\begin{eqnarray}\label{rdex}
   &&\hskip-7mmB_{s}(t,Y)=\sum_{n=0}^s\,\frac{1}{(s-n)!}\sum_{j_1\neq\ldots\neq j_{s-n}=1}^s\,\,
        \sum\limits_{\substack{Z\subset X}}\,(-1)^{|X \backslash Z|}\,
        \mathcal{G}_{s-n+|Z|}(t,Y\backslash X\cup Z)\,B_{s-n}^0(Y\backslash X).
\end{eqnarray}
Introducing the operator $\mathfrak{a}^{+}$ (an analog of the creation operator \cite{BG}):
\begin{eqnarray}\label{oper_znuw}
   &&(\mathfrak{a}^{+}g)_{s}(Y)\doteq\sum_{j=1}^s\,g_{s-1}(Y\backslash (j))
\end{eqnarray}
defined on $\mathfrak{L}_{\gamma}(\mathcal{F}_\mathcal{H})$, as a result of the symmetry
property of the Maxwell-Boltzmann statistics, expression \eqref{rdex} can be rewritten
in the following compact form
\begin{eqnarray*}
   &&B(t)=\sum\limits_{n=0}^{\infty}\frac{1}{n!}\,\sum\limits_{k=0}^{n}\,(-1)^{n-k}\,
      \frac{n!}{k!(n-k)!}\,(\mathfrak{a}^{+})^{n-k}\mathcal{G}(t)(\mathfrak{a}^{+})^{k}B^0=
      e^{-\mathfrak{a}^{+}}\mathcal{G}(t)e^{\mathfrak{a}^{+}}B^0.
\end{eqnarray*}

We can obtain one more representation for a solution of the initial-value problem of the dual quantum
BBGKY hierarchy, if we express the cumulants $\mathfrak{A}_{1+n}(t),\,n\geq1,$ of groups of operators
\eqref{grG} with respect to the $1st$-order and $2nd$-order cumulants. In fact, it holds
\begin{eqnarray*}
   &&\mathfrak{A}_{1+n}(t,\{(Y\backslash X)\},X)=
      \sum\limits_{\substack{Z\subset X,\\Z\neq \emptyset}}\mathfrak{A}_{2}(t,\{Y\backslash X\},\{Z\})
      \sum\limits_{\mathrm{P}:\,X \backslash Z ={\bigcup\limits}_i X_i}
      (-1)^{|\mathrm{P}|}\,|\mathrm{P}|!\,\prod_{i=1}^{|\mathrm{P}|} \mathfrak{A}_{1}(t,\{X_{i}\}),
\end{eqnarray*}
where ${\sum\limits}_{\substack{Z\subset X,\\Z\neq\emptyset}}$ is a sum over all nonempty subsets
$Z\subset X$ of the set $X$. Then taking into account identity \eqref{id} and equality
\eqref{e}, we get the following representation for solution expansion \eqref{sdh} of the
dual quantum BBGKY hierarchy
\begin{eqnarray*}
    &&B _{s}(t,Y)=\mathfrak{A}_{1}(t,Y)B_{s}^0(Y)+\\
    &&+\sum_{n=1}^s\,\frac{1}{(s-n)!}\,\sum_{j_1\neq\ldots\neq j_{s-n}=1}^s\,\,\,
       \sum\limits_{\substack{Z\subset X,\\Z\neq \emptyset}}\,(-1)^{|X\backslash Z|}\,\,
       \mathfrak{A}_{2}(t,\{Y\backslash X\},\{Z\})\,B_{s-n}^0(Y\backslash X),
\end{eqnarray*}
where $Y\equiv(1,\ldots,s)$, $X\equiv Y\backslash (j_1,\ldots,j_{s-n})$,
i.e. $Y\backslash X=(j_1,\ldots,j_{s-n})$.

\subsection{Marginal density operators}
As stated above \eqref{avmar} the mean value of the marginal observable
$B(t)\in\mathfrak{L}_{\gamma}(\mathcal{F}_\mathcal{H})$ at $t\in \mathbb{R}$ in the initial
marginal state $F(0)=(1,F_{1}^{0}(1),\ldots,F_{s}^{0}(1,\ldots,s),\ldots)\in
\mathfrak{L}_{\alpha}^{1}(\mathcal{F}_\mathcal{H})$ is defined by the functional
\begin{eqnarray}\label{avmar-1}
   &&(B(t),F(0))=\sum\limits_{s=0}^{\infty}\,\frac{1}{s!}\,
      \mathrm{Tr}_{\mathrm{1,\ldots,s}}\,B_{s}(t,1,\ldots,s)F_{s}^{0}(1,\ldots,s).
\end{eqnarray}
According to estimate \eqref{es}, functional \eqref{avmar-1} exists under the condition
that $\gamma<e^{-1}$, and the following estimate holds
\begin{eqnarray*}
    &&\big|(B(t),F(0))\big|\leq e^2(1-\gamma e)^{-1}
      \big\|B(0)\big\|_{\mathfrak{L}_{\gamma}(\mathcal{F}_\mathcal{H})}
      \big\|F(0)\big\|_{\mathfrak{L}^{1}_{{\gamma}^{-1}}(\mathcal{F}_\mathcal{H})}.
\end{eqnarray*}
In consequence of the validity for functional \eqref{avmar-1} of the following equalities
\begin{eqnarray*}
    &&(I,D(0))^{-1}(A(t),D(0))=(I,D(t))^{-1}(A(0),D(t))=(B(0),F(t))=\\
    &&=\sum\limits_{s=0}^{\infty}\,\frac{1}{s!}\,
       \mathrm{Tr}_{\mathrm{1,\ldots,s}}\,B_{s}^{0}(1,\ldots,s)F_{s}(t,1,\ldots,s),
\end{eqnarray*}
where the marginal density operator $F_{s}(t,1,\ldots,s)$ is defined by means of
density operators \eqref{rozv_fon-N} according to formula \eqref{ms} and the marginal observable
$B_{s}^{0}(1,\ldots,s)$ is defined by means of \eqref{H-N12} according to formula \eqref{mo}
respectively, it is possible to describe the evolution within the framework of the marginal state evolution.

According to \eqref{averagegs}, marginal density operators \eqref{ms} can be defined in terms of the
correlation operators of clusters of particles \eqref{gClusters} by the expansion
\begin{eqnarray}\label{FClusters}
    &&F_{s}(t,Y)\doteq\sum\limits_{n=0}^{\infty}\frac{1}{n!}\,
       \mathrm{Tr}_{s+1,\ldots,s+n}\,\,g_{1+n}(t,\{Y\},s+1,\ldots,s+n),\quad s\geq1,
\end{eqnarray}
where the correlation operator $g_{1+n}(t)$ is given by expansion \eqref{gClusters}. Starting from
an alternative approach to the description of states by the von Neumann hierarchy, we define the
marginal density operators by the use of solutions of the Cauchy problem of the von Neumann hierarchy
for correlation operators of clusters of particles \eqref{rozvNF-Nclusters}. As is obvious from what
will be said the cumulant structure of the von Neumann hierarchy solution \eqref{rozvNF-Nclusters}
induces the cumulant structure of solution expansion \eqref{GUG(0)} of initial-value problem of the
quantum BBGKY hierarchy for marginal density operators that are adopted for the description of
infinite-particle systems, i.e. infinite-particle dynamics is generated by the dynamics of correlations.

\subsection{The quantum BBGKY hierarchy}
The evolution of states is described by the sequences $F(t)=(1,F_{1}(t,1),\ldots,$ $F_{s}(t,1,\ldots,s),\ldots)$
of the marginal density operators that satisfy the Cauchy problem of the quantum BBGKY hierarchy
\begin{eqnarray}
 \label{BBGKY}
   &&\frac{d}{dt}F_{s}(t,Y)=\big(\sum\limits_{j=1}^{s}(-\mathcal{N}(j))+
       \sum\limits_{j_1<j_2=1}^{s}(-\mathcal{N}_{\mathrm{int}}(j_1,j_2))\big)F_{s}(t,Y)+\nonumber \\
   &&+\sum\limits_{j=1}^{s}\mathrm{Tr}_{s+1}(-\mathcal{N}_{\mathrm{int}}(j,s+1))F_{s+1}(t,Y,s+1),\\
       \nonumber \\
 \label{BBGKYi}
   &&F_{s}(t,Y)\mid_{t=0}=F_{s}^{0}(Y),\quad s\geq 1,
\end{eqnarray}
where $Y\equiv(1,\ldots,s)$ and on $\mathfrak{L}_{0}^1(\mathcal{H}_n)\subset\mathfrak{L}^1(\mathcal{H}_n)$
the operators $(-\mathcal{N}(j))$ and $(-\mathcal{N}_{\mathrm{int}}(j_1,j_{2}))$ are
correspondingly defined by formulas
\begin{eqnarray}\label{comst}
   &&(-\mathcal{N}(j))f_n\doteq -i(K(j)f_n-f_n K(j)),\\
   &&(-\mathcal{N}_{\mathrm{int}})(j_1,j_{2})f_n\doteq -
       i(\Phi(j_1,j_{2})f_n-f_n\Phi(j_1,j_{2})).\nonumber
\end{eqnarray}
In case of the space $\mathcal{H}=L^{2}(\mathbb{R}^\nu)$, hierarchy of evolution equations \eqref{BBGKY}
for kernels of the operators $F_{s}(t),\, s\geq 1$, (the marginal or $s$-particle density matrix) are given by
\begin{eqnarray*}
    &&i\frac{\partial}{\partial t}F_{s}(t,q_{1},\ldots,q_{s};q^{'}_{1},\ldots,q^{'}_{s})=
       \Big(-\frac{1}{2}\sum\limits_{i=1}^{s}(\Delta_{q_{i}}-\Delta_{q^{'}_{i}})+\\
    &&+\sum\limits_{i<j=1}^{s}\big(\Phi(q_{i}-q_{j})-\Phi(q^{'}_{i}-q^{'}_{j})\big)\Big)
       F_{s}(t,q_{1},\ldots,q_{s};q^{'}_{1},\ldots,q^{'}_{s})+\\
    &&+\sum\limits_{i=1}^{s}\int d q_{s+1}\big(\Phi(q_{i}-q_{s+1})-\Phi(q^{'}_{i}-q_{s+1})\big)
       F_{s+1}(t,q_{1},\ldots,q_{s},q_{s+1};q^{'}_{1},\ldots,q^{'}_{s},q_{s+1}).
\end{eqnarray*}
We note that in case of many-body interaction potentials \eqref{Hn} hierarchy \eqref{BBGKY}
has the form \cite{G}
\begin{eqnarray}\label{hn}
   &&\frac{d}{dt}F_{s}(t,Y)=-\mathcal{N}_{s}(Y)F_{s}(t,Y)+\\
   &&+\sum\limits_{n=1}^{\infty}\frac{1}{n!}\,
        \mathrm{Tr}_{s+1,\ldots,s+n}\hskip-2mm\sum\limits_{\mbox{\scriptsize $\begin{array}{c}
       {Z\subset Y},\\Z\neq\emptyset \end{array}$}}\big(-\mathcal{N}_{\mathrm{int}}^{(|Z|+n)}\big)(Z,s+1,\ldots,s+n)
        F_{s+n}(t),\quad s\geq 1\nonumber,
\end{eqnarray}
where the operator $\mathcal{N}^{(n)}_{\mathrm{int}}$ is defined on
$\mathfrak{L}^{1}_{0}(\mathcal{H}_n)\subset \mathfrak{L}^{1}(\mathcal{H}_n)$
by formula \eqref{gennd}.

For finitely many particles the quantum BBGKY hierarchy \eqref{BBGKY} is rigorously derived on basis of
the von Neumann equation \eqref{vonNeumannEqn} according to the definition of marginal density operators
\eqref{ms}. The rigorous derivation of the quantum BBGKY hierarchy from the von Neumann hierarchy is
given in \cite{PP09}. Another way of looking to the derivation of the quantum BBGKY hierarchy consists in
the construction of the adjoint (dual) equations to the dual quantum BBGKY hierarchy \eqref{dh} with respect
to bilinear form \eqref{avmar-1}.

On the space $\mathfrak{L}_{\alpha}^{1}(\mathcal{F}_\mathcal{H})$ for abstract initial-value problem
\eqref{BBGKY}-\eqref{BBGKYi} the following statement holds \cite{GerS06}.

\begin{theorem}
If $F(0)\in\mathfrak{L}_{\alpha}^{1}(\mathcal{F}_\mathcal{H})$ and $\alpha>e$, then for $t\in\mathbb{R}$
a unique solution of the Cauchy problem \eqref{BBGKY}-\eqref{BBGKYi} of the quantum BBGKY hierarchy
exists and is given by the expansion
\begin{eqnarray}\label{RozvBBGKY}
   &&F_{s}(t,Y)=\sum\limits_{n=0}^{\infty}\frac{1}{n!}\,\mathrm{Tr}_{s+1,\ldots,{s+n}}\,
       \mathfrak{A}_{1+n}(-t,\{Y\},\, X\backslash Y)F_{s+n}^{0}(X), \quad s\geq1,
\end{eqnarray}
where $Y\equiv(1,\ldots,s)$, $X\equiv(1,\ldots,s+n)$, and the evolution operator
\begin{eqnarray}\label{cumulant1+n}
   &&\hskip-5mm\mathfrak{A}_{1+n}(-t,\{Y\},\,X\backslash Y)=\sum\limits_{\mathrm{P}\,:(\{Y\},\,X\setminus Y)=
      {\bigcup\limits}_i X_i}(-1)^{|\mathrm{P}|-1}(|\mathrm{P}|-1)!
      \prod_{X_i\subset\mathrm{P}}\mathcal{G}_{|\theta(X_i)|}(-t,\theta(X_i))
\end{eqnarray}
is the $(1+n)th$-order cumulant of groups of operators \eqref{groupG}, ${\sum\limits}_\mathrm{P}$
is the sum over all possible partitions $\mathrm{P}$ of the set $(\{Y\},\,X\setminus Y)$ into
$|\mathrm{P}|$ nonempty mutually disjoint subsets $X_i\subset(\{Y\},\,X\setminus Y)$.
For initial data $F(0)\in\mathfrak{L}^{1}_{\alpha,0}\subset\mathfrak{L}^{1}_{\alpha}(\mathcal{F}_\mathcal{H})$
it is a strong solution and for arbitrary initial data from the space
$\mathfrak{L}_{\alpha}^{1}(\mathcal{F}_\mathcal{H})$ it is a weak solution.
\end{theorem}

Owing to estimate \eqref{est}, i.e.
\begin{eqnarray*}
   &&\big\|\mathfrak{A}_{1+n}(-t)f_{s+n}\big\|_{\mathfrak{L}^{1}(\mathcal{H}_{s+n})}
       \leq n!e^{n+2}\big\|f_{s+n}\big\|_{\mathfrak{L}^{1}(\mathcal{H}_{s+n})},
\end{eqnarray*}
for $F^0\in\mathfrak{L}_{\alpha}^{1}(\mathcal{F}_ \mathcal{H})$ series \eqref{RozvBBGKY}
converges on the norm of the space $\mathfrak{L}_{\alpha}^{1}(\mathcal{F}_\mathcal{H})$
provided that $\alpha>e$, and the inequality holds
\begin{eqnarray*}
    &&\|F(t)\|_{\mathfrak{L}_{\alpha}^{1}(\mathcal{F}_\mathcal{H})}\leq
        c_{\alpha}\|F(0)\|_{\mathfrak{L}_{\alpha}^{1}(\mathcal{F}_\mathcal{H})},
\end{eqnarray*}
where $c_{\alpha}=e^{2}(1-\frac{e}{\alpha})^{-1}$. The parameter $\alpha$ can be interpreted
as the value inverse to the average number of particles.

Let us indicate some properties of cumulants \eqref{cumulant1+n}. If $n=0$, for
$f_{s}\in\mathfrak{L}_0^{1}(\mathcal{H}_s)\subset\mathfrak{L}^{1}(\mathcal{H}_s)$ in the sense
of the norm convergence of the space $\mathfrak{L}^{1}(\mathcal{H}_s)$ the generator of
first-order cumulant \eqref{cumulant1+n} is given by operator \eqref{infOper}
\begin{eqnarray*}
    &&\lim\limits_{t\rightarrow 0}\frac{1}{t}(\mathfrak{A}_{1}(-t,\{Y\})-I)f_{s}(Y)=
       -\mathcal{N}_{s}f_{s}(Y).
\end{eqnarray*}
In case of $n=1$ for $f_{s+1}\in\mathfrak{L}_0^{1}(\mathcal{H}_{s+1})\subset\mathfrak{L}^{1}(\mathcal{H}_{s+1})$
we obtain the following equality in the same sense
\begin{eqnarray}\label{ic}
   &&\lim\limits_{t\rightarrow 0}\frac{1}{t}\,\mathfrak{A}_{2}(-t,\{Y\},s+1)f_{s+1}(Y,s+1)
       =\sum_{i=1}^{s}(-\mathcal{N}_{\mathrm{int}}(i,s+1))f_{s+1}(Y,s+1),
\end{eqnarray}
and for $n\geq2$ as a consequence of the fact that we consider a system of particles interacting
by a two-body potential \eqref{H}, it holds
\begin{eqnarray*}
   &&\lim\limits_{t\rightarrow 0}\frac{1}{t}\,\mathfrak{A}_{1+n}(-t,\{Y\},X\backslash Y)f_{s+n}(X)=0.
\end{eqnarray*}

With a view to generality we consider the case of many-body interaction potentials \eqref{Hn}.
On the space $\mathfrak{L}_{\alpha}^{1}(\mathcal{F} _\mathcal{H})$ a solution of the initial-value
problem of the quantum BBGKY hierarchy is defined a one-parametric mapping (the adjoint mapping
to mapping \eqref{mapdual} in the sense of bilinear form \eqref{average})
\begin{eqnarray*}\label{mapBBGKY}
   &&\mathbb{R}^1\ni t\mapsto(U(-t)f)_{s}(Y)\doteq
      \sum\limits_{n=0}^{\infty}\frac{1}{n!}\mathrm{Tr}_{s+1,\ldots,{s+n}}\,
      \mathfrak{A}_{1+n}(-t,\{Y\},X \backslash Y)f_{s+n}(X)
\end{eqnarray*}
with the following properties \cite{G}. If $f\in\mathfrak{L}_{\alpha}^{1}(\mathcal{F} _\mathcal{H})$
and $\alpha>e$, then one-parametric mapping \eqref{mapBBGKY} is a $C_{0}$-group \cite{G}. On the
subspace $\mathfrak{L}^{1}_{\alpha, 0}\subset\mathfrak{L}^{1}_{\alpha}(\mathcal{F}_\mathcal{H})$
the infinitesimal generator $\mathcal{B}={\bigoplus\limits}_{n=0}^{\infty}\mathcal{B}_{n}$ of this
group is defined by the operator
\begin{eqnarray*}\label{genBBGKY}
   &&(\mathcal{B}f)_{s}(Y)\doteq-\mathcal{N}_{s}(Y)f_{s}(Y)+\\
   &&+\sum\limits_{k=1}^{s}\frac{1}{k!}\sum\limits_{i_1\neq\ldots\neq i_{k}=1}^{s}
      \,\sum\limits_{n=1}^{\infty}\frac{1}{n!}\mathrm{Tr}_{s+1,\ldots,s+n}
      (-\mathcal{N}_{\mathrm{int}}^{(k+n)})(i_1,\ldots,i_{k},X \backslash Y)
      f_{s+n}(X),\quad s\geq1,\nonumber
\end{eqnarray*}
where on $\mathfrak{L}_{0}^{1}(\mathcal{H}_{s+n})\subset\mathfrak{L}^{1}(\mathcal{H}_{s+n})$
the operator $(-\mathcal{N}^{(k+n)}_{\mathrm{int}})$ is defined by formula \eqref{Nintn}.

We indicate that nonperturbative solution \eqref{RozvBBGKY} of the quantum BBGKY hierarchy
is  transformed to the form of the perturbation (iteration) series as a result of applying
of analogs of the Duhamel equation \cite{BanArl} to cumulants \eqref{cumulant1+n} of groups of
operators. To reduce expansion \eqref{RozvBBGKY} to the BBGKY hierarchy iteration series we
put groups of operators in the expression of cumulant \eqref{cumulant1+n} into a new order
with respect to the groups of operators which act on the variables $Y$
\begin{eqnarray}\label{peregr}
    &&\mathfrak{A}_{1+n}(-t,\{Y\},X\backslash Y)=\\
    &&=\sum\limits_{Z\subset\, X\backslash Y}\mathcal{G}_{|Y\cup\,Z|}(-t,Y\cup\,Z)
        \sum\limits_{\mathrm{P}\,:(X\backslash(Y\cup Z))={\bigcup\limits}_i X_i}
        (-1)^{|\mathrm{P}|}|\mathrm{P}|!\prod_{X_i\subset\mathrm{P}}
        \mathcal{G}_{|X_{i}|}(-t,X_{i}).\nonumber
\end{eqnarray}
If $X_{i}\subset X\backslash Y$, then for the trace class operator $F_{|X|}^{0}$ and the unitary
group of operator \eqref{groupG} the equality is valid
\begin{eqnarray*}
    &&\mathrm{Tr}_{s+1,\ldots,s+n}\prod_{X_i\subset \mathrm{P}}\mathcal{G}_{|X_{i}|}(-t;X_{i})
        F_{|X|}^{0}(X)=\mathrm{Tr}_{s+1,\ldots,s+n}F_{|X|}^{0}(X).
\end{eqnarray*}
Then, taking into account the equality
\begin{eqnarray*}
    &&\sum\limits_{\mathrm{P}:\,(X\backslash (Y\cup\,Z))={\bigcup\limits}_i X_i}
       (-1)^{|\mathrm{P}|}|\mathrm{P}|!=(-1)^{|X\backslash (Y\cup\, Z)|},
\end{eqnarray*}
from expression \eqref{peregr} for solution expansion \eqref{RozvBBGKY} we obtain
\begin{eqnarray}\label{cherez1}
    &&F_{s}(t,Y)=\sum\limits_{n=0}^{\infty}\frac{1}{n!}\,
       \mathrm{Tr}_{s+1,\ldots,s+n}\,U_{1+n}(-t,\{Y\},X\backslash Y)F_{|X|}^{0}(X),
\end{eqnarray}
where $U_{1+n}(-t)$ is the $(1+n)th$-order reduced cumulant of groups of operators \eqref{groupG}
\begin{eqnarray*}\label{rc}
    &&U_{1+n}(-t,\{Y\},X\backslash Y)=\sum\limits_{Z\subset X\backslash Y}(-1)^{|X\backslash Y\cup Z)|}
       \mathcal{G}_{|Y\cup Z|}(-t,Y\cup Z).\nonumber
\end{eqnarray*}
Using the symmetry property of particles obeying the Maxwell-Boltzmann statistics, the equalities valid
\begin{eqnarray*}
    &&\mathrm{Tr}_{s+1,\ldots,s+n}\sum\limits_{Z\subset X\backslash Y}
       (-1)^{|X\backslash (Y\cup Z)|}\mathcal{G}_{|Y\cup Z|}(-t,Y\cup Z)F_{|X|}^0(X)=\nonumber\\
    &&=\mathrm{Tr}_{s+1,\ldots,s+n}\sum\limits_{k=0}^{n}(-1)^{k}\sum\limits_{i_{1}<\ldots<i_{n-k}=s+1}^{s+n}
       \mathcal{G}_{|Y|+n-k}(-t,Y,i_{1},\ldots,i_{n-k})F_{s+n}^0(X)=\nonumber\\
    &&=\mathrm{Tr}_{s+1,\ldots,{s+n}}\sum\limits_{k=0}^{n}(-1)^{k}\frac{n!}{k!(n-k)!}
       \mathcal{G}_{|Y|+n-k}(-t,Y,s+1,\ldots,s+n-k)F_{s+n}^0(X).\nonumber
\end{eqnarray*}
Hence for the evolution operator $U_{1+n}(-t)$ it holds
\begin{eqnarray}\label{rc}
    &&U_{1+n}(-t,\{Y\},X\backslash Y)=\sum\limits_{k=0}^{n}(-1)^{k}\frac{n!}{k!(n-k)!}
       \mathcal{G}_{s+n-k}(-t,Y,s+1,\ldots,s+n-k),
\end{eqnarray}
and consequently, we derive the representation which is written down in terms of the operator
$\mathfrak{a}$ (an analog of the annihilation operator \cite{CGP97})
\begin{eqnarray}\label{opann}
    &&(\mathfrak{a}f)_n\doteq\mathrm{Tr}_{n+1}f_{n+1},
\end{eqnarray}
as the following expansion
\begin{eqnarray}\label{rcexp}
    &&F(t)=\sum\limits_{n=0}^{\infty}\frac{1}{n!}\sum\limits_{k=0}^{n}(-1)^{k}\frac{n!}{k!(n-k)!}
         \mathfrak{a}^{n-k}\mathcal{G}(-t)\mathfrak{a}^{k}F^{0}.
\end{eqnarray}
We remark that this representation for the solution expansion for the first time is obtain in \cite{Pe95}
by another method in the form
\begin{eqnarray*}
   &&F(t)=e^\mathfrak{a}\mathcal{G}(-t)e^{-\mathfrak{a}}F(0).
\end{eqnarray*}
Finally, in view of the validity of the equality
\begin{eqnarray*}
   &&\frac{d}{d\tau}\mathcal{G}(-t+\tau)\mathfrak{a}\mathcal{G}(-\tau)F(0)=\mathcal{G}(-t+\tau)
       \big[\mathcal{N},\mathfrak{a}\big]\mathcal{G}(-\tau)F(0),
\end{eqnarray*}
where $\big[\cdot,\cdot\big]$ is the commutator of operators, namely in componentwise form
\begin{eqnarray*}
   &&\big(\big[\mathcal{N},\mathfrak{a}\big]f\big)_s(Y)=
      \sum\limits_{j=1}^{s}\mathrm{Tr}_{s+1}(-\mathcal{N}_{\mathrm{int}}(j,s+1))f_{s+1}(Y,s+1),
\end{eqnarray*}
expansion \eqref{rcexp} is represented in the form of the perturbation series
of the quantum BBGKY hierarchy \eqref{BBGKY}
\begin{eqnarray*}\label{iter}
   &&F(t)=\sum\limits_{n=0}^{\infty}\,\int\limits_{0}^{t} dt_{1}\ldots
        \int\limits_{0}^{t_{n-1}} dt_{n}\mathcal{G}(-t+t_{1})\big[\mathcal{N},\mathfrak{a}\big]
        \mathcal{G}(-t_1+t_2)\ldots\\
   &&\ldots\mathcal{G}(-t_{n-1}+t_n)\big[\mathcal{N},\mathfrak{a}\big]\mathcal{G}(-t_{n})F(0),\nonumber
\end{eqnarray*}
or in componentwise form
\begin{eqnarray}\label{iter}
   &&\hskip-7mmF_s(t,Y)=\sum\limits_{n=0}^{\infty}\,\int\limits_{0}^{t} dt_{1}\ldots
       \int\limits_{0}^{t_{n-1}} dt_{n}\mathrm{Tr}_{s+1,\ldots,s+n}\mathcal{G}_s(-t+t_{1})
       \sum\limits_{j_1=1}^{s}(-\mathcal{N}_{\mathrm{int}}(j_1,s+1))\times\\
   &&\hskip-7mm\times\mathcal{G}_{s+1}(-t_1+t_2)\ldots\mathcal{G}_{s+n-1}(-t_{n-1}+t_n)
       \sum\limits_{j_n=1}^{s+n-1}(-\mathcal{N}_{\mathrm{int}}(j_n,s+n))
       \mathcal{G}_{s+n}(-t_{n})F_{s+n}^0(X).\nonumber
\end{eqnarray}

Recurrence relations \eqref{groupKlast} underlie of the classification of possible solution
representations of the Cauchy problem \eqref{BBGKY}-\eqref{BBGKYi} of the quantum BBGKY hierarchy.
In fact, using cluster expansion \eqref{groupKlast} of the group of operators \eqref{groupG}
it is possible to construct other representations. For example, solving the recurrence relations
\eqref{groupKlast} with respect to the cumulants of first and second order, we have
\begin{eqnarray}\label{kymyl_2}
   &&\mathfrak{A}_{1+n}(-t,\{Y\},X\backslash Y)=\\
   &&=\sum\limits_{\substack{Z\,\subset{X \backslash Y},\\Z\neq\,\emptyset}}
       \mathfrak{A}_{2}(-t,\{X\},\{Z\})\sum\limits_{\mathrm{P}:\,(X \backslash (Y\cup\,Z))=
       {\bigcup\limits}_i X_i}(-1)^{|\mathrm{P}|}|\mathrm{P}|!
       \prod\limits_{X_{i}\subset\mathrm{P}}\mathfrak{A}_{1}(-t,\{X_{i}\}).\nonumber
\end{eqnarray}
Summing up the relevant terms of expression \eqref{kymyl_2} in expansion \eqref{cumulant1+n}
similarly to the case of \eqref{cherez1}, we obtain the expansion representation of initial-value
problem \eqref{BBGKY}-\eqref{BBGKYi} through the second order cumulant
\begin{eqnarray}\label{l_5}
    &&F_s(t,Y)=\sum\limits_{n=0}^{\infty}\frac{1}{n!}\mathrm{Tr}_{s+1,\ldots,s+n}
        \sum\limits_{\substack{Z \subset{X \backslash Y},\\Z \neq \,\emptyset}}(-1)^{|X\backslash (Y\cup\,Z)|}
        \mathfrak{A}_{2}(-t,\{X\},\{Z\})F_{s+n}^{0}(X).
\end{eqnarray}
For $F(0)\in\mathfrak{L}_{\alpha}^{1}(\mathcal{F}_ \mathcal{H})$ series \eqref{l_5} converges
on the norm of the space $\mathfrak{L}_{\alpha}^{1}(\mathcal{F}_\mathcal{H})$ and the estimate holds
\begin{eqnarray*}
    &&\big\|F(t)\big\|_{\mathfrak{L}^{1}_{\alpha}(\mathcal{F}_\mathcal{H})}\leq
       2e^{2}\big\|F(0)\big\|_{\mathfrak{L}^{1}_{\alpha}(\mathcal{F}_\mathcal{H})}.
\end{eqnarray*}
The statement that for initial data from the space $\mathfrak{L}_{\alpha}^{1}(\mathcal{F}_\mathcal{H})$
solution \eqref{l_5} of the Cauchy problem of the quantum BBGKY hierarchy \eqref{BBGKY}-\eqref{BBGKYi}
actually follows from the equivalence of different representations and the existence statement
of solution \eqref{RozvBBGKY}.

\subsection{The nonlinear quantum BBGKY hierarchy for marginal correlation operators}
In view of the definition of mean-value functional \eqref{averagegs}, for example,
the dispersion of an additive-type observable is determined by the functional
\begin{eqnarray}\label{dispg}
    &&\langle(A^{(1)}-\langle A^{(1)}\rangle(t))^2\rangle(t)=\\
    &&=\sum\limits_{n=0}^{\infty}\frac{1}{n!}\,\mathrm{Tr}_{1,\ldots,1+n}\,(a_1^2(1)-
       \langle A^{(1)}\rangle^2(t))g_{1+n}(t,1,\ldots,1+n)+\nonumber\\
    &&+\sum\limits_{n=0}^{\infty}\frac{1}{n!}
       \,\mathrm{Tr}_{1,\ldots,2+n}\,a_{1}(1)a_{1}(2)g_{2+n}(t,1,\ldots,2+n),\nonumber
\end{eqnarray}
where $\langle A^{(1)}\rangle(t)$ is defined by expression \eqref{averagegs} and the operators $g_{s+n}(t)$
are defined by expansions \eqref{gfromDFB}. For $A^{(1)}\in\mathfrak{L}(\mathcal{F}_\mathcal{H})$
and $g\in\mathfrak{L}^{1}(\mathcal{F}_\mathcal{H})$ functional \eqref{dispg} exists.
Following to formula \eqref{dispg} we introduce the marginal correlation operators by the series
\begin{eqnarray}\label{Gexpg}
   &&G_{s}(t,1,\ldots,s)\doteq\sum\limits_{n=0}^{\infty}\frac{1}{n!}\,
      \mathrm{Tr}_{s+1,\ldots,s+n}\,\,g_{s+n}(t,1,\ldots,s+n),\quad s\geq1,
\end{eqnarray}
where the operator $g_{s+n}(t,1,\ldots,s+n)$ is defined by expansion \eqref{gfromDFB} over
solutions \eqref{rozv_fon-N} of the von Neumann equations \eqref{vonNeumannEqn}. According
to estimate (\ref{gEstimate}), series (\ref{Gexpg}) exists and the estimate holds:
$\big\|G_s(t)\big\|_{\mathfrak{L}^{1}(\mathcal{H}_{s})}\leq s!(2e^2)^s\emph{c}^s\sum_{n=0}^{\infty}(2e^2)^n\emph{c}^n$.

Then the dispersion of an additive-type observable is defined within framework of the marginal
correlation operators \eqref{Gexpg} as follows \cite{BQ}
\begin{eqnarray*}
    &&\langle(A^{(1)}-\langle A^{(1)}\rangle(t))^2\rangle(t)=\\
    &&=\mathrm{Tr}_{1}\,(a_1^2(1)-\langle A^{(1)}\rangle^2(t))G_{1}(t,1)+
       \mathrm{Tr}_{1,2}\,a_{1}(1)a_{1}(2)G_{2}(t,1,2).
\end{eqnarray*}
Thus, macroscopic characteristics of fluctuations of observable are determined by marginal correlation
operators \eqref{Gexpg} on the microscopic level.

Assuming as a basis an alternative approach to the description of the state evolution within framework of
the von Neumann hierarchy, we shall define the marginal correlation operators by the use of solutions of the
Cauchy problem \eqref{vNh}-\eqref{vNhi} of the von Neumann hierarchy for correlation operators by formula
\eqref{Gexpg}. We note that every term of marginal correlation operator expansion \eqref{Gexpg} is determined
by the $(s+n)$-particle correlation operator \eqref{rozvNh} as contrasted to marginal density operator expansion
\eqref{FClusters} which is defined by the $(1+n)$-particle correlation operator \eqref{rozvNF-Nclusters}.

Traditionally marginal correlation operators are introduced by means of the cluster expansions of the
marginal density operators \eqref{FClusters} governed by the quantum BBGKY hierarchy \eqref{BBGKY}
\begin{eqnarray}\label{FG}
   &&F_{s}(t,Y)=\sum\limits_{\mbox{\scriptsize $\begin{array}{c}\mathrm{P}:Y=\bigcup_{i}X_{i}\end{array}$}}
      \prod_{X_i\subset \mathrm{P}}G_{|X_i|}(t,X_i),\quad s\geq1,
\end{eqnarray}
where ${\sum\limits}_{\mathrm{P}:Y=\bigcup_{i} X_{i}}$ is the sum over all possible partitions $\mathrm{P}$
of the set $Y\equiv(1,\ldots,s)$ into $|\mathrm{P}|$ nonempty mutually disjoint subsets $X_i\subset Y$.
Hereupon solutions of cluster expansions \eqref{FG}
\begin{eqnarray}\label{gBigfromDFB}
   &&G_{s}(t,Y)=\sum\limits_{\mbox{\scriptsize $\begin{array}{c}\mathrm{P}:Y=\bigcup_{i}X_{i}\end{array}$}}
      (-1)^{|\mathrm{P}|-1}(|\mathrm{P}|-1)!\,\prod_{X_i\subset \mathrm{P}}F_{|X_i|}(t,X_i), \quad s\geq1,
\end{eqnarray}
are interpreted as the operators that describe correlations of many-particle systems. Thus, marginal
correlation operators \eqref{gBigfromDFB} are cumulants (semi-invariants) of the marginal density operators.
It is obvious that definition \eqref{Gexpg} follows from \eqref{gBigfromDFB} in consequence of definition
\eqref{FClusters} and relations \eqref{grel} between correlation operators of particle clusters and correlation
operators of particles.

The evolution of all possible states of quantum many-particle systems obeying the Maxwell-Boltzmann
statistics with the Hamiltonian \eqref{H} can be described within the framework of marginal correlation
operators governed by the nonlinear quantum BBGKY hierarchy
\begin{eqnarray}
 \label{gBigfromDFBa}
    &&\frac{d}{dt}G_s(t,Y)=\mathcal{N}(Y\mid G(t))+
        \mathrm{Tr}_{s+1}\sum_{i\in Y}(-\mathcal{N}_{\mathrm{int}}(i,s+1))\big(G_{s+1}(t,Y,s+1)+\\
    &&+\sum_{\mbox{\scriptsize$\begin{array}{c}\mathrm{P}:(Y,s+1)=X_1\bigcup X_2,\\i\in
        X_1;s+1\in X_2\end{array}$}}G_{|X_1|}(t,X_1)G_{|X_2|}(t,X_2)\big),\nonumber\\ \nonumber\\
 \label{gBigfromDFBai}
    &&G_{s}(t,Y)\big|_{t=0}=G_{s}^0(Y),\quad s\geq1.
\end{eqnarray}
In equation \eqref{gBigfromDFBa} the operator $\mathcal{N}(Y|G(t))$ is generator of the von Neumann
hierarchy \eqref{vonNeumannTwoBody} defined by
\begin{eqnarray}\label{Nnl}
    &&\mathcal{N}(Y\mid G(t))\doteq-\mathcal{N}_{s}(Y)G_s(t,Y)+\\
    &&+\sum\limits_{\mathrm{P}:\,Y=X_{1}\bigcup X_2}\,\sum\limits_{i_{1}\in X_{1}}\sum\limits_{i_{2}\in X_{2}}
       (-\mathcal{N}_{\mathrm{int}}(i_{1},i_{2}))G_{|X_{1}|}(t,X_{1})G_{|X_{2}|}(t,X_{2}),\nonumber
\end{eqnarray}
where ${\sum\limits}_{\mathrm{P}:\,Y=X_{1}\bigcup X_2}$ is the sum over all possible partitions
$\mathrm{P}$ of the set $Y\equiv(1,\ldots,s)$ into two nonempty mutually disjoint subsets $X_1\subset Y$
and $X_2\subset Y$, and $\sum_{\mbox{\scriptsize$\begin{array}{c}\mathrm{P}:(Y,s+1)=X_1\bigcup X_2,
\\i\in X_1; s+1\in X_2\end{array}$}}$ is the sum over all possible partitions of the set $(Y,s+1)$ into
two mutually disjoint subsets $X_1$ and $X_2$ such that $i$ particle  belongs to the subset $X_1$ and
$s+1$ particle belongs to $X_2$.

We note that in case of many-body interaction potentials \eqref{Hn} hierarchy \eqref{gBigfromDFBa}
has the form
\begin{eqnarray}\label{hn}
   &&\hskip-5mm\frac{d}{dt}G_s(t,Y)=\mathcal{N}(Y\mid G(t))+\\
   &&\hskip-5mm+\sum_{n=1}^\infty\sum_{k=1}^n\sum_{1=j_1<\ldots<j_k}^s \mathrm{Tr}_{s+1,\ldots,s+1+n-k}
     (-\mathcal{N}_{\mathrm{int}}^{(n+1)}(j_1,\ldots,j_k,s+1,\ldots,s+1+n-k))\times\nonumber\\
   &&\hskip-5mm\times\sum_{\mbox{\scriptsize$\begin{array}{c}\mathrm{P}:(1,\ldots,s+1+n-k)=\bigcup_{i=1}X_i,
     \\X_i\not \subseteq Y\backslash(j_1,\ldots,j_k),\,|\mathrm{P}|\leq n+1\end{array}$}}
     \prod_{X_{i}\subset\mathrm{P}}G_{|X_i|}(t,X_i),\nonumber
\end{eqnarray}
where we use notations accepted above.

The nonlinear quantum BBGKY hierarchy \eqref{gBigfromDFBa} is derived on basis of the
von Neumann hierarchy for correlation operators \eqref{vNh} according to definition \eqref{Gexpg}
or on basis of the quantum BBGKY hierarchy for marginal density operators \eqref{BBGKY} according to
cluster expansions \eqref{FG}. The evolution of marginal correlation operators of both finitely and
infinitely many quantum particles is described by initial-value problem of the nonlinear quantum BBGKY
hierarchy \eqref{gBigfromDFBa}. For finitely many particles the nonlinear quantum BBGKY hierarchy
is equivalent to the von Neumann hierarchy \eqref{vNh}.

To construct a nonperturbative solution of the Cauchy problem \eqref{gBigfromDFBa}-\eqref{gBigfromDFBai}
of the nonlinear quantum BBGKY hierarchy we first consider in capacity of initial data the marginal
correlation operators satisfying a chaos property. In terms of marginal correlation operators \eqref{Gexpg}
a chaos property means that
\begin{eqnarray}\label{inG}
   &&G_s^0(1,\ldots,s)=g_1^0(1)\delta_{s,1},\quad s\geq1,
\end{eqnarray}
where $\delta_{s,1}$ is a Kronecker symbol. Taking into account the structure of a solution
\eqref{rozvChaosN} of the von Neumann hierarchy \eqref{vNh} in case of initial data
\eqref{posl_g(0)}, from expansion \eqref{Gexpg} we obtain
\begin{eqnarray*}\label{Gg(0)}
   &&G_{s}(t,1,\ldots,s)=\sum\limits_{n=0}^{\infty}\frac{1}{n!}
       \,\mathrm{Tr}_{s+1,\ldots, s+n}\,\mathfrak{A}_{s+n}(-t,1,\ldots,s+n)
       \prod_{i=1}^{s+n}g_{1}^0(i),,\quad s\geq 1,
\end{eqnarray*}
where $\mathfrak{A}_{s+n}(-t)$ is the $(s+n)th$-order cumulant \eqref{cumulants} of groups of operators
\eqref{groupG}. In consequence of equality \eqref{inG} we finally derive
\begin{eqnarray}\label{GUG(0)}
   &&G_{s}(t,1,\ldots,s)=\sum\limits_{n=0}^{\infty}\frac{1}{n!}
       \,\mathrm{Tr}_{s+1,\ldots, s+n}\,\mathfrak{A}_{s+n}(-t,1,\ldots,s+n)
       \prod_{i=1}^{s+n}G_{1}^0(i).
\end{eqnarray}
Since estimate \eqref{est} holds, series \eqref{GUG(0)} converges provided that
$\|G_{1}(0)\|_{\mathfrak{L}^{1}(\mathcal{H})}\leq (2e)^{-1}$.

Thus, the cumulant structure of solution \eqref{rozvNF-N_F} of the von Neumann hierarchy
\eqref{vNh} induces the cumulant structure of expansion solution \eqref{GUG(0)} of the
initial-value problem of the nonlinear quantum  BBGKY hierarchy for marginal correlation operators.

We emphasize that in case of chaos initial data solution expansion (\ref{RozvBBGKY}) of the quantum
BBGKY hierarchy (\ref{BBGKY}) for marginal density operators differs from solution expansion (\ref{GUG(0)})
of the nonlinear quantum BBGKY hierarchy (\ref{gBigfromDFBa}) for marginal correlation operators only by
the order of the cumulants of the groups of operators of the von Neumann equations\cite{DP},\cite{GP}
\begin{eqnarray}\label{FUg}
   &&F_{s}(t,Y)=\sum\limits_{n=0}^{\infty}\frac{1}{n!}
       \,\mathrm{Tr}_{s+1,\ldots, s+n}\,\mathfrak{A}_{1+n}(-t,\{Y\},X\setminus Y)
       \,\prod_{i=1}^{s+n}F_{1}^0(i), \quad s\geq 1,
\end{eqnarray}
where the operator $\mathfrak{A}_{1+n}(t)$ is $(1+n)th$-order cumulant (\ref{cumulant1+n}).
Series (\ref{FUg}) converges under the condition that: $\|F_{1}(0)\|_{\mathfrak{L}^{1}(\mathcal{H})}\leq e^{-1}$.

Let us construct a nonperturbative solution expansion of the Cauchy problem \eqref{gBigfromDFBa}-
\eqref{gBigfromDFBai} in the case of general initial data. According to definition
\eqref{Gexpg} of the marginal correlation operators, i.e.
\begin{eqnarray*}
    &&G(t)=e^\mathfrak{a}g(t),
\end{eqnarray*}
where the operator $e^\mathfrak{a}$ is defined by \eqref{opann} and the sequence $g(t)$ is a solution of
the von Neumann hierarchy defined by group \eqref{rozvNh}, i.e.
\begin{eqnarray}\label{rNF-N_F}
    &&g(t)=\mathcal{G}(t\mid g(0)),
\end{eqnarray}
and the equality: $g(0)=e^\mathfrak{-a}G(0)$, we finally derive
\begin{eqnarray}\label{s}
    &&G(t)=e^\mathfrak{a}\mathcal{G}(t\mid e^\mathfrak{-a}G(0)).
\end{eqnarray}

Thus, a solution of the Cauchy problem of the nonlinear quantum BBGKY hierarchy for marginal correlation
operators is defined by the following one-parametric mapping
\begin{eqnarray*}
    &&\mathbb{R}\ni t\rightarrow \mathcal{U}(t\mid f)=e^\mathfrak{a}\mathcal{G}(t\mid e^\mathfrak{-a}f),
\end{eqnarray*}
which is defined on the space $\mathfrak{L}^{1}(\mathcal{F}_\mathcal{H})$ and has the group property.
On the subspaces $\mathfrak{L}^{1}_{0}(\mathcal{H}_{s})\subset \mathfrak{L}^{1}(\mathcal{H}_{s})$, $s\geq1$,
the infinitesimal generator $\mathcal{B}(Y|f)$ of this group is defined by the operator
\begin{eqnarray*}\label{gNnl}
    &&\mathcal{B}(Y\mid f)\doteq \mathcal{N}(Y\mid f)+
        \mathrm{Tr}_{s+1}\sum_{i\in Y}(-\mathcal{N}_{\mathrm{int}}(i,s+1))\big(f_{s+1}(t,Y,s+1)+\\
    &&+\sum_{\mbox{\scriptsize$\begin{array}{c}\mathrm{P}:(Y,s+1)=X_1\bigcup X_2,\\i\in
        X_1;s+1\in X_2\end{array}$}}f_{|X_1|}(t,X_1)f_{|X_2|}(t,X_2)\big),
\end{eqnarray*}
where the same notations as for formula \eqref{gBigfromDFBa} have been used.

To set down formula \eqref{s} in componentwise form, we observe that the following equality holds
\begin{eqnarray*}
    &&\prod\limits_{X_i \subset \mathrm{P}}(e^\mathfrak{-a}G(0))_{|X_i|}(X_i)=
       \sum\limits_{k=0}^{\infty}\frac{(-1)^k}{k!}\mathrm{Tr}_{s+n+1,\ldots,s+n+k}
       \sum\limits_{k_1=0}^{k}\frac{k!}{k_{1}!(k-k_{1})!}\ldots\\
    &&\ldots\sum\limits_{k_{|\mathrm{P}|-1}=0}^{k_{|\mathrm{P}|-2}}
       \frac{k_{|\mathrm{P}|-2}!}{k_{|\mathrm{P}|-1}!(k_{|\mathrm{P}|-2}-k_{|\mathrm{P}|-1})!}
       G_{|X_1|+k-k_1}^{0}(X_1,s+n+1,\ldots,s+n+k-k_1)\ldots\\
    &&\ldots\times G_{|X_{|\mathrm{P}|}|+k_{|\mathrm{P}|-1}}^{0}(X_{|\mathrm{P}|},
       s+n+k-k_{|\mathrm{P}|-1}+1,\ldots,s+n+k).
\end{eqnarray*}
Then in view of definitions \eqref{opann} and \eqref{rozvNh}, we have
\begin{eqnarray*}
    &&G_{s}(t,Y)=\sum\limits_{n=0}^{\infty}\frac{1}{n!}
        \,\mathrm{Tr}_{s+1,\ldots, s+n}\,\sum\limits_{\mathrm{P}:\,(Y,s+1,\ldots, s+n)=\bigcup_i X_i}
        \mathfrak{A}_{|\mathrm{P}|}(-t,\{X_1\},\ldots,\{X_{|\mathrm{P}|}\})\times\\
    &&\times\prod\limits_{X_i \subset \mathrm{P}}(e^\mathfrak{-a}G(0))_{|X_i|}(X_i),
\end{eqnarray*}
and as a result we derive a solution expansion of the nonlinear quantum BBGKY hierarchy \cite{GP11}
\begin{eqnarray}\label{sss}
    &&G_{s}(t,Y)=\sum\limits_{n=0}^{\infty}\frac{1}{n!}
        \,\mathrm{Tr}_{s+1,\ldots, s+n}\,U_{1+n}(t;\{Y\},s+1,\ldots,s+n\mid G(0)),\quad s\geq1,
\end{eqnarray}
where the $(1+n)th$-order reduced cumulant $U_{1+n}(t)$ of groups \eqref{rozvNh} has been introduced
\begin{eqnarray}\label{ssss}
    &&U_{1+n}(t;\{Y\},s+1,\ldots,s+n \mid G(0))\doteq\\
    &&\doteq\sum\limits_{k=0}^{n}(-1)^k \frac{n!}{k!(n-k)!}\,\sum\limits_{\mathrm{P}:\,
        (1,\ldots,s+n-k)=\bigcup_i X_i}
        \mathfrak{A}_{|\mathrm{P}|}(-t,\{X_1\},\ldots,\{X_{|\mathrm{P}|}\})\times\nonumber\\
    &&\times\sum\limits_{k_1=0}^{k}\frac{k!}{k_{1}!(k-k_{1})!}\ldots
        \sum\limits_{k_{|\mathrm{P}|-1}=0}^{k_{|\mathrm{P}|-2}}\frac{k_{|\mathrm{P}|-2}!}{k_{|\mathrm{P}|-1}!
        (k_{|\mathrm{P}|-2}-k_{|\mathrm{P}|-1})!}G_{|X_1|+k-k_1}^{0}(X_1,\nonumber\\
    &&s+n-k+1,\ldots,s+n-k_1)\ldots G_{|X_{|\mathrm{P}|}|+k_{|\mathrm{P}|-1}}^{0}
        (X_{|\mathrm{P}|},s+n-k_{|\mathrm{P}|-1}+1,\ldots,s+n).\nonumber
\end{eqnarray}

Let us indicate some properties of reduced cumulants \eqref{ssss} of groups of operators \eqref{rozvNh}.
In case of $n=0$, first order reduced cumulant \eqref{ssss} has the form
\begin{eqnarray*}
    &&U_{1}(t;\{Y\}\mid f)=\sum\limits_{\mathrm{P}:\,Y=\bigcup_i X_i}
        \mathfrak{A}_{|\mathrm{P}|}(-t,\{X_1\},\ldots,\{X_{|\mathrm{P}|}\})
        \prod\limits_{X_i\subset\mathrm{P}}f_{|X_i|}(X_{i}),
\end{eqnarray*}
i.e. it is the group of operators \eqref{rozvNh}. Its infinitesimal generator coincides with generator
of the von Neumann hierarchy \eqref{vonNeumannTwoBody}
\begin{eqnarray*}
    &&\lim\limits_{t\rightarrow 0}\frac{1}{t}(U_{1}(t;\{Y\}\mid f)-f_{s}(Y))=\mathcal{N}(Y\mid f), \quad s\geq1,
\end{eqnarray*}
for $f\in\mathfrak{L}^{1}(\mathcal{F}_\mathcal{H})$ in the sense of the norm convergence of the
space $\mathfrak{L}^{1}(\mathcal{H}_s)$, where the operator $\mathcal{N}(Y\mid f)$ is defined by
formula \eqref{Nnl}.

In case of $n=1$ for second order reduced cumulant \eqref{ssss} in the same sense we obtain
the following equality
\begin{eqnarray*}
   &&\mathrm{Tr}_{s+1}\lim\limits_{t\rightarrow 0}\frac{1}{t}\,U_{2}(t;\{Y\},s+1\mid f)
       =\sum_{i\in Y}\mathrm{Tr}_{s+1}(-\mathcal{N}_{\mathrm{int}}(i,s+1))\big(f_{s+1}(t,Y,s+1)+\\
   &&+\sum_{\mbox{\scriptsize$\begin{array}{c}\mathrm{P}:(Y,s+1)=X_1\bigcup X_2,\\i\in
        X_1;s+1\in X_2\end{array}$}}f_{|X_1|}(t,X_1)f_{|X_2|}(t,X_2)\big),
\end{eqnarray*}
where notations are used as above for hierarchy \eqref{gBigfromDFBa}, and for $n\geq2$ as a
consequence of the fact that we consider a system of particles interacting by a two-body potential,
it holds
\begin{eqnarray*}
   &&\mathrm{Tr}_{s+1,\ldots,s+n}\lim\limits_{t\rightarrow 0}\frac{1}{t}\,U_{1+n}(t;\{Y\},s+1,\ldots,s+n\mid f)=0.
\end{eqnarray*}

According to the estimate
\begin{eqnarray*}
  &&\big\|\mathfrak{A}_{|\mathrm{P}|}(-t,\{X_1\},\ldots,
     \{X_{|\mathrm{P}|}\})f_n\big\|_{\mathfrak{L}^{1}(\mathcal{H}_{n})}
     \leq |\mathrm{P}|!\,e^{|\mathrm{P}|}\big\|f_n\big\|_{\mathfrak{L}^{1}(\mathcal{H}_{n})},
\end{eqnarray*}
on the space $\mathfrak{L}^{1}(\mathcal{H}_{s})$, series \eqref{sss} converges provided that
$\max_{n\geq1}\big\|G_n^0\big\|_{\mathfrak{L}^{1}(\mathcal{H}_{n})}<(2e^{3})^{-1}$.

For abstract initial-value problem for hierarchy (\ref{gBigfromDFBa}) on the space of sequences of
trace-class operators $\mathfrak{L}^{1}(\mathcal{F}_\mathcal{H})$ the following statement is true \cite{GP11}.

\begin{theorem}
If $\max_{n\geq1}\big\|G_n^0\big\|_{\mathfrak{L}^{1}(\mathcal{H}_{n})}<(2e^{3})^{-1}$,
then for $t\in\mathbb{R}$ a solution of the initial-value problem \eqref{gBigfromDFBa}-\eqref{gBigfromDFBai}
of the nonlinear quantum BBGKY hierarchy is determined by expansion \eqref{sss}.
If $G_{n}^{0}\in\mathfrak{L}^{1}_{0}(\mathcal{H}_{n})\subset\mathfrak{L}^{1}(\mathcal{H}_{n})$
it is a strong (classical) solution and for arbitrary initial data
$G_{n}^{0}\in\mathfrak{L}^{1}(\mathcal{H}_{n})$ it is a weak (generalized) solution.
\end{theorem}

We note also that in case of many-particle systems obeying quantum statistics, i.e.
many-particle systems of fermions or bosons the nonlinear quantum BBGKY hierarchy for
marginal correlation operators has the form
\begin{eqnarray*}
   &&\frac{d}{dt}G_s(t,Y)=\mathcal{N}(Y|G(t))+
      \mathrm{Tr}_{s+1}\sum_{i\in Y}(-\mathcal{N}_{\mathrm{int}}(i,s+1))\big(G_{s+1}(t,Y,s+1)+\\
   &&+\sum_{\mbox{\scriptsize$\begin{array}{c}\mathrm{P}:(Y,s+1)=X_1\bigcup X_2,\\i\in
      X_1;s+1\in X_2\end{array}$}}\mathcal{S}_{s+1}^{\pm}G_{|X_1|}(t,X_1)G_{|X_2|}(t,X_2)\big),\quad s\geq1,
\end{eqnarray*}
where $\mathcal{N}(Y|G(t))$ is generator \eqref{Nnlfb} of the von Neumann hierarchy of fermions
or bosons and the operator $\mathcal{S}_{s+1}^{\pm}$ is defined by formula \eqref{Sn}.

We emphasize that the evolution of marginal correlation operators of both finitely and infinitely many
quantum particles is described by initial-value problem of the nonlinear quantum BBGKY hierarchy
\eqref{gBigfromDFBa}. For finitely many particles the nonlinear quantum BBGKY hierarchy is equivalent
to the von Neumann hierarchy \eqref{vNh}).


\section{The origin of the quantum kinetic evolution}
It is well known \cite{BQ,CGP97} that in certain situations the collective behavior of many-particle
systems can be adequately described by the kinetic equations, i.e. by evolution equations for a
one-particle marginal density operator.
In this section we discuss the problem of potentialities inherent in the description of the evolution
of states of many-particle systems in terms of a one-particle density operator or more exactly the
problem of an equivalence of the hierarchies of quantum evolution equations and quantum kinetic equations.
We demonstrate that in fact, if initial data is completely defined by a one-particle marginal density
operator, then all possible states of infinite-particle systems at arbitrary moment of time can be
described within the framework of a one-particle density operator and the marginal functionals of such
states without any approximations \cite{GT},\cite{G11}. By other words the state described in terms of
the sequence $F(t)=(1,F_1(t),\ldots,F_s(t),\ldots)$ of marginal density operators can be described within
the framework of the sequence $F(t\mid F_{1}(t))=(1,F_1(t),F_2(t\mid F_{1}(t)),\ldots,F_s(t\mid F_{1}(t)),\ldots)$
of explicitly defined marginal functionals $F_s(t\mid F_{1}(t)),\,s\geq2$, of the solution $F_1(t)$ of the
generalized quantum kinetic equation.

Thus, in this section the origin of the microscopic description of quantum many-particle systems by
means of kinetic equations is considered.

\subsection{The generalized quantum kinetic equation}
We consider the Cauchy problem of the quantum BBGKY hierarchy \eqref{BBGKY} with initial data
$F(t)|_{t=0}=F^{(c)}\equiv(F_1^0(1),\ldots,{\prod}_{i=1}^s F_1^0(i),\ldots)$, which is intrinsic
for the kinetic description of many-particle systems because in this case all possible states are
characterized by means of a one-particle marginal density operator.
Then we deal with initial value problem of the quantum BBGKY hierarchy \eqref{BBGKY} which
is not completely well-defined Cauchy problem, because the generic initial data $F^{(c)}$, is not
independent for every unknown marginal density operator $F_{s}(t,1,\ldots,s),\,s\geq1$, from the
hierarchy of equations. Consequently such initial-value problem can be naturally reformulated as the new
Cauchy problem for a one-particle density operator, that corresponds to the independent initial
one-particle density operator and the sequence of explicitly defined marginal functionals of the
state $F_{s}\big(t,1,\ldots,s\mid F_{1}(t)\big),\,s\geq 2$, of the solution $F_{1}(t)$
of such Cauchy problem \cite{GT}.

Let us formulate the restated Cauchy problem and a sequence of the marginal functionals of the state
which describe the evolution of all possible states of quantum many particles in an equivalent way as
compared with the quantum BBGKY hierarchy.

The one-particle density operator $F_{1}(t)$ is governed by the following initial-value
problem of the generalized quantum kinetic equation \cite{GT}
\begin{eqnarray}
  \label{gke}
    &&\frac{d}{dt}F_{1}(t,1)=-\mathcal{N}(1)F_{1}(t,1)+\\
    &&+\mathrm{Tr}_{2}(-\mathcal{N}_{\mathrm{int}}(1,2))
        \sum\limits_{n=0}^{\infty}\frac{1}{n!}\mathrm{Tr}_{3,\ldots,n+2}\,
        \mathfrak{V}_{1+n}(t,\{1,2\},3,\ldots,n+2)\prod _{i=1}^{n+2} F_{1}(t,i),\nonumber\\
        \nonumber\\
  \label{2}
    &&F_1(t,1)|_{t=0}= F_1^0(1),
\end{eqnarray}
where the operator $\mathcal{N}_{\mathrm{int}}(1,2)$ is defined by formula \eqref{comst},
and the $(1+n)th$-order generated evolution operator $\mathfrak{V}_{1+n}(t),\,n\geq0$, is defined as
follows (in case of $s=2$ for $Y\equiv(1,\ldots,s)$ and $X\setminus Y\equiv(s+1,\ldots,{s+n})$)
\begin{eqnarray}\label{skrr}
   &&\mathfrak{V}_{1+n}(t,\{Y\},X\setminus Y)\doteq
      \sum_{k=0}^{n}\,(-1)^k\,\sum_{n_1=1}^{n}\ldots\sum_{n_k=1}^{n-n_1-\ldots-n_{k-1}}
      \frac{n!}{(n-n_1-\ldots-n_k)!}\times\\
   &&\times\widehat{\mathfrak{A}}_{1+n-n_1-\ldots-n_k}(t,\{Y\},s+1,\ldots,s+n-n_1-\ldots-n_k)
      \times\nonumber
\end{eqnarray}
\begin{eqnarray*}
   &&\times\prod_{j=1}^k\sum\limits_{\mbox{\scriptsize$\begin{array}{c}\mathrm{D}_{j}:Z_j=\bigcup_{l_j} X_{l_j},\\
      |\mathrm{D}_{j}|\leq s+n-n_1-\dots-n_j\end{array}$}}\frac{1}{|\mathrm{D}_{j}|!}\sum_{i_1\neq\ldots\neq i_{|\mathrm{D}_{j}|}=1}^{s+n-n_1-\ldots-n_j}\prod_{X_{l_j}\subset \mathrm{D}_{j}}\,\frac{1}{|X_{l_j}|!}\,\,
      \widehat{\mathfrak{A}}_{1+|X_{l_j}|}(t,i_{l_j},X_{l_j}).\nonumber
\end{eqnarray*}
In expansion \eqref{skrr} we denote by $\sum_{\mathrm{D}_{j}:Z_j=\bigcup_{l_j} X_{l_j}}$ is the sum over
all possible dissections of the linearly ordered set $Z_j\equiv(s+n-n_1-\ldots-n_j+1,\ldots,s+n-n_1-\ldots-n_{j-1})$
on no more than $s+n-n_1-\ldots-n_j$ linearly ordered subsets and $\widehat{\mathfrak{A}}_{1+n}(t)$ is the
$(1+n)th$-order cumulant
\begin{eqnarray*}
   &&\widehat{\mathfrak{A}}_{1+n}(t,\{Y\},X\setminus Y)=\sum\limits_{\mathrm{P}\,:(\{Y\},\, X\setminus Y)=
      {\bigcup\limits}_i X_i}(-1)^{|\mathrm{P}|-1}(|\mathrm{P}|-1)!
      \prod_{X_i\subset\mathrm{P}}\widehat{\mathcal{G}}_{|\theta(X_i)|}(t,\theta(X_i)),
\end{eqnarray*}
of the scattering operators \eqref{so}
\begin{eqnarray*}
   &&\widehat{\mathcal{G}}_{n}(t)=\mathcal{G}_{n}(-t,1,\ldots,n)
      \prod _{i=1}^{n}\mathcal{G}_{1}(t,i), \quad n\geq1.
\end{eqnarray*}
The series of collision integral \eqref{gke} converges under the condition:
$\|F_{1}(t)\|_{\mathfrak{L}^{1}(\mathcal{H})}<e^{-8}$ \cite{GG11}.

The marginal functionals of the state $F_{s}\big(t,1,\ldots,s\mid F_{1}(t)\big),\,s\geq 2$, are
represented by the following expansions over products of the solution $F_{1}(t)$ of the Cauchy
problem \eqref{gke}-\eqref{2}
\begin{eqnarray}\label{f}
   &&F_{s}(t,Y\mid F_{1}(t))\doteq\sum _{n=0}^{\infty}\frac{1}{n!}\,
      \mathrm{Tr}_{s+1,\ldots,{s+n}}\,\mathfrak{V}_{1+n}(t,\{Y\},X\setminus Y)
      \prod _{i=1}^{s+n}F_{1}(t,i),
\end{eqnarray}
where the $(1+n)th$-order generated evolution operator $\mathfrak{V}_{1+n}(t),\,n\geq0$, is defined
by expansion \eqref{skrr}. The marginal functionals of the state are represented by converged series
\eqref{f} under the condition that: $\|F_{1}(t)\|_{\mathfrak{L}^{1}(\mathcal{H})}<e^{-(3s+2)}$ \cite{GG11}.

We observe that the kinetic dynamics of states is described in terms of cumulants of scattering operators
\eqref{so} in contrast to the evolution of states described by the quantum BBGKY hierarchy \eqref{BBGKY}.
We give a few examples of the generated evolution operators $\mathfrak{V}_{n}$, $n\geq1$, of the lower orders
\begin{eqnarray}\label{skrre}
   &&\mathfrak{V}_{1}(t,\{Y\})=\widehat{\mathfrak{A}}_{1}(t,\{Y\}),\\
   &&\mathfrak{V}_{2}(t,\{Y\},s+1)=\widehat{\mathfrak{A}}_{2}(t,\{Y\},s+1)-
      \widehat{\mathfrak{A}}_{1}(t,\{Y\})\sum_{i=1}^s \widehat{\mathfrak{A}}_{2}(t,i,s+1).\nonumber
\end{eqnarray}

It should be emphasized that in case under consideration, i.e. in case of the absence of correlations at initial time,
the correlations generated by the dynamics of a system are completely governed by evolution operators \eqref{skrr}.

Typical properties for the kinetic description of the evolution of constructed marginal functionals
of the state \eqref{f} are induced by the properties of generated evolution operators \eqref{skrr}.
Let us indicate some intrinsic properties of the evolution operators $\mathfrak{V}_{1+n}(t),\,n\geq0$,
representative for cumulants (semi-invariants) of groups of operators.

First of all we observe that they are solutions of the following recursive relations (the kinetic
cluster expansions, which are in some sense analog of equilibrium virial expansions)\cite{GT}
\begin{eqnarray}\label{kce}
   &&\mathfrak{A}_{1+n}(-t,\{Y\},s+1,\ldots,s+n)=\\
   &&=\sum_{n_1=0}^{n}\frac{n!}{(n-n_1)!}\,
     \mathfrak{V}_{1+n-n_1}(t,\{Y\},s+1,\ldots,s+n-n_1)
     \sum\limits_{\mbox{\scriptsize $\begin{array}{c}\mathrm{D}:Z=\bigcup_k X_k,\\|\mathrm{D}|
     \leq s+n-n_1\end{array}$}}\frac{1}{|\mathrm{D}|!}\times\nonumber\\
   &&\times\,\sum_{i_1\neq\ldots\neq i_{|\mathrm{D}|}=1}^{s+n-n_1}\,
     \prod_{X_{k}\subset \mathrm{D}}\,\frac{1}{|X_k|!}\,\mathfrak{A}_{1+|X_{k}|}(-t,i_k,X_{k})
     \prod\limits_{\mbox{\scriptsize$\begin{array}{c}{m=1},
     \\m\neq i_1,\ldots,i_{|\mathrm{D}|}\end{array}$}}^{s+n-n_1}\mathfrak{A}_1(-t,m),\nonumber
\end{eqnarray}
where $\sum_{\mathrm{D}:Z=\bigcup_l X_l,\,|\mathrm{D}|\leq s+n-n_1}$ is the sum over all possible
dissections $\mathrm{D}$ of the linearly ordered set $Z\equiv(s+n-n_1+1,\ldots,s+n)$ on no more than
$s+n-n_1$ linearly ordered subsets.

Since in case of a system of non-interacting particles for scattering operators \eqref{so}
the equality holds: $\widehat{\mathcal{G}}_{n}(t)=I$, where $I$ is a unit operator, then we have
\begin{eqnarray*}
    &&\mathfrak{V}_{1+n}(t)=I\delta_{n,0},
\end{eqnarray*}
where $\delta_{n,0}$ is a Kronecker symbol. Similarly, at initial time $t=0$ it is true:
$\mathfrak{V}_{1+n}(0)=I\delta_{n,0}$.

The infinitesimal generator of the first-order generated evolution operator \eqref{skrr} is defined by
the following limit in the sense of the norm convergence in the space $\mathfrak{L}^{1}(\mathcal{H}_{n})$
\begin{eqnarray*}
    &&\lim\limits_{t\rightarrow 0}\frac{1}{t}(\mathfrak{V}_{1}(t,\{1,\ldots,n\})-I)f_{n}
       =\sum_{i<j=1}^n(-\mathcal{N}_{\mathrm{int}}(i,j))f_{n},
\end{eqnarray*}
where the operator $(-\mathcal{N}_{\mathrm{int}}(i,j))$ is defined by formula \eqref{com}.
In general case, i.e. $n\geq2$, in the sense of the norm convergence on the space
$\mathfrak{L}^{1}(\mathcal{H}_{n})$ for the $n$-order generated evolution operator \eqref{skrr} it holds
\begin{eqnarray*}
    &&\lim\limits_{t\rightarrow 0}\frac{1}{t}\mathfrak{V}_{n}(t,1,\ldots,n) f_{n}=0.
\end{eqnarray*}

Before constructing a solution of initial-value problem \eqref{gke}-\eqref{2} in the space
$\mathfrak{L}^{1}(\mathcal{H})$ we generalize kinetic equation \eqref{gke} for particles interacting
via many-body interaction potentials $\Phi^{(n)},\,n\geq1$. In this case the generalized quantum kinetic
equation has the form
\begin{eqnarray}\label{gkeN}
    &&\frac{d}{dt}F_{1}(t,1)=-\mathcal{N}(1)F_{1}(t,1)+
       \sum\limits_{n=1}^{\infty}\sum _{k=1}^{n}\frac{1}{(n-k)!}\frac{1}{k!}
       \,\mathrm{Tr}_{2,\ldots,n+1}(-\mathcal{N}_{\mathrm{int}}^{(k+1)})(1,\\
    &&\ldots,k+1)\mathfrak{V}_{1+n-k}(t,\{1,\ldots,k+1\},k+2,\ldots,n+1)
       \prod _{i=1}^{n+1} F_{1}(t,i),\nonumber
\end{eqnarray}
where $\mathfrak{V}_{1+n-k}(t)$, is the $(1+n-k)th$-order generated evolution operator \eqref{skrr}
and notations \eqref{com},\eqref{gennd} are used. The collision integral in the generalized quantum
kinetic equation \eqref{gkeN} is defined by the convergent series under condition that
$\|F_{1}(t)\|_{\mathfrak{L}^{1}(\mathcal{H})}<e^{-8}$ \cite{GT}.

For the sake of a comparison of the structure of various collision integral components
in \eqref{gkeN} we give expressions of the collision integral term describing a two-body
interaction and three particle correlations
\begin{eqnarray*}
    &&\mathrm{Tr}_{2,3}(-\mathcal{N}_{\mathrm{int}}^{(2)})(1,2)
       \mathfrak{V}_{2}(t,\{1,2\},3)F_{1}(t,1)F_{1}(t,2)F_{1}(t,3),
\end{eqnarray*}
and the collision integral term describing a three-body interaction
\begin{eqnarray*}
    &&\frac{1}{2!}\mathrm{Tr}_{2,3}(-\mathcal{N}_{\mathrm{int}}^{(3)})(1,2,3)
       \mathfrak{V}_{1}(t,\{1,2,3\})F_{1}(t,1)F_{1}(t,2)F_{1}(t,3),
\end{eqnarray*}
where the evolution operators $\mathfrak{V}_{2}(t,\{1,2\},3)$ and $\mathfrak{V}_{1}(t,\{1,2,3\})$
are defined by \eqref{skrr}.

For the Cauchy problem \eqref{gkeN}-\eqref{2} (or \eqref{gke}-\eqref{2}) on the space
$\mathfrak{L}^{1}(\mathcal{H})$ the following statement is true.

\begin{theorem}
The global in time solution of initial-value problem \eqref{gkeN}-\eqref{2} is determined by
the following expansion
\begin{eqnarray}\label{ske}
   &&F_{1}(t,1)= \sum\limits_{n=0}^{\infty}\frac{1}{n!}\,\mathrm{Tr}_{2,\ldots,{1+n}}\,\,
      \mathfrak{A}_{1+n}(-t,1,\ldots,n+1)\prod _{i=1}^{n+1}F_{1}^0(i),
\end{eqnarray}
where the cumulants $\mathfrak{A}_{1+n}(-t),\, n\geq0,$ are defined by formula \eqref{cumulant1+n}.
If $\|F_1^0\|_{\mathfrak{L}^{1}(\mathcal{H})}<(e(1+e^{9}))^{-1}$, then for
$F_1^0\in\mathfrak{L}^{1}_{0}(\mathcal{H})$ it is a strong (classical) solution and for an arbitrary
initial data $F_1^{0}\in\mathfrak{L}^{1}(\mathcal{H})$ it is a weak (generalized) solution.
\end{theorem}

In section 4 the relationships of the evolution of observables and quantum states
described in terms of marginal density operators have been considered in the general case.
In case of initial states specified by a one-particle marginal density operator, the dual BBGKY hierarchy
describes the dual picture of the evolution to the picture of the evolution of states governed by the
generalized quantum kinetic equation and an infinite sequence of explicitly defined functionals of a
solution of such evolution equation. In fact, the following equality is true
\begin{eqnarray}\label{w}
   &&(B(t),F^{(c)})=\big(B(0),F(t\mid F_{1}(t))\big),
\end{eqnarray}
where the initial state $F^c$ is given as above, the sequence $B(t)$ defined by expansion \eqref{sdh}
and $F(t\mid F_{1}(t))=(1,F_1(t),F_2(t\mid F_{1}(t)),\ldots,F_s(t\mid F_{1}(t)),\ldots)$ is a sequence of marginal
functionals of the state \eqref{f} with the first element $F_1(t)$ given by series \eqref{ske}.

To verify equality \eqref{w} we transform functional $\langle B(t)|F^{(c)}\rangle$ as follows
\begin{eqnarray}\label{tf}
   &&(B(t),F^{(c)})=\sum_{s=0}^{\infty}\,\frac{1}{s!}\,
     \mathrm{Tr}_{\mathrm{1,\ldots,s}}\,B_{s}^0(1,\ldots,s)\times\\
   &&\times\sum\limits_{n=0}^{\infty}\frac{1}{n!}\,\mathrm{Tr}_{s+1,\ldots,s+n}\,
     \mathfrak{A}_{1+n}(-t,\{Y\},s+1,\ldots,s+n)\prod _{i=1}^{s}F_{1}^0(i),\nonumber
\end{eqnarray}
where the $(1+n)th$-order cumulant $\mathfrak{A}_{1+n}(-t,\{Y\},s+1,\ldots, s+n)$ is defined by
\eqref{cumulant1+n}. For $F_1^{0}\in\mathfrak{L}^{1}(\mathcal{H})$ and $B_{s}^{0}\in\mathfrak{L}(\mathcal{H}_s)$
obtained functional \eqref{tf} exists under the condition $\|F_1^0\|_{\mathfrak{L}^{1}(\mathcal{H})}<e^{-7}$.
Then we expand the cumulants $\mathfrak{A}_{1+n}(-t)$ over the new evolution operators
$\mathfrak{V}_{1+n}(t),\,n\geq0$, into the kinetic cluster expansions \eqref{kce}.
Representing series over the summation index $n$ and the sum over the summation index $n_1$
in functional \eqref{tf} as the two-fold series and identifying the series over the summation
index $n_1$ with the products of one-particle density operators
\begin{eqnarray}\label{prod}
  &&\sum_{n_1=0}^{\infty}\mathrm{Tr}_{s+n+1,\ldots,s+n+n_1}
     \sum\limits_{\mbox{\scriptsize $\begin{array}{c}\mathrm{D}:Z=
     \bigcup_k X_k,\\|\mathrm{D}|\leq s+n\end{array}$}}
     \,\,\sum_{i_1<\ldots<i_{|\mathrm{D}|}=1}^{s+n}\,
     \prod_{X_{k}\subset \mathrm{D}}\,\frac{1}{|X_k|!}\times\\
  &&\times\mathfrak{A}_{1+|X_{k}|}(-t,i_k,X_{k})\prod\limits_{\mbox{\scriptsize$\begin{array}{c}{l=1},
     \\l\neq i_1,\ldots, i_{|\mathrm{D}|}\end{array}$}}^{s+n}
     \mathfrak{A}_1(-t,l)\prod_{j=1}^{n+s+n_1}F_1^0(j)=\prod _{i=1}^{s+n} F_{1}(t,i),\nonumber
\end{eqnarray}
we transform functional \eqref{tf} to the form in terms of marginal functionals of the state \eqref{f}.
Thus, equality \eqref{w} holds.

In a particular case of the additive-type marginal observables $B^{(1)}(t)$ given by \eqref{af}
equality \eqref{w} is reduced to the form
\begin{eqnarray*}
   &&(B^{(1)}(t),F^{(c)})=\mathrm{Tr}_{1}\,a_1(1)F_{1}(t,1),
\end{eqnarray*}
where the one-particle marginal density operator $F_{1}(t)$ is represented by the expansion
\begin{eqnarray*}
   &&F_{1}(t,1)=\sum\limits_{n=0}^{\infty}\frac{1}{n!}\,\mathrm{Tr}_{2,\ldots,{1+n}}\,\,
      \mathfrak{A}_{1+n}(-t,1,\ldots,n+1)\prod _{i=1}^{n+1}F_{1}^0(i),
\end{eqnarray*}
which coincides with solution \eqref{ske} of the Cauchy problem \eqref{gke}-\eqref{2}. Hence for
additive-type marginal observables the generalized quantum kinetic equation \eqref{gke} is dual
to the dual quantum BBGKY hierarchy \eqref{dh} with respect to bilinear form \eqref{avmar-1}.

Let us derive the evolution equation, which satisfies obtained expansion for the one-particle
marginal density operator $F_{1}(t)$. Taking into account equality \eqref{infOper} and observing
the validity of equality \eqref{ic} for cumulants of groups \eqref{groupG}, we differentiate over
the time variable in the sense of pointwise convergence on the space $\mathfrak{L}^{1}(\mathcal{H})$
obtained expansion for $F_{1}(t)$. As result it holds
\begin{eqnarray}\label{de}
  &&\frac{d}{dt}F_{1}(t,1)=-\mathcal{N}_{1}(1)F_{1}(t,1)+\\
  &&+\mathrm{Tr}_{2}(-\mathcal{N}_{\mathrm{int}}(1,2))
       \sum\limits_{n=0}^{\infty}\frac{1}{n!}\mathrm{Tr}_{3,\ldots,n+2}
       \,\mathfrak{A}_{1+n}(t,\{1,2\},3,\ldots,n+2)\prod_{i=1}^{n+2} F_1^0(i).\nonumber
\end{eqnarray}
In the second summand in the right-hand side of equality \eqref{de} we expand cumulants \eqref{cumulant1+n} of groups
\eqref{groupG} into kinetic cluster expansions \eqref{kce} and represent series over the summation index $n$
and the sum over the summation index $n_1$ as the two-fold series. Then the following equalities take place
\begin{eqnarray*}
  &&\hskip-8mm\sum\limits_{n=0}^{\infty}\frac{1}{n!}\,\mathrm{Tr}_{2,\ldots,n+2}(-\mathcal{N}_{\mathrm{int}}(1,2))
     \mathfrak{A}_{1+n}(t,\{1,2\},3,\ldots,n+2)\prod_{i=1}^{n+2}F_1^0(i)=\\
  &&\hskip-8mm=\sum\limits_{n=0}^{\infty}\frac{1}{n!}\,\mathrm{Tr}_{2,\ldots,n+2}(-\mathcal{N}_{\mathrm{int}}(1,2))
     \sum_{n_1=0}^{n}\frac{n!}{(n-n_1)!}\mathfrak{V}_{1+n-n_1}(t,\{1,2\},3,\ldots,n+2-n_1)\times
\end{eqnarray*}
\begin{eqnarray*}
  &&\hskip-8mm\times\sum_{\mathrm{D}:Z=\bigcup_l X_l}\frac{1}{|\mathrm{D}|!}
      \sum_{i_1\neq\ldots\neq i_{|\mathrm{D}|}=1}^{n+2-n_1}\,
      \,\prod_{X_{l}\subset\mathrm{D}}\frac{1}{|X_l|!}\,\mathfrak{A}_{1+|X_{l}|}(t,i_l,X_{l})
      \prod\limits_{\mbox{\scriptsize$\begin{array}{c}{m=1},\\m\neq i_1,\ldots,i_{|\mathrm{D}|}\end{array}$}}^{2+n-n_1}
      \mathfrak{A}_1(t,m)\prod_{i=1}^{n+2} F_1^0(i)=\\
  &&\hskip-8mm=\mathrm{Tr}_{2}(-\mathcal{N}_{\mathrm{int}}(1,2))\sum\limits_{n=0}^{\infty}
      \frac{1}{n!}\,\mathrm{Tr}_{3,\ldots,n+2}\mathfrak{V}_{1+n}(t,\{1,2\},3,\ldots,n+2)
      \sum_{n_1=0}^{\infty}\,\sum_{\mathrm{D}:Z^{'}=\bigcup_l X_l}\frac{1}{|\mathrm{D}|!}\times\\
  &&\hskip-8mm\times\sum_{i_1\neq\ldots\neq i_{|\mathrm{D}|}=1}^{n+2}
      \,\prod_{X_{l}\subset\mathrm{D}}\frac{1}{|X_l|!}\,\mathfrak{A}_{1+|X_{l}|}(t,i_l,X_{l})
      \prod\limits_{\mbox{\scriptsize$\begin{array}{c}{m=1},\\m\neq i_1,\ldots,i_{|\mathrm{D}|}\end{array}$}}^{n+2}
      \mathfrak{A}_1(t,m)\prod_{i=1}^{n+2+n_1}F_1^0(i),
\end{eqnarray*}
where $Z\equiv(n+3-n_1,\ldots,n+2)$ and $Z^{'}\equiv(n+3,\ldots,n+2+n_1)$ are linearly ordered sets
and the notations accepted above are used. Consequently, in case of $s=2$ applying to the obtained
expression formula \eqref{prod}, from equality
\eqref{de} we derive
\begin{eqnarray*}\label{eq}
  &&\frac{d}{dt}F_{1}(t,1)=-\mathcal{N}_{1}(1)F_{1}(t,1)+\\
  &&+\mathrm{Tr}_{2}(-\mathcal{N}_{\mathrm{int}}(1,2))\sum_{n=0}^{\infty}\frac{1}{n!}\,\mathrm{Tr}_{3,\ldots,{n+2}}
     \,\mathfrak{V}_{1+n}(t,\{1,2\},3,\ldots,n+2)\prod _{i=1}^{n+2} F_{1}(t,i).\nonumber
\end{eqnarray*}
Constructed identity for a one-particle (marginal) density operator defined by expansion
\eqref{ske} we treat as the evolution equation, which governs the one-particle states of
many-particle quantum systems.

We remark that one more approach to the derivation of the generalized quantum kinetic equation consists
in its construction on the basis of dynamics of correlations \cite{GP},\cite{DP}.

Thus, in case of initial data $F^{(c)}\equiv(F_1^0(1),\ldots,{\prod}_{i=1}^s F_1^0(i),\ldots)$, which is
completely characterized by the one-particle marginal
density operator $F_{1}^0$, solution \eqref{sdh} of the Cauchy problem \eqref{dh}-\eqref{dhi} of the dual
quantum BBGKY hierarchy for marginal observables and a solution of the Cauchy problem of the generalized
kinetic equation \eqref{gke}-\eqref{2} together with marginal functionals of the state \eqref{f} give two
equivalent approaches to the description of the evolution of quantum many-particle systems.

We note also that in case of many-particle systems obeying the Fermi-Dirac and Bose-Einstein statistics
the generalized quantum kinetic equation has the form \cite{GT10}
\begin{eqnarray*}
     &&\frac{d}{dt}F_{1}(t,1)=-\mathcal{N}(1)F_{1}(t,1)+\\
     &&+\mathrm{Tr}_{2}(-\mathcal{N}_{\mathrm{int}}(1,2))\sum\limits_{n=0}^{\infty}\frac{1}{n!}\mathrm{Tr}_{3,\ldots,n+2}\,
        \mathfrak{V}_{1+n}(t,\{1,2\},3,\ldots,n+2)\mathcal{S}^{\pm}_{n+2}\prod _{i=1}^{n+2}F_{1}(t,i),
\end{eqnarray*}
where the operator $\mathcal{S}^{\pm}_{s}$ is defined by \eqref{Sn}, and the marginal functionals of the
states are represented by the series
\begin{eqnarray*}
    &&F_{s}(t,Y\mid F_{1}(t))\doteq\sum _{n=0}^{\infty}\frac{1}{n!}\,
        \mathrm{Tr}_{s+1,\ldots,{s+n}}\,\mathfrak{V}_{1+n}(t,\{Y\},\,X\setminus Y)
        \mathcal{S}^{\pm}_{s+n}\prod _{i=1}^{s+n}F_{1}(t,i),
\end{eqnarray*}
where the $(1+n)th$-order generated evolution operator $\mathfrak{V}_{1+n}(t),\,n\geq0$, is determined
by expansion \eqref{skrr}.

\subsection{The marginal functionals of correlations}
Within the framework of the description of states in terms of marginal correlation operators
\eqref{gBigfromDFB} the marginal correlation functionals $G_{s}\big(t,Y\mid F_{1}(t)\big),\,s\geq2$,
are represented by the expansions similar to \eqref{f}, namely
\begin{eqnarray}\label{cf}
    &&G_{s}(t,Y\mid F_{1}(t))=\\
    &&=\sum\limits_{n=0}^{\infty}\frac{1}{n!}\,\mathrm{Tr}_{s+1,\ldots,s+n}
        \mathfrak{V}_{1+n}(t,\theta(\{Y\}),s+1,\ldots,s+n)\prod _{i=1}^{s+n}F_{1}(t,i),\nonumber
\end{eqnarray}
where it is used the notion of declasterization mapping \eqref{Theta}. Hence in contrast to expansion
\eqref{f} the $n$ term of expansions \eqref{cf} of marginal correlation functionals
$G_{s}(t,Y\mid F_{1}(t))$ is governed by $(1+n)th$-order evolution operator \eqref{skrr} of
the $(s+n)th$-order cumulants of the scattering operators, for example, as compared with \eqref{skrre} the
lower order evolution operators $\mathfrak{V}_{1+n}(t,\theta(\{Y\}),s+1,\ldots,s+n),\,n\geq0$, have the form
\begin{eqnarray*}
   &&\mathfrak{V}_{1}(t,\theta(\{Y\}))=\widehat{\mathfrak{A}}_{s}(-t,\theta(\{Y\}),\\
   &&\mathfrak{V}_{2}(t,\theta(\{Y\}),s+1)=\\
   &&=\widehat{\mathfrak{A}}_{s+1}(-t,\theta(\{Y\}),s+1)-
      \widehat{\mathfrak{A}}_{s}(-t,\theta(\{Y\}))\sum_{i=1}^s\widehat{\mathfrak{A}}_{2}(-t,i,s+1),
\end{eqnarray*}
and in case of $s=2$, it holds
\begin{eqnarray*}
    &&\mathfrak{V}_{1}(t,\theta(\{1,2\}))=\widehat{\mathcal{G}}_{2}(t,1,2)-I,
\end{eqnarray*}
where $\widehat{\mathcal{G}}_{2}(t,1,2)$ is scattering operator \eqref{so}.

We indicate that expansions \eqref{f} of marginal functionals of the state are nonequilibrium analog
of the Mayer-Ursell expansions over powers of the density of equilibrium marginal density operators.

Within the framework of the description of states by marginal functionals of the state \eqref{f} the
average value, for example, of the additive-type marginal observable $B^{(1)}=(0,a_{1}(1),0,\ldots)$
is given by the functional
\begin{eqnarray}\label{averageg}
    &&\big\langle B^{(1)}\big\rangle(t)=\mathrm{Tr}_{1}\,a_{1}(1)F_{1}(t,1),
\end{eqnarray}
i.e. it is defined by a solution of the generalized quantum kinetic equation \eqref{gke}, or in general
case of the $s$-ary marginal observable $B^{(s)}=(0,\ldots,0,a_{s}(1,\ldots,s),0,\ldots)$ by the functional
\begin{eqnarray*}
    &&\big\langle B^{(s)}\big\rangle(t)=\frac{1}{s!}\,\mathrm{Tr}_{1,\ldots,s}\,a_{s}(1,\ldots,s)
       F_{s}(t,1,\ldots,s\mid F_{1}(t)), \quad s\geq2,
\end{eqnarray*}
where $F_{s}(t,1,\ldots,s\mid F_{1}(t))$ is the marginal functional of the state \eqref{f}. For
$B^{(s)}\in\mathfrak{L}(\mathcal{F}_\mathcal{H})$ and $F_1(t)\in\mathfrak{L}^{1}(\mathcal{H})$
these functionals exist.

The dispersion of an additive-type observable is defined by a solution of the generalized quantum
kinetic equation \eqref{gke} and marginal correlation functionals \eqref{cf} as follows
\begin{eqnarray*}
    &&\big\langle\big(B^{(1)}-\big\langle B^{(1)}\big\rangle(t)\big)^2\big\rangle(t)=\\
    &&=\mathrm{Tr}_{1}\,\big(a_1^2(1)-\big\langle B^{(1)}\big\rangle^2(t)\big)F_{1}(t,1)+
       \mathrm{Tr}_{1,2}\,a_{1}(1)a_{1}(2)G_{2}(t,1,2\mid F_{1}(t)),
\end{eqnarray*}
where the functional $\langle B^{(1)}\rangle(t)$ is determined by expression \eqref{averageg}.

We note that the dispersion of observables is minimal for states characterized by marginal correlation
functionals \eqref{cf} equals to zero, i.e. from macroscopic point of view the evolution of
many-particle states with the minimal dispersion is the Markovian kinetic evolution.
In fact functionals \eqref{cf} or \eqref{f} characterize the correlations of states
of quantum many-particle systems. We illustrate close links of functionals \eqref{cf}
and \eqref{f} in the following way
\begin{eqnarray*}
    &&F_{2}(t,1,2\mid F_{1}(t))=F_{1}(t,1)F_{1}(t,2)+G_{2}(t,1,2\mid F_{1}(t)).
\end{eqnarray*}
Basically this equality gives the classification of all possible currently in use scaling limits.
In the scaling limits it is assumed that chaos property of initial state preserves in
time, i.e. the scaling limit means such limit of dimensionless parameters of a system in which
the marginal correlation functional $G_{2}(t,1,2\mid F_{1}(t))$ vanishes. According to definition
\eqref{cf}, it is possible, if every finite particle cluster moves without collisions.

Finally we point out the relationship of generalized quantum kinetic equation \eqref{gke} and the specific
quantum kinetic equations. The last can be derived from the generalized quantum kinetic equation in the
appropriate scaling limits \cite{Sh} or as a result of certain approximations. Let us consider first two
terms of expansion \eqref{f}. If an interaction potential in \eqref{H} is a bounded operator and
$f_{s+1}\in\mathfrak{L}^{1}(\mathcal{H}_{s+1})$, then for the second-order cumulant of scattering
operators \eqref{so} an analog of the Duhamel equation holds
\begin{eqnarray*}
  &&\widehat{\mathfrak{A}}_{2}(t,\{Y\},s+1)f_{s+1}=\int_0^t d\tau\,
     \mathcal{G}_{s}(-\tau,Y)\mathcal{G}_{1}(-\tau,s+1)\sum\limits_{i_1=1}^{s}
     (-\mathcal{N}_{\mathrm{int}}(i_1,s+1))\times\\
  &&\times\widehat{\mathcal{G}}_{s+1}(\tau-t,Y,s+1)
     \prod_{i_2=1}^{s+1}\mathcal{G}_{1}(\tau,i_2)f_{s+1},
\end{eqnarray*}
and, consequently, for the second-order evolution operator $\mathfrak{V}_{2}(t,\{Y\},s+1)$ we have
\begin{eqnarray*}
   &&\mathfrak{V}_{2}(t,\{Y\},s+1)f_{s+1}\doteq\big(\widehat{\mathfrak{A}}_{2}(t,\{Y\},s+1)-
       \widehat{\mathfrak{A}}_{1}(t,\{Y\})\sum_{i_1=1}^s \widehat{\mathfrak{A}}_{2}(t,i_1,s+1)\big)f_{s+1}=\\
   &&=\int_0^t d\tau\,\mathcal{G}_{s}(-\tau,Y)\mathcal{G}_{1}(-\tau,s+1)\big(\sum\limits_{i_1=1}^{s}
       (-\mathcal{N}_{\mathrm{int}}(i_1,s+1))\widehat{\mathcal{G}}_{s+1}(\tau-t,Y,s+1)-\\
   &&-\widehat{\mathcal{G}}_{s}(\tau-t,Y)\sum\limits_{i_1=1}^{s}
       (-\mathcal{N}_{\mathrm{int}}(i_1,s+1))
       \widehat{\mathcal{G}}_{2}(\tau-t,i_1,s+1)\big)\prod_{i_2=1}^{s+1}\mathcal{G}_{1}(\tau,i_2)f_{s+1}.
\end{eqnarray*}
Observing that in the kinetic (macroscopic) scale of the variation of variables \cite{CIP} the groups
of operators \eqref{grG} of finitely many particles depend on microscopic time variable $\varepsilon^{-1}t$,
where $\varepsilon\geq0$ is a scale parameter, the dimensionless marginal functionals of the state are
represented in the form: $F_{s}\big(\varepsilon^{-1}t,Y\mid F_{1}(t)\big)$. Then in the limit
$\varepsilon\rightarrow0$ the first two terms of the dimensionless marginal functional expansions \eqref{f}
\begin{eqnarray*}
   &&\widehat{\mathcal{G}}_{s}(\varepsilon^{-1}t,Y)\prod _{i=1}^{s} F_{1}(t,i)+
\end{eqnarray*}
\begin{eqnarray*}
   &&+\int_0^{\varepsilon^{-1}t} d\tau\,\mathcal{G}_{s}(-\tau,Y)\mathrm{Tr}_{s+1}
      \big(\sum\limits_{i_1=1}^{s}(-\mathcal{N}_{\mathrm{int}}(i_1,s+1))
      \widehat{\mathcal{G}}_{s+1}(\varepsilon^{-1}t,Y,s+1)-\\
   &&-\widehat{\mathcal{G}}_{s}(\varepsilon^{-1}t,Y)
      \sum\limits_{i_1=1}^{s}(-\mathcal{N}_{\mathrm{int}}(i_1,s+1))
      \widehat{\mathcal{G}}_{2}(\varepsilon^{-1}t,i_1,s+1)\big)
      \prod_{i_2=1}^{s+1}\mathcal{G}_{1}(\tau,i_2)F_{1}(t,i_2)
\end{eqnarray*}
coincide with corresponding terms constructed by the perturbation method with the use of the
weakening of correlation condition by Bogolyubov \cite{CGP97}. Thus, in the kinetic scale the
collision integral of the generalized kinetic equation \eqref{gke} takes the form of Bogolyubov's
collision integral \cite{CGP97} and we observe that in a space homogeneous case the collision
integral of the first approximation has a more general form than the quantum Boltzmann collision integral.

\subsection{On quantum kinetic equations in case of correlated initial data}
We have proved that in case of initial data which is completely defined by a one-particle
density operator, all possible states of infinite-particle systems of bosons or fermions at
an arbitrary moment of time can be described within the framework of a one-particle density
operator and explicitly defined functionals of such operator without any approximations. One
of the advantages of this approach is the possibility to construct the kinetic equations in
scaling limits in the presence of correlations of particle states at initial time, for instance,
correlations characterizing the condensed states of interacting particles obeying Fermi-Dirac or
Bose-Einstein statistics \cite{LSSY}.

Let us consider initial data
\begin{eqnarray*}
   &&F(t)|_{t=0}=\big(F_1^0(1),h_{2}(1,2)F_1^0(1)F_1^0(2),\ldots,h_{n}(1,...,n)
       {\prod}_{i=1}^{n}F_1^0(i),\ldots\big),
\end{eqnarray*}
where the operators $h_{n}\in\mathfrak{L}^{1}(\mathcal{H}_n),\,n\geq2$, are specified initial
correlations. Such initial data is typical for the condensed states of quantum gases, for
example, the equilibrium state of the Bose condensate satisfies the weakening of correlation
condition with the correlations which characterize the condensed state \cite{BQ}.

In the case under consideration the kinetic cluster expansions of cumulants of groups, i.e. recurrence
relations \eqref{kce}, take the form
\begin{eqnarray}\label{kcec}
   &&\hskip-8mm\mathfrak{A}_{1+n}(-t,\{Y\},s+1,\ldots,s+n)h_{1+n}(\{Y\},s+1,\ldots,s+n)
      \prod_{i=1}^{s+n}\mathfrak{A}_{1}(t,i)=\\
   &&\hskip-8mm=\sum_{n_1=0}^{n}\frac{n!}{(n-n_1)!}
      \mathfrak{G}_{1+n-n_1}(t,\{Y\},s+1,\ldots,s+n-n_1)\sum_{\mathrm{D_{s+n}}:Z=\bigcup_i
      X_i}\,\,\sum_{i_1<i_2<\ldots<i_{|\mathrm{D_{s+n}}|}=1}^{s+n-n_1}\nonumber\\
   &&\hskip-8mm\prod_{k=1}^{|\mathrm{D_{s+n}}|}\frac{1}{|X_k|!}\,
      \mathfrak{A}_{1+|X_{k}|}(t,i_k,X_{k})\prod_{k=1}^{|\mathrm{D_{s+n}}|}h_{1+|X_{k}|}(i_k,X_{k})\mathfrak{A}_{1}(t,i_k)
      \prod_{j\in Z}\mathfrak{A}_{1}(t,j),\nonumber
\end{eqnarray}
where the notations of formula \eqref{kce} are used. In terms of new evolution operators
(the $(1+n)th$-order scattering cumulants)
\begin{eqnarray*}
   &&\breve{\mathfrak{A}}_{1+n}(t,\{Y\},X\backslash Y)=
       \mathfrak{A}_{1+n}(-t,\{Y\},X\backslash Y)h_{1+n}(\{Y\},X\backslash Y)
       \prod_{i=1}^{s+n}\mathfrak{A}_{1}(t,i),
\end{eqnarray*}
the solutions $\mathfrak{G}_{1+n}(t,\{Y\},X\backslash Y),\,n\geq0$, of recurrence relations
\eqref{kcec} are represented by the expansions
\begin{eqnarray}\label{skrrc}
   &&\hskip-7mm\mathfrak{G}_{1+n}(t,\{Y\},X\setminus Y )\doteq n!\,
       \sum_{k=0}^{n}\,(-1)^k\,\sum_{n_1=1}^{n}\ldots
       \sum_{n_k=1}^{n-n_1-\ldots-n_{k-1}}\frac{1}{(n-n_1-\ldots-n_k)!}\times\\
   &&\hskip-7mm\times\breve{\mathfrak{A}}_{1+n-n_1-\ldots-n_k}(t,\{Y\},s+1,\ldots,
       s+n-n_1-\ldots-n_k)\times\nonumber\\
   &&\hskip-7mm\times\prod_{j=1}^k\,\sum\limits_{\mbox{\scriptsize$\begin{array}{c}
       \mathrm{D}_{j}:Z_j=\bigcup_{l_j}X_{l_j},\\
       |\mathrm{D}_{j}|\leq s+n-n_1-\dots-n_j\end{array}$}}\frac{1}{|\mathrm{D}_{j}|!}
       \sum_{i_1\neq\ldots\neq i_{|\mathrm{D}_{j}|}=1}^{s+n-n_1-\ldots-n_j}\,\,
       \prod_{X_{l_j}\subset \mathrm{D}_{j}}\,\frac{1}{|X_{l_j}|!}\,\,
       \breve{\mathfrak{A}}_{1+|X_{l_j}|}(t,i_{l_j},X_{l_j}),\nonumber
\end{eqnarray}
where $\sum_{\mathrm{D}_{j}:Z_j=\bigcup_{l_j} X_{l_j}}$ is the sum over all possible dissections of
the linearly ordered set $Z_j\equiv(s+n-n_1-\ldots-n_j+1,\ldots,s+n-n_1-\ldots-n_{j-1})$ on no more
than $s+n-n_1-\ldots-n_j$ linearly ordered subsets. For example,
\begin{eqnarray*}
   &&\mathfrak{G}_{1}(t,\{Y\})=\breve{\mathfrak{A}}_{1}(t,\{Y\})=\mathfrak{A}_{1}(-t,\{Y\})h_{1}(\{Y\})
       \prod_{i=1}^{s}\mathfrak{A}_{1}(t,i).
\end{eqnarray*}

Therefore the one-particle density operator $F_{1}(t)$ is governed by the following generalized
quantum kinetic equation
\begin{eqnarray}
  \label{gkec}
    &&\frac{d}{dt}F_{1}(t,1)=-\mathcal{N}(1)F_{1}(t,1)+\\
    &&+\mathrm{Tr}_{2}(-\mathcal{N}_{\mathrm{int}}(1,2))
        \sum\limits_{n=0}^{\infty}\frac{1}{n!}\mathrm{Tr}_{3,\ldots,n+2}\,
        \mathfrak{G}_{1+n}(t,\{1,2\},3,\ldots,n+2)\prod _{i=1}^{n+2} F_{1}(t,i),\nonumber
\end{eqnarray}
where the operator $\mathcal{N}_{\mathrm{int}}(1,2)$ is defined by formula \eqref{comst},
and the $(1+n)th$-order generated evolution operator $\mathfrak{G}_{1+n}(t),\,n\geq0$, is defined by
expansion \eqref{skrrc}. Correspondingly the marginal functionals of the state are represented by the
following expansions
\begin{eqnarray}\label{cfc}
   &&F_{s}(t,Y\mid F_{1}(t))\doteq\sum _{n=0}^{\infty}\frac{1}{n!}\,
      \mathrm{Tr}_{s+1,\ldots,{s+n}}\,\mathfrak{G}_{1+n}(t,\{Y\},X\setminus Y)
      \prod _{i=1}^{s+n}F_{1}(t,i).
\end{eqnarray}

Thus, the coefficients of generalized quantum kinetic equation \eqref{gkec} and generated evolution
operators \eqref{skrrc} of marginal functionals \eqref{cfc} are determined by the operators of initial
correlations.


\section{On scaling limits of hierarchy solutions}
The current point of view on the problem of the derivation of quantum kinetic equations from
underlaying many-particle dynamics consists in the following. Since the evolution of states
of infinitely many quantum particles is generally described by a sequence of marginal density
operators governed by the quantum BBGKY hierarchy, then the evolution of states can be effectively
described by a one-particle density operator governed by the quantum kinetic equation in a suitable
scaling limit \cite{Sh,Shb}. In this section the mean field (self-consistent field) scaling asymptotic
behavior of stated above solutions is established. The constructed asymptotics are governed by the
quantum Vlasov hierarchy for limit marginal density operators and limit solution of the generalized
quantum kinetic equation \cite{GerUJP}, the nonlinear quantum Vlasov hierarchy for limit marginal
correlation operators and the dual quantum Vlasov hierarchy for limit marginal observables, respectively
\cite{G11}. In case of initial data satisfying a chaos property \cite{CGP97}, which means the absence
of correlations at initial time, the constructed asymptotics are governed by the Vlasov quantum kinetic
equation. Moreover, within the framework of the description of the evolution by the dual quantum BBGKY
hierarchy or by the generalized quantum kinetic equation it is possible to construct the kinetic equations
in scaling limits in case of presence of correlations at initial time, for instance, correlations which
characterize the condensate states \cite{GT11}.

\subsection{A mean field limit of a solution of the dual quantum BBGKY hierarchy}
We consider the $n$-particle system with the Hamiltonian
\begin{eqnarray}\label{Hs}
    &&H_{n}=\sum\limits_{i=1}^{n}K(i)+\epsilon\sum\limits_{i_{1}<i_{2}=1}^{n}\Phi(i_{1},i_{2}),
\end{eqnarray}
where $\epsilon>0$ is a scaling parameter. At first we construct the mean field scaling limit
of a solution of initial-value problem \eqref{dh}-\eqref{dhi} of the dual quantum BBGKY hierarchy.

Let for initial data $B_{s}^0\in\mathfrak{L}(\mathcal{H}_{s})$ there exists the scaling limit
$b_{s}^0\in\mathfrak{L}(\mathcal{H}_{s})$, i.e.
\begin{eqnarray}\label{asumdin}
   &&\mathrm{w^{\ast}-}\lim\limits_{\epsilon\rightarrow 0}(\epsilon^{-s}B_{s}^0-b_{s}^0)=0,
\end{eqnarray}
then for arbitrary finite time interval there exists the mean field limit of solution \eqref{sdh}
of the Cauchy problem \eqref{dh}-\eqref{dhi} of the dual quantum BBGKY hierarchy in the sense of
the $\ast$-weak convergence of the space $\mathfrak{L}(\mathcal{H}_s)$
\begin{eqnarray}\label{asymt}
    &&\mathrm{w^{\ast}-}\lim\limits_{\epsilon\rightarrow 0}(\epsilon^{-s}B_{s}(t)-b_{s}(t))=0,
       \quad s\geq1,
\end{eqnarray}
which is defined by the expansion
\begin{eqnarray}\label{Iterd}
    &&b_{s}(t,Y)=\sum\limits_{n=0}^{s-1}\,\int\limits_0^tdt_{1}\ldots\int\limits_0^{t_{n-1}}dt_{n}
       \,\mathcal{G}_{s}^{0}(t-t_{1})\sum\limits_{i_{1}\neq j_{1}=1}^{s}
       \mathcal{N}_{\mathrm{int}}(i_{1},j_{1})\,\mathcal{G}_{s-1}^{0}(t_{1}-t_{2})\\
    &&\ldots\,\, \mathcal{G}_{s-n+1}^{0}(t_{n-1}-t_{n})
       \sum\limits^{s}_{\mbox{\scriptsize $\begin{array}{c}i_{n}\neq j_{n}=1,\\
       i_{n},j_{n}\neq (j_{1},\ldots,j_{n-1})\end{array}$}}
       \mathcal{N}_{\mathrm{int}}(i_{n},j_{n})\times\nonumber\\
    &&\times \mathcal{G}_{s-n}^{0}(t_{n})b_{s-n}^0(Y\backslash (j_{1},\ldots,j_{n})),\nonumber
\end{eqnarray}
where the following notation of the group of operators \eqref{grG} of noninteracting particles is used
\begin{eqnarray*}
   &&\mathcal{G}_{s-n+1}^{0}(t_{n-1}-t_{n})
      \equiv\mathcal{G}_{s-n+1}^{0}(t_{n-1}-t_{n},Y \backslash (j_{1},\ldots,j_{n-1}))=\\
   &&=\prod\limits_{j\in Y \backslash (j_{1},\ldots,j_{n-1})}\mathcal{G}_{1}(t_{n-1}-t_{n},j).
\end{eqnarray*}

Before to prove this statement we give some comments. If $b(0)\in\mathfrak{L}(\mathcal{F}_\mathcal{H})$,
the sequence $b(t)=(b_0,b_1(t),\ldots,b_{s}(t),\ldots)$ of limit marginal observables \eqref{Iterd}
is a generalized global solution of the initial-value problem of the dual quantum Vlasov hierarchy
\begin{eqnarray}
  \label{vdh}
    &&\frac{d}{dt}b_{s}(t,Y)=\sum\limits_{i=1}^{s}\mathcal{N}(i)\,b_{s}(t,Y)+
       \sum_{j_1\neq j_{2}=1}^s\mathcal{N}_{\mathrm{int}}(j_1,j_{2})\,b_{s-1}(t,Y\backslash(j_1)),\\
       \nonumber\\
  \label{vdhi}
    &&b_{s}(t)\mid_{t=0}=b_{s}^0,\quad s\geq1.
\end{eqnarray}
This fact is proved similar to the case of an iteration series of the dual quantum BBGKY hierarchy \cite{BG}.
It should be noted that equations set \eqref{vdh} has the structure of recurrence evolution equations.
We make a few examples of the dual quantum Vlasov hierarchy \eqref{vdh} in terms of operator kernels of
the limit marginal observables
\begin{eqnarray*}
    &&i\,\frac{\partial}{\partial t}b_{1}(t,q_1;q'_1)=-\frac{1}{2}(-\Delta_{q_1}+\Delta_{q'_1})b_{1}(t,q_1;q'_1),\\
    &&i\,\frac{\partial}{\partial t}b_{2}(t,q_1,q_2;q'_1,q'_2)=
       -\frac{1}{2}\sum\limits_{i=1}^2(-\Delta_{q_i}+\Delta_{q'_i})b_{2}(t,q_1,q_2;q'_1,q'_2)+\\
    &&+\big(\Phi(q'_1-q'_2)-\Phi(q_1-q_2)\big)\big(b_{1}(t,q_1;q'_1)+b_{1}(t,q_2;q'_2)\big).
\end{eqnarray*}

Let us consider a particular case of observables, namely the mean field limit of the additive-type
marginal observables. In this case solution \eqref{sdh} of the dual quantum BBGKY hierarchy \eqref{dh}
has the form \eqref{af}. If for the additive-type observables $B^{(1)}(0)=(0,a_{1}(1),0,\ldots)$
condition \eqref{asumdin} is satisfied, i.e. it holds
\begin{eqnarray*}
   &&\mathrm{w^{\ast}-}\lim\limits_{\epsilon\rightarrow 0}( \epsilon^{-1}a_{1}(1)-b_{1}^{0}(1))=0,
\end{eqnarray*}
then according to statement \eqref{asymt}, we have
\begin{eqnarray*}
   &&\mathrm{w^{\ast}-}\lim\limits_{\epsilon\rightarrow 0}(\epsilon^{-s}B_{s}^{(1)}(t)-b_{s}^{(1)}(t))=0,
\end{eqnarray*}
where the limit operator $b_{s}^{(1)}(t)$ is defined by the expression
\begin{eqnarray}\label{itvad}
   &&b_{s}^{(1)}(t,Y)=\int\limits_0^t dt_{1}\ldots\int\limits_0^{t_{s-2}}dt_{s-1}
      \,\mathcal{G}_{s}^{0}(t-t_{1})\sum\limits_{i_{1}\neq j_{1}=1}^{s}
      \mathcal{N}_{\mathrm{int}}(i_{1},j_{1})\,\mathcal{G}_{s-1}^{0}(t_{1}-t_{2})\\
   &&\ldots \,\,\mathcal{G}_{2}^{0}(t_{s-2}-t_{s-1})
      \sum\limits^{s}_{\mbox{\scriptsize $\begin{array}{c}i_{s-1}\neq j_{s-1}=1,\\
      i_{s-1},j_{s-1}\neq (j_{1},\ldots,j_{s-2})\end{array}$}}
      \mathcal{N}_{\mathrm{int}}(i_{s-1},j_{s-1})\times\nonumber\\
   &&\times \mathcal{G}_{1}^{0}(t_{s-1})\,b_{1}^{0}(Y\backslash (j_{1},\ldots,j_{s-1})),\quad s\geq1,\nonumber
\end{eqnarray}
as a special case of expansion \eqref{Iterd}. We give examples of expressions \eqref{itvad}
\begin{eqnarray*}
   &&b_{1}^{(1)}(t,1)=\mathcal{G}_{1}(t,1)\,b_{1}^{0}(1),\\
   &&b_{2}^{(1)}(t,1,2)=\int\limits_0^t d\tau\,\prod\limits_{i=1}^{2}\mathcal{G}_{1}(t-\tau,i)\,
      \mathcal{N}_{\mathrm{int}}(1,2)\sum\limits_{j=1}^{2}\mathcal{G}_{1}(\tau,j)\,b_{1}^{0}(j).
\end{eqnarray*}

To establish the relationship of constructed mean field asymptotic behavior of marginal observables
with asymptotic behavior of marginal states we consider initial data satisfying the factorization
property or a chaos property \cite{CGP97}, which means the absence of correlations at initial time.
For a system of identical particles, obeying the Maxwell-Boltzmann statistics, we have
\begin{eqnarray*}\label{h2}
    &&F(t)|_{t=0}= F^{(c)}\equiv(F_1^0(1),\ldots,\prod_{i=1}^s F_1^0(i),\ldots).
\end{eqnarray*}
This assumption about initial data is intrinsic for the kinetic description of a gas, because in
this case all possible states are characterized only by a one-particle marginal density operator. Let
\begin{eqnarray}\label{lh2}
    &&\lim\limits_{\epsilon\rightarrow 0}\big\|\,\epsilon F_1^0-f_1^0
       \,\big\|_{\mathfrak{L}^{1}(\mathcal{H})}=0,
\end{eqnarray}
then the limit of initial state satisfies a chaos property too
\begin{eqnarray}\label{lih2}
    &&f^{(c)}\equiv(f_1^0(1),\ldots,\prod\limits_{i=1}^{s}f_{1}^0(i),\ldots).
\end{eqnarray}
For $b(t)\in\mathfrak{L}_{\gamma}(\mathcal{F}_\mathcal{H})$ and $f_1^0\in\mathfrak{L}^{1}(\mathcal{H})$,
under the condition $\|f_1^0\|_{\mathfrak{L}^{1}(\mathcal{H})}<\gamma$, the mean field limit
of mean value functional \eqref{avmar-1} exists and it is determined by the expansion
\begin{eqnarray*}
    &&(b(t),f^{(c)})=\sum\limits_{s=0}^{\infty}\,\frac{1}{s!}\,
        \mathrm{Tr}_{1,\ldots,s}\,b_{s}(t,1,\ldots,s)\prod \limits_{i=1}^{s} f_1^0(i).
\end{eqnarray*}
In consequence of the validity of the following equality for the limit additive-type marginal
observables \eqref{itvad} 
\begin{eqnarray}\label{avmar-2}
    &&(b^{(1)}(t),f^{(c)})=\sum\limits_{s=0}^{\infty}\,\frac{1}{s!}\,
       \mathrm{Tr}_{1,\ldots,s}\,b_{s}^{(1)}(t,1,\ldots,s)\prod\limits_{i=1}^{s}f_{1}^0(i)=\\
    &&=\mathrm{Tr}_{1}\,b_{1}^{0}(1)f_{1}(t,1),\nonumber
\end{eqnarray}
where the operator $b_{s}^{(1)}(t)$ is given by expansion \eqref{itvad} and the one-particle
limiting density operator $f_{1}(t,1)$ is determined by the series
\begin{eqnarray}\label{viter}
    &&f_{1}(t,1)=\sum\limits_{n=0}^{\infty}\int\limits_0^tdt_{1}\ldots\int\limits_0^{t_{n-1}}dt_{n}\,
        \mathrm{Tr}_{2,\ldots,n+1}\prod\limits_{i_1=1}^{1}\mathcal{G}_{1}(-t+t_{1},i_1)\times\\
    &&\times(-\mathcal{N}_{\mathrm{int}}(1,2))\prod\limits_{j_1=1}^{2}
        \mathcal{G}_{1}(-t_{1}+t_{2},j_1)\ldots\prod\limits_{i_{n}=1}^{n}
        \mathcal{G}_{1}(-t_{n}+t_{n},i_{n})\times\nonumber\\
    &&\times\sum\limits_{k_{n}=1}^{n}(-\mathcal{N}_{\mathrm{int}}(k_{n},n+1))
        \prod\limits_{j_n=1}^{n+1}\mathcal{G}_{1}(-t_{n},j_n)\prod\limits_{i=1}^{n+1}f_1^0(i),\nonumber
\end{eqnarray}
we establish that operator \eqref{viter} is a solution of the initial-value problem of the Vlasov
quantum kinetic equation
\begin{eqnarray}
 \label{Vlasov1}
    &&\frac{d}{dt}f_{1}(t,1)=-\mathcal{N}(1)f_{1}(t,1)+
        \mathrm{Tr}_{2}(-\mathcal{N}_{\mathrm{int}}(1,2))f_{1}(t,1)f_{1}(t,2),\\
        \nonumber\\
  \label{Vlasovi}
    &&f_1(t)|_{t=0}= f_1^0.
\end{eqnarray}

Correspondingly, a chaos property in the Heisenberg picture of the evolution of quantum many-particle
systems is fulfilled. This fact follows from the equality for the limit $k$-ary  marginal observables,
i.e. $b^{(k)}(0)=(0,\ldots,b_{k}^{0}(1,\ldots,k),0,\ldots)$,
\begin{eqnarray}\label{dchaos}
    &&(b^{(k)}(t),f^{(c)})=\sum\limits_{s=0}^{\infty}\,\frac{1}{s!}\,
       \mathrm{Tr}_{1,\ldots,s}\,b_{s}^{(k)}(t,1,\ldots,s)\prod\limits_{i=1}^{s} f_1^0(i)=\\
    &&=\frac{1}{k!}\mathrm{Tr}_{1,\ldots,k}\,b_{k}^{0}(1,\ldots,k)
       \prod\limits_{i=1}^{k}f_{1}(t,i),\quad k\geq2,\nonumber
\end{eqnarray}
where the limit one-particle marginal density operator $f_{1}(t,i)$ is defined by expansion
\eqref{viter} and therefore it is governed by the Cauchy problem \eqref{Vlasov1}-\eqref{Vlasovi}.

Thus, in the mean field scaling limit an equivalent approach to the description of the kinetic evolution
of quantum many-particle systems in terms of the Cauchy problem \eqref{Vlasov1}-\eqref{Vlasovi} of the
Vlasov kinetic equation is given by the Cauchy problem \eqref{vdh}-\eqref{vdhi} of the dual
quantum Vlasov hierarchy for the additive-type marginal observables. In case of the $k$-ary  marginal
observables a solution of the dual quantum Vlasov hierarchy \eqref{vdh} is equivalent to the preservation
of a chaos property for $k$-particle marginal density operators in the sense of equality \eqref{dchaos}.

\subsection{A mean field limit of a solution of the quantum BBGKY hierarchy}
Within the framework of the description of the evolution in terms of states the scaling mean field limit
of a nonperturbative solution of the Cauchy problem \eqref{BBGKY}-\eqref{BBGKYi} of the quantum BBGKY
hierarchy in case of arbitrary initial data is stated as follows.

Let for initial data $F_{s}^0, f_{s}^0\in\mathfrak{L}^{1}(\mathcal{H}_{s})$ it holds
\begin{eqnarray}\label{asum}
    &&\lim\limits_{\epsilon\rightarrow 0}\big\| \epsilon^{s}
       F_{s}^0(1,\ldots,s)-f_{s}^0(1,\ldots,s)\big\|_{\mathfrak{L}^{1}(\mathcal{H}_{s})}=0,
\end{eqnarray}
then for any finite time interval for solution \eqref{RozvBBGKY} there exists the mean field limit
\begin{eqnarray}\label{ls}
    &&\lim\limits_{\epsilon\rightarrow 0}\big\|\epsilon^{s}
       F_{s}(t,1,\ldots,s)-f_{s}(t,1,\ldots,s)\big\|_{\mathfrak{L}^{1}(\mathcal{H}_{s})}=0,
\end{eqnarray}
where the limiting operator $f_s(t)\in\mathfrak{L}^{1}(\mathcal{H}_{s})$ is determined by the series
\begin{eqnarray}\label{Iter2}
   &&f_{s}(t,1,\ldots,s)=\sum\limits_{n=0}^{\infty}\int\limits_0^tdt_{1}\ldots
       \int\limits_0^{t_{n-1}}dt_{n}\mathrm{Tr}_{s+1,\ldots,s+n}
       \prod\limits_{j_{1}=1}^{s}\mathcal{G}_{1}(-t+t_{1},j_{1})\times
\end{eqnarray}
\begin{eqnarray*}
   &&\times\sum\limits_{i_{1}=1}^{s}(-\mathcal{N}_{\mathrm{int}}(i_{1},s+1))
       \prod\limits_{l_{1}=1}^{s+1}\mathcal{G}_{1}(-t_{1}+t_{2},l_{1})\ldots \nonumber\\
   &&\prod\limits_{j_{n}=1}^{s+n-1}\mathcal{G}_{1}(-t_{n-1}+t_{n},j_{n})
       \sum\limits_{i_{n}=1}^{s+n-1}(-\mathcal{N}_{\mathrm{int}}(i_{n},s+n))
       \prod\limits_{l_{n}=1}^{s+n}\mathcal{G}_{1}(-t_{n},l_{n})f_{s+n}^0,\nonumber
\end{eqnarray*}
which is norm convergent on the space $\mathfrak{L}^{1}(\mathcal{F}_\mathcal{H})$ on finite time
interval. Limiting marginal operators \eqref{Iter2} are governed by the limiting BBGKY hierarchy
known as the quantum Vlasov hierarchy
\begin{eqnarray}\label{BBGKYlim}
   &&\frac{d}{dt}f_{s}(t)=\sum\limits_{i=1}^{s}(-\mathcal{N}(i))f_{s}(t)
      +\sum\limits_{i=1}^{s}\mathrm{Tr}_{s+1}(-\mathcal{N}_{\mathrm{int}}(i,s+1))
      f_{s+1}(t), \quad s\geq1.
\end{eqnarray}

The validity of this statement is the consequence of the following formulas on an asymptotic
perturbation of cumulants of groups. If $f_{s}\in\mathfrak{L}^{1}(\mathcal{H}_{s})$, then for the
first-order cumulant $\mathfrak{A}_{1}(-t,\{Y\})$ the equality holds
\begin{eqnarray}\label{lemma}
    &&\lim\limits_{\epsilon\rightarrow 0}\big\|\big(\mathfrak{A}_{1}(-t,\{Y\})-
        \prod\limits_{j=1}^{s}\mathcal{G}_{1}(-t,j)\big)f_{s}\big\|_{\mathfrak{L}^{1}(\mathcal{H}_{s})}=0,
\end{eqnarray}
and in general case for $n\geq1$ we have
\begin{eqnarray}\label{Duam2}
    &&\hskip-12mm\lim\limits_{\epsilon\rightarrow 0}\big\|\big(\frac{1}{\epsilon^n}\frac{1}{n!}\,
       \mathfrak{A}_{1+n}(-t,\{Y\},s+1,\ldots,s+n)-\\
    &&\hskip-12mm-\int\limits_0^tdt_{1}\ldots\int\limits_0^{t_{n-1}}dt_{n} \prod\limits_{j_1=1}^{s}
       \mathcal{G}_{1}(-t+t_{1},j_1)\sum\limits_{i_{1}=1}^{s}(-\mathcal{N}_{\mathrm{int}}(i_{1},s+1))
       \prod\limits_{l_1=1}^{s+1}\mathcal{G}_{1}(-t_{1}+t_{2},l_1)\ldots\nonumber\\
    &&\hskip-12mm\ldots\prod\limits_{j_{n}=1}^{s+n-1}\mathcal{G}_{1}(-t_{n-1}+t_{n},j_{n})
       \sum\limits_{i_{n}=1}^{s+n-1}(-\mathcal{N}_{\mathrm{int}}(i_{n},s+n))
       \prod\limits_{l_n=1}^{s+n}\mathcal{G}_{1}(-t_{n},l_n)\big)
       f_{s+n}\big\|_{\mathfrak{L}^{1}(\mathcal{H}_{s+n})}=0.\nonumber
\end{eqnarray}
for finite time interval.

Let us consider initial data satisfying the factorization property or a chaos property \eqref{lih2},
which means the absence of correlations at initial time. We observe that, if the initial data satisfy
the chaos property for particles obeying the Maxwell-Boltzmann statistics, then solution expansion
\eqref{Iter2} of initial-value problem \eqref{BBGKYlim}-\eqref{lih2} of the Vlasov hierarchy is
possessed of the same property
\begin{eqnarray}\label{chaosv}
    &&f_{s}(t,1,\ldots,s)=\prod\limits_{j=1}^{s}f_{1}(t,j), \quad s\geq 2,
\end{eqnarray}
where $f_{1}(t,j)$ is defined by series \eqref{Iter2} in case of $s=1$ and initial data \eqref{lih2},
i.e. \eqref{viter}, which for bounded interaction potential is the norm convergent on the space
$\mathfrak{L}^{1}(\mathcal{H})$ under the condition
\begin{eqnarray*}
    &&t<t_0\equiv(2\,\|\Phi\|_{\mathfrak{L}(\mathcal{H}_{2})}
         \|f_1^0\|_{\mathfrak{L}^{1}(\mathcal{H})})^{-1}.
\end{eqnarray*}
In other words, under the condition on initial data $F_{s}^0\in\mathfrak{L}^{1}(\mathcal{H}_{s})$
\begin{eqnarray*}
    &&\lim\limits_{\epsilon\rightarrow 0}\big\|\epsilon^{s} F_{s}^0(1,\ldots,s)-\prod\limits_{j=1}^{s}
         f_{1}^0(j)\big\|_{\mathfrak{L}^{1}(\mathcal{H}_{s})}=0,
\end{eqnarray*}
on finite time interval for solution \eqref{RozvBBGKY} of the quantum BBGKY hierarchy it holds
\begin{eqnarray*}
    &&\lim\limits_{\epsilon\rightarrow 0}\big\|\epsilon^{s} F_{s}(t,1,\ldots,s)-\prod\limits_{j=1}^{s}
         f_{1}(t,j)\big\|_{\mathfrak{L}^{1}(\mathcal{H}_{s})}=0,
\end{eqnarray*}
where the one-particle limiting density operator $f_{1}(t,1)$ is determined by series
\eqref{viter} which is a solution of the Cauchy problem \eqref{Vlasov1}-\eqref{Vlasovi}
of the Vlasov quantum kinetic equation.

In case of pure states, i.e. if the limit one-particle density operator $f_{1}(t)=|\psi_{t}\rangle\langle\psi_{t}|$
is a one-dimensional projector onto a unit vector $|\psi_{t}\rangle$, we have
\begin{eqnarray*}
    &&\lim\limits_{\epsilon\rightarrow 0}\big\|\,\epsilon^{s}F_{s}(t)-
       |\psi_{t}\rangle\langle\psi_{t}|^{\otimes s}\,\big\|_{\mathfrak{L}^{1}(\mathcal{H}_{s})}=0,
\end{eqnarray*}
where $|\psi_{t}\rangle$ is a solution of the Hartree equation, which in terms of the kernel
$f_{1}(t,q,q')=\psi(t,q)\psi(t,q')$ of the marginal operator $f_{1}(t)$ has the form
\begin{eqnarray*}
    &&i\frac{\partial}{\partial t} \psi(t,q)=-\frac{1}{2}\Delta_{q}\psi(t,q)+ \int d q'
       \Phi(q-q')|\psi(t,q')|^{2}\psi(t,q).
\end{eqnarray*}
If the kernel of the interaction potential is the Dirac measure $\Phi(q)=\delta(q)$, then
we derive the cubic nonlinear Schr\"{o}dinger equation
\begin{eqnarray*}
    &&i\frac{\partial}{\partial t}\psi(t,q)=
       -\frac{1}{2}\Delta_{q}\psi(t,q)+|\psi(t,q)|^{2}\psi(t,q).
\end{eqnarray*}
We observe that in the general case of many-body potentials with the scaled Hamiltonian
\begin{eqnarray}\label{Hns}
    &&H_{n}=\sum\limits_{i=1}^{n}K(i)+\sum\limits_{k=2}^{n}\epsilon^{k-1}
        \sum\limits_{i_{1}<\ldots<i_{k}=1}^{n}\Phi^{(k)}(i_{1},\ldots,i_{k}),
\end{eqnarray}
the Hartree equation takes on form
\begin{eqnarray*}
    &&i\frac{\partial}{\partial t}\psi(t,q_1)=
      -\frac{1}{2}\Delta_{q_1}\psi(t,q_1)+\\
    &&+\sum\limits_{n=1}^{\infty}\frac{1}{n!}\int d q_2\ldots d q_{n+1}
      \Phi^{(n+1)}(q_1,\ldots,q_{n+1})\prod\limits_{i=2}^{n+1}|\psi(t,q_i)|^{2}\psi(t,q_1).
\end{eqnarray*}

We note also that in case of many-particle systems obeying quantum statistics statement \eqref{ls}
is true for operators from the corresponding spaces $\mathfrak{L}^{1}(\mathcal{H}_{s}^{\pm})$.
For initial data satisfying a chaos property
\begin{eqnarray*}\label{lih2fd}
    &&f^{(c)}\equiv(f_1^0(1),\ldots,\mathcal{S}^{\pm}_{s}{\prod\limits}_{i=1}^{s}f_{1}^0(i),\ldots),
\end{eqnarray*}
where the operator $\mathcal{S}^{\pm}_{s}$ is defined by \eqref{Sn}, the Vlasov quantum kinetic
equation has the form
\begin{eqnarray*}
     &&\frac{d}{dt}f_{1}(t,1)=-\mathcal{N}(1)f_{1}(t,1)+
        \mathrm{Tr}_{2}(-\mathcal{N}_{\mathrm{int}}(1,2))
        \mathcal{S}^{\pm}_{2}f_{1}(t,1)f_{1}(t,2),
\end{eqnarray*}
and consequently for pure states of fermions we derive the Hartree-Fock equation.

\subsection{On a mean field asymptotics of dynamics of correlations}
We give some comments on the existence of mean field scaling limits of the constructed solution.
The mean field limit $g_s(t,1,\ldots,s),\,s\geq1$, of marginal correlation operators \eqref{sss}
exists
\begin{eqnarray*}\label{asymp}
   &&\lim\limits_{\epsilon\rightarrow 0}\big\|\epsilon^{s}G_{s}(t)-
      g_{s}(t)\big\|_{\mathfrak{L}^{1}(\mathcal{H}_s)}=0, \quad s\geq1,
\end{eqnarray*}
provided that
\begin{eqnarray}\label{asic}
   &&\lim\limits_{\epsilon\rightarrow 0}\big\|\epsilon^{s}G_{s}^0-
      g_{s}^0\big\|_{\mathfrak{L}^{1}(\mathcal{H}_s)}=0, \quad s\geq1,
\end{eqnarray}
and it is governed by the nonlinear Vlasov quantum hierarchy
\begin{eqnarray}\label{gBigfromDFBa_lim}
   &&\frac{\partial}{\partial t}g_s(t,Y)=\sum_{i\in Y}(-\mathcal{N}(i))g_{s}(t,Y) +
      \mathrm{Tr}_{s+1}\sum_{i\in Y}(-\mathcal{N}_{\mathrm{int}}(i,s+1))\times\\
   &&\times\big(g_{s+1}(t,Y,s+1)+\sum_{\mbox{\scriptsize
      $\begin{array}{c}\mathrm{P}:(Y,s+1)=X_1\bigcup X_2,\\i\in X_1;s+1\in X_2\end{array}$}}
      g_{|X_1|}(t,X_1)g_{|X_2|}(t,X_2)\big),\nonumber
\end{eqnarray}
where we use the same notations as for hierarhy \eqref{gBigfromDFBa}.

If initial data satisfies chaos property, then we establish
\begin{eqnarray}\label{Gcid}
   &&\lim\limits_{\epsilon\rightarrow 0}\big\|\epsilon^{s}G_{s}(t)\big\|_{\mathfrak{L}^{1}(\mathcal{H}_s)}=0,
      \quad s\geq2,
\end{eqnarray}
since solution expansions \eqref{GUG(0)} for marginal correlation operators are defined by the $(s+n)th$-order
cumulants as contrasted to solution expansions \eqref{RozvBBGKY} for marginal density operators which defined
by the $(1+n)th$-order cumulants and in the same way as statement \eqref{Duam2} the equality holds
\begin{eqnarray*}
     &&\lim\limits_{\epsilon\rightarrow0}\big\|\frac{1}{\epsilon^{n}}\,
        \mathfrak{A}_{s+n}(-t,1,\ldots,s+n)f_{s+n}\big\|_{\mathfrak{L}^{1}(\mathcal{H}_{s+n})}=0.
\end{eqnarray*}
In the case of $s=1$ provided that \eqref{asic} we have
\begin{eqnarray*}
   &&\lim\limits_{\epsilon\rightarrow 0}\big\|\epsilon G_{1}(t)-
      g_{1}(t)\big\|_{\mathfrak{L}^{1}(\mathcal{H})}=0,
\end{eqnarray*}
where for finite time interval the limiting one-particle marginal correlation operator $g_1(t,1)$ is
given by the norm convergent on the space $\mathfrak{L}^{1}(\mathcal{H})$ series
\begin{eqnarray}\label{1mco}
   &&\hskip-10mm g_{1}(t,1)=\\
   &&\hskip-10mm =\sum\limits_{n=0}^{\infty}\int\limits_0^tdt_{1}\ldots
      \int\limits_0^{t_{n-1}}dt_{n}\,\mathrm{Tr}_{2,\ldots,n+1}\mathcal{G}_{1}(-t+t_{1},1)
      (-\mathcal{N}_{\mathrm{int}}(1,2))\prod\limits_{j_1=1}^{2}
      \mathcal{G}_{1}(-t_{1}+t_{2},j_1)\ldots\nonumber\\
   &&\hskip-10mm \ldots\prod\limits_{i_{n}=1}^{n}\mathcal{G}_{1}(-t_{n}+t_{n},i_{n})
      \sum\limits_{k_{n}=1}^{n}(-\mathcal{N}_{\mathrm{int}}(k_{n},n+1))
      \prod\limits_{j_n=1}^{n+1}\mathcal{G}_{1}(-t_{n},j_n)\prod\limits_{i=1}^{n+1}g_1^0(i),\nonumber
\end{eqnarray}
which obviously coincides with iteration series \eqref{viter} of the Vlasov quantum kinetic equation. In view
of the validity of limit \eqref{Gcid} from the Vlasov nonlinear quantum hierarchy \eqref{gBigfromDFBa_lim}
we also conclude that limit one-particle marginal correlation operator \eqref{1mco} is governed by the Cauchy
problem of the Vlasov quantum kinetic equation \eqref{Vlasov1}-\eqref{Vlasovi}.

Therefore the Vlasov nonlinear quantum hierarchy \eqref{gBigfromDFBa_lim} describes the evolution of
initial correlations.

\subsection{On scaling limits of the generalized quantum kinetic equation}
In this section we consider the relationship of the specific quantum kinetic equations with the
generalized quantum kinetic equation. First we construct the mean field (self-consistent field)
asymptotics of solution \eqref{ske} of initial-value problem of the generalized quantum kinetic
equation for a system with the Hamiltonian \eqref{Hs}, i.e.
\begin{eqnarray}
  \label{gkeNs}
    &&\frac{d}{dt}F_{1}(t,1)=-\mathcal{N}(1)F_{1}(t,1)+\\
    &&+\varepsilon\mathrm{Tr}_{2}(-\mathcal{N}_{\mathrm{int}}(1,2))
        \sum\limits_{n=0}^{\infty}\frac{1}{n!}\mathrm{Tr}_{3,\ldots,n+2}\,
        \mathfrak{V}_{1+n}(t,\{1,2\},3,\ldots,n+2)\prod _{i=1}^{n+2}F_{1}(t,i),\nonumber\\
        \nonumber\\
  \label{2N}
    &&F_1(t,1)|_{t=0}= F_1^0(1),
\end{eqnarray}
and marginal correlation functionals \eqref{cf}.

If there exists the mean field limit $f_{1}^0\in\mathfrak{L}^{1}(\mathcal{H})$ of initial data
\eqref{2N} in the following sense
\begin{eqnarray*}
   &&\lim\limits_{\epsilon\rightarrow 0}\big\|\epsilon\,F_{1}^0-
       f_{1}^0\big\|_{\mathfrak{L}^{1}(\mathcal{H})}=0,
\end{eqnarray*}
then for arbitrary finite time interval there exists the limit $f_{1}(t)$ of solution \eqref{ske}
of the generalized quantum kinetic equation \eqref{gkeNs}
\begin{eqnarray}\label{1lim}
   &&\lim\limits_{\epsilon\rightarrow 0}\big\|\epsilon\,F_{1}(t)-
       f_{1}(t)\big\|_{\mathfrak{L}^{1}(\mathcal{H}_{1})}=0,
\end{eqnarray}
where the one-particle limiting density operator $f_{1}(t,1)$ is determined by series \eqref{viter}
which is a strong solution of the Cauchy problem of the Vlasov quantum kinetic equation
\begin{eqnarray*}
 \label{Vlasov1n}
   &&\frac{d}{dt}f_{1}(t,1)=-\mathcal{N}(1)f_{1}(t,1)+
      \mathrm{Tr}_{2}(-\mathcal{N}_{\mathrm{int}})(1,2)f_{1}(t,1)f_{1}(t,2),\nonumber\\
   &&\nonumber\\
 \label{Vlasovin}
   &&f_1(t)|_{t=0}= f_1^0.
\end{eqnarray*}

Taking into account equality \eqref{1lim}, for marginal functionals of the state \eqref{f} we establish
\begin{eqnarray*}
   &&\lim\limits_{\epsilon\rightarrow 0}\big\|\epsilon^{s} F_{s}(t,Y\mid F_{1}(t))-
      \prod\limits_{j=1}^{s}f_{1}(t,j)\big\|_{\mathfrak{L}^{1}(\mathcal{H}_{s})}=0,
\end{eqnarray*}
where $f_{1}(t)$ is defined by series \eqref{viter}. This equality means that in the mean
field scaling limit initial chaos property preserves in time.

The validity of this limit statement is the consequence of formulas \eqref{lemma} and
\eqref{Duam2} on an asymptotic perturbation of cumulants of groups and definition \eqref{skrr}
of the generated evolution operators $\mathfrak{V}_{1+n}(t),\,n\geq0$, of marginal functionals
of the state \eqref{f}. Indeed, if $f_{s}\in\mathfrak{L}^{1}(\mathcal{H}_{s})$, then the equality
is correct
\begin{eqnarray*}
  &&\lim\limits_{\epsilon\rightarrow 0}\big\|(\widehat{\mathcal{G}}_{s}(t,Y)-I)
     f_{s}\big\|_{\mathfrak{L}^{1}(\mathcal{H}_{s})}=0.
\end{eqnarray*}
Further for the first-order generated evolution operator $\mathfrak{V}_{1}(t,\{Y\})$ the following
equality holds
\begin{eqnarray*}
  &&\lim\limits_{\epsilon\rightarrow 0}\big\|\big(\mathfrak{V}_{1}(t,\{Y\})-
     I\big)f_{s}\big\|_{\mathfrak{L}^{1}(\mathcal{H}_{s})}=0,
\end{eqnarray*}
for any finite time interval, and in the general case $n\geq1$, we have
\begin{eqnarray*}\label{limf}
  &&\lim\limits_{\epsilon\rightarrow 0}\big\|\frac{1}{\epsilon^n}\,\mathfrak{V}_{1+n}(t,\{Y\},X\setminus Y)
     f_{s+n}\big\|_{\mathfrak{L}^{1}(\mathcal{H}_{s+n})}=0.
\end{eqnarray*}
According to this formula on an asymptotic perturbation of evolution operators \eqref{skrr},
we establish the mean field asymptotics of marginal correlation functionals \eqref{cf}
\begin{eqnarray*}\label{asymp}
  &&\lim\limits_{\epsilon\rightarrow 0}\big\|\epsilon^{s}G_{s}(t,Y\mid F_{1}(t))
     \big\|_{\mathfrak{L}^{1}(\mathcal{H}_s)}=0, \quad s\geq2.
\end{eqnarray*}

In case of initial states involving correlations for evolution operators \eqref{skrrc}
in the mean field limit the equality holds
\begin{eqnarray*}
  &&\lim\limits_{\epsilon\rightarrow 0}\big\|\mathfrak{G}_{1+n}(t,\{Y\},s+1,\ldots,s+n)
     f_{s+n}\big\|_{\mathfrak{L}^{1}(\mathcal{H}_{s+n})}=0, \quad n\geq1,
\end{eqnarray*}
and in case of the first-order generated evolution operator \eqref{skrrc} we have respectively
\begin{eqnarray*}
  &&\lim\limits_{\epsilon\rightarrow 0}\big\|\big(\mathfrak{G}_{1}(t,\{Y\})-
     \prod_{i_1=1}^{s}\mathcal{G}_{1}(-t,i_1)h_{1}(\{Y\})\prod_{i_2=1}^{s}\mathcal{G}_{1}(t,i_2)\big)
     f_{s}\big\|_{\mathfrak{L}^{1}(\mathcal{H}_{s})}=0.
\end{eqnarray*}

Thus, the mean field asymptotics of marginal functionals \eqref{cf} has the form
\begin{eqnarray*}
  &&\lim\limits_{\epsilon\rightarrow 0}\big\|\epsilon^{s}F_{s}(t,Y\mid F_{1}(t))-
     \prod_{i_1=1}^{s}\mathcal{G}_{1}(-t,i_1)h_{1}(\{Y\})
     \prod_{i_2=1}^{s}\mathcal{G}_{1}(t,i_2)
     \prod\limits_{j=1}^{s}f_{1}(t,j)\big\|_{\mathfrak{L}^{1}(\mathcal{H}_{s})}=0,
\end{eqnarray*}
that means the propagation of initial correlations in time in the mean field limit, and the
limit one-particle density operator satisfies the modified Vlasov quantum kinetic equation
\begin{eqnarray}\label{mVe}
  &&\frac{d}{dt}f_{1}(t,1)=-\mathcal{N}(1)f_{1}(t,1)+\\
  &&+\mathrm{Tr}_{2}(-\mathcal{N}_{\mathrm{int}})(1,2)
     \prod _{i_1=1}^{2}\mathcal{G}_{1}(-t,i_1)h_{1}(\{1,2\})
     \prod_{i_2=1}^{2}\mathcal{G}_{1}(t,i_2)f_{1}(t,1)f_{1}(t,2).\nonumber
\end{eqnarray}
If the kernel of the interaction potential is the Dirac measure $\Phi(q)=\delta(q)$, then the
sufficient equation for the description of pure state evolution governed by kinetic equation
\eqref{mVe} is the Gross-Pitaevskii-type equation
\begin{eqnarray*}
  &&i\frac{\partial}{\partial t}\psi(t,q)=-\frac{1}{2}\Delta_{q}\psi(t,q)+
     \int d q'd q''\mathfrak{b}(t,q,q;q',q'')\psi(t,q'')\psi^{\ast}(t,q)\psi(t,q'),
\end{eqnarray*}
where the coupling ratio $\mathfrak{b}(t,q,q;q',q'')$ is the kernel of the scattering amplitude operator
$\prod _{i_1=1}^{2}\mathcal{G}_{1}(-t,i_1)b_{1}(\{1,2\})\prod_{i_2=1}^{2}\mathcal{G}_{1}(t,i_2)$.
Observing that in the kinetic (macroscopic) scale of the variation of variables the groups of operators
(\ref{grG}) of finitely many particles depend on microscopic time variable $\varepsilon^{-1}t$,
where $\varepsilon\geq0$ is a scale parameter, the dimensionless marginal functionals of the state are
represented in the form: $F_{s}\big(\varepsilon^{-1}t,Y\mid F_{1}(t)\big)$. Then in the limit
$\varepsilon\rightarrow0$ we obtain the Markovian kinetic evolution with the coefficient
$\mathfrak{b}(\infty,q,q;q',q'')$.


\section{Conclusion and outlook}
The concept of cumulants \eqref{cumulant} of the groups of operators forms the basis of the nonperturbative
solution expansions for hierarchies of quantum evolution equations, namely the dual quantum BBGKY hierarchy
for marginal observables \eqref{dh}, the quantum BBGKY hierarchy for marginal density operators \eqref{BBGKY},
the von Neumann hierarchy \eqref{vNh} for correlation operators and the nonlinear quantum BBGKY hierarchy
for marginal correlation operators \eqref{gBigfromDFBa}, and as well as it underlies of the kinetic evolution
\eqref{gke}. The nonperturbative hierarchy solutions are represented in the form of an expansion over the
particle clusters which evolution is governed by the corresponding order cumulant \eqref{cumulant} of the
groups of operators \eqref{grG} of the Heisenberg equations \eqref{H-N1} or the groups of operators
\eqref{groupG} of the von Neumann equations \eqref{vonNeumannEqn}.

We emphasize that intensional Banach spaces for the description of states of infinite-particle systems,
which are suitable for the description of the kinetic evolution or equilibrium states, are different from
the exploit spaces \cite{CGP97},\cite{Pe95}. Thus, marginal density operators or correlation operators
from the space $\mathfrak{L}^{1}(\mathcal{F}_\mathcal{H})$ describe finitely many quantum particles. In
order to describe the evolution of infinitely many particles we have to construct solutions for initial
data from more general Banach spaces than the space of sequences of trace class operators. For example,
it can be the space of sequences of bounded translation invariant operators which contains the marginal
density operators of equilibrium states \cite{Pe95}. In that case every term of the solution expansion
\eqref{RozvBBGKY} of the quantum BBGKY hierarchy or the nonlinear quantum BBGKY hierarchy \eqref{sss}
and correspondingly the generalized quantum kinetic equation \eqref{ske} as well as mean-value functional
\eqref{avmar} in case of the dual quantum BBGKY hierarchy \eqref{sdh} contains the divergent traces, which
can be renormalized due to the cumulant structure of the solution expansions.

The origin of the microscopic description of non-equilibrium correlations of Bose and Fermi many-particle
systems was considered in section 3. For the correlation operators that give an alternative description of
the quantum state evolution of Bose and Fermi many-particle systems, the von Neumann hierarchy of nonlinear
evolution equations \eqref{vNfb} was introduced. In particular, it was established that in case of absence
of correlations in the system at initial time, the correlations generated by the dynamics of a system
\eqref{rozvChaosN} are completely determined by cumulants \eqref{cumulants} of the groups of operators
\eqref{groupG} of the von Neumann equations.

The links of constructed solution of the von Neumann hierarchy both with the solution of the quantum BBGKY
hierarchy \eqref{FClusters} and with the nonlinear quantum BBGKY hierarchy for marginal correlation operators
\eqref{Gexpg} were discussed. The cumulant structure of solution \eqref{rozvNF-N_F} of the von Neumann
hierarchy \eqref{vNfb} induces the cumulant structure of solution expansions both the initial-value problem
of the nonlinear quantum BBGKY hierarchy for marginal correlation operators \eqref{sss} and the quantum BBGKY
hierarchy for marginal density operators \eqref{RozvBBGKY}. Thus, the dynamics of infinite-particle systems is
generated by the dynamics of correlations. We note that along with the definition within the framework of the
non-equilibrium grand canonical ensemble the marginal density operators can be defined within the framework of
dynamics of correlations \eqref{FClusters} that allows to give the rigorous meaning to the states for more
general classes of operators than trace-class operators.

We remark that the rigorous results on the evolution equations in functional derivatives for generating
functionals of states and observables of classical many-particle systems, namely the BBGKY hierarchy
and the dual BBGKY hierarchy in functional derivatives respectively, are presented in \cite{FG}.

We developed also an approach to a description of the evolution of states by means of the quantum kinetic
equations. It was demonstrated that in fact, if initial data is completely defined by a one-particle density
operator, then all possible states of infinite-particle systems at arbitrary moment of time can be described
within the framework of a one-particle density operator without any approximations. In other words, for
mentioned states the evolution of states governed by the quantum BBGKY hierarchy \eqref{BBGKY} can be completely
described by the generalized quantum kinetic equation \eqref{gke}. It should be emphasized that the kinetic
evolution is an inherent property of infinite-particle systems. In spite of the fact that in terms of a one-particle
marginal density operator from the space of trace-class operators can be described a system with the finite average
number of particles, the generalized quantum kinetic equation has been derived on the basis of the formalism of
nonequilibrium grand canonical ensemble since its framework is adopted to the description of infinite-particle systems
in suitable Banach spaces as well.

We note that constructed marginal functionals of the state \eqref{f} or \eqref{cf} characterize the correlations
of states of quantum many-particle systems. Owing to that from macroscopic point of view the evolution of
many-particle states with the minimal dispersion is the Markovian kinetic evolution, then from microscopic point
of view such evolution is characterized by marginal correlation functionals \eqref{cf} which equal to zero.

An approach to the description of kinetic evolution of quantum many-particle systems in terms of the evolution
of marginal observables is also developed. In the mean field limit the evolution of marginal observables is
governed by the dual quantum Vlasov hierarchy \eqref{vdh}. One of the advantages of such approach as well as 
developed approach of the generalized quantum kinetic equation \eqref{gke} is the possibility to construct the 
kinetic equations in scaling limits in case of the presence of correlations at initial time \eqref{mVe}, for 
instance, correlations which characterize the condensate states of particles \cite{BQ}.


\addcontentsline{toc}{section}{References}
\renewcommand{\refname}{References}


\begin{thebibliography}{99}

\bibitem{CGP97}
     \newblock C. Cercignani, V.I. Gerasimenko and D.Ya. Petrina,
     \newblock ``Many-Particle Dynamics and Kinetic Equations",
     \newblock Kluwer Acad. Publ., 1997.

\bibitem{CIP}
     \newblock C. Cercignani, R. Illner and M. Pulvirenti,
     \newblock ``The Mathematical Theory of Dilute Gases",
     \newblock  Springer-Verlag, 1994.

\bibitem{BQ}
    \newblock M.M. Bogolyubov,
    \newblock ``Lectures on Quantum Statistics. Problems of Statistical Mechanics of Quantum Systems", (Ukrainian)
    \newblock Rad. Shkola, 1949.

\bibitem{Pe95}
    \newblock D.Ya. Petrina,
    \newblock ``Mathematical Foundations of Quantum Statistical Mechanics. Continuous Systems",
    \newblock Kluwer Acad. Publ., 1995.

\bibitem{DauL}
    \newblock R. Dautray and J.L. Lions,
    \newblock ``Mathematical Analysis and Numerical Methods for Science and Technology", \textbf{5},
    \newblock Springer-Verlag, 1992.

\bibitem{BR}
    \newblock O. Bratelli and D.W. Robinson,
    \newblock ``Operator Algebras and Quantum Statistical Mechanics", \textbf{2},
    \newblock Springer, 1997.

\bibitem{BanArl}
    \newblock J. Banasiak and L. Arlotti,
    \newblock ``Perturbations of Positive Semigroups with Applications",
    \newblock Springer, 2006.

\bibitem{AGT}
    \newblock R. Adami, F. Golse and A. Teta,
    \newblock \emph{Rigorous derivation of the cubic NLS in dimension one},
    \newblock J. Stat. Phys., \textbf{127} (6) (2007), 1193-1220.

\bibitem{AA}
    \newblock A. Arnold,
    \newblock \emph{Mathematical properties of quantum evolution equations},
    \newblock Lecture Notes in Math., \textbf{1946} (2008), 45-109.

\bibitem{BGGM2}
    \newblock C. Bardos, F. Golse, A.D. Gottlieb and N.J. Mauser,
    \newblock \emph{Mean field dynamics of fermions and the time-dependent Hartree-Fock equation},
    \newblock J. Math. Pures Appl., \textbf{82} (2003), 665-683.

\bibitem{FL}
    \newblock J. Fr\"{o}hlich, S. Graffi and S. Schwarz,
    \newblock \emph{Mean-field and classical limit of many-body Schr\"{o}dinger dynamics for bosons},
    \newblock Comm. Math. Phys., \textbf{271} (2007), 681-697.

\bibitem{ESchY2}
    \newblock L. Erd\"{o}s, B. Schlein and H.-T. Yau,
    \newblock \emph{Derivation of the cubic nonlinear Schr\"{o}dinger equation from quantum dynamics of many-body systems},
    \newblock Invent. Math., \textbf{167} (3) (2007), 515-614.

\bibitem{EShY10}
    \newblock L. Erd\"{o}s, B. Schlein and H.-T. Yau,
    \newblock \emph{Derivation of the Gross-Pitaevskii Equation for the Dynamics of Bose-Einstein Condensate},
    \newblock Ann. of Math., \textbf{172} (2010), 291-370.

\bibitem{LSSY}
    \newblock E.H. Lieb, R. Seiringer, J.P. Solovej and J. Yngvason,
    \newblock ``The mathematics of the Bose gas and its condensation",
    \newblock Birkh\"{a}user, 2005.

\bibitem{M10}
    \newblock A. Michelangeli,
    \newblock \emph{Strengthened convergence of marginals to the cubic nonlinear Schr\"{o}dinger equation},
    \newblock Kinet. Relat. Models, \textbf{3} (2010), 457-471.

\bibitem{PP09}
    \newblock F. Pezzotti and M. Pulvirenti,
    \newblock \emph{Mean-field limit and semiclassical expansion of quantum particle system},
    \newblock Ann. Henri Poincar\'{e}, \textbf{10} (2009), 145-187.

\bibitem{CP}
     \newblock T. Chen and N. Pavlovic,
     \newblock \emph{The quintic NLS as the mean field limit of a Boson gas with three-body interactions},
     \newblock J. Funct. Anal., \textbf{260} (4) (2011), 959-997.

\bibitem{GMM}
     \newblock M. Grillakis, M. Machedon and D. Margetis,
     \newblock \emph{Second-order corrections to mean field evolution of weakly interacting bosons. I},
     \newblock Comm. Math. Phys., \textbf{294} (1) (2010), 273-301.

\bibitem{S-R}
     \newblock L. Saint-Raymond,
     \newblock \emph{Kinetic models for superfluids: a review of mathematical results},
     \newblock C. R. Physique, \textbf{5} (2004), 65–75.

\bibitem{BCEP3}
    \newblock D. Benedetto, F. Castella, R. Esposito and M. Pulvirenti,
    \newblock \emph{A short review on the derivation of the nonlinear quantum Boltzmann equations},
    \newblock Commun. Math. Sci., \textbf{5} (2007), 55-71.

\bibitem{ESY}
    \newblock L. Erd\"{o}s, M. Salmhofer and H.-T. Yau,
    \newblock \emph{On quantum Boltzmann equation},
    \newblock J. Stat. Phys., \textbf{116} (2004), 367-380.

\bibitem{Sh}
    \newblock H. Spohn,
    \newblock \emph{Kinetic equations from Hamiltonian dynamics: Markovian limits},
    \newblock Rev. Modern Phys., \textbf{52} (3) (1980), 569-615.

\bibitem{Shb}
    \newblock H. Spohn,
    \newblock ``Large Scale Dynamics of Interacting Particles",
    \newblock Springer-Verlag, 1991.

\bibitem{Gre56}
    \newblock M.S. Green,
    \newblock \emph{Boltzmann equation from the statistical mechanical point of view},
    \newblock J. Chem. Phys., \textbf{ 25} (5) (1956), 836-855.

\bibitem{GP}
    \newblock M.S. Green and R.A. Piccirelly,
    \newblock \emph{Basis of the functional assumption in the theory of the Boltzmann equation},
    \newblock Phys. Rev. 1963, \textbf{132} (3), 1388-1410.

\bibitem{C68}
     \newblock E.G.D. Cohen,
     \newblock The kinetic theory of dence gases.
     \newblock In: Cohen, E.G.D., editor. \emph{Fundamental Problem in Statistical Mechanics}. North-Holand Publ.;
               \textbf{2}; 1968; 228-275.
\bibitem{B47}
     \newblock N.N. Bogolyubov and K.P. Gurov,
     \newblock \emph{Kinetic equations in quantum mechanics},
     \newblock J. Exp. Theor. Phys. 1947, \textbf{17}, 614-628.

\bibitem{DP}
    \newblock D.O. Polishchuk,
    \newblock \emph{The BBGKY hierarchy and dynamics of correlations},
    \newblock Ukrainian J. Phys., \textbf{55} (5) (2010), 593-598.

\bibitem{BGer}
    \newblock G. Borgioli and V.I. Gerasimenko,
    \newblock \emph{The dual BBGKY hierarchy for the evolution of observables},
    \newblock Riv. Mat. Univ. Parma, \textbf{4} (2001), 251-267.

\bibitem{GerRS}
    \newblock V.I. Gerasimenko, T.V. Ryabukha and M.O. Stashenko,
    \newblock \emph{On the structure of expansions for the BBGKY hierarchy solutions},
    \newblock J. Phys. A: Math. Gen., \textbf{37} (2004), 9861-9872.

\bibitem{GerS06}
    \newblock V.I. Gerasimenko and V.O. Shtyk,
    \newblock \emph{Initial-value problem of the Bogolyubov hierarchy for quantum systems of particles},
    \newblock Ukrain. Math. J., \textbf{58} (9) (2006), 1175-1191.

\bibitem{GerS}
    \newblock V.I. Gerasimenko and V.O. Shtyk,
    \newblock \emph{Evolution of correlations of quantum many-particle systems},
    \newblock J. Stat. Mech. Theory Exp., \textbf{3} (2008), P03007.

\bibitem{G}
    \newblock V.I. Gerasimenko,
    \newblock \emph{Groups of operators for evolution equations of quantum many-particle systems},
    \newblock Oper. Theory Adv. Appl., \textbf{191} (2) (2009), 341-355.

\bibitem{GerUJP}
    \newblock V.I. Gerasimenko,
    \newblock \emph{Approaches to derivation of quantum kinetic equations},
    \newblock Ukr. Phys. J., \textbf{54} (8-9) (2009), 834-846.

\bibitem{GP97}
     \newblock V.I. Gerasimenko and D.Ya. Petrina,
     \newblock \emph{On the generalized kinetic equation},
     \newblock Reports of NAS of Ukraine, \textbf{7} (1997), 7-12.

\bibitem{GT}
    \newblock V.I. Gerasimenko and Zh.A. Tsvir,
    \newblock \emph{A description of the evolution of quantum states by means of the kinetic equation},
    \newblock J. Phys. A: Math. Theor., \textbf{43} (48) (2010), 485203.

\bibitem{GT10}
    \newblock V.I. Gerasimenko and Zh.A. Tsvir,
    \newblock \emph{Generalized quantum kinetic equation for interacting particles with quantum statistics},
    \newblock Math. Bulletin Sh. Sci. Soc., \textbf{7} (2010), 351-367.

\bibitem{GT11}
    \newblock V.I. Gerasimenko and Zh.A. Tsvir,
    \newblock \emph{Quantum kinetic equations of many-particle systems in condensed states},
    \newblock arXiv:1109.1998, (2011), 10p.

\bibitem{GG11}
    \newblock V.I. Gerasimenko and I.V. Gapyak,
    \newblock \emph{On rigorous derivation of the Enskog kinetic equation},
    \newblock arXiv:1107.5572, (2011), 28p.

\bibitem{BG}
    \newblock G. Borgioli and V.I. Gerasimenko,
    \newblock \emph{Initial-value problem of the quantum dual BBGKY hierarchy},
    \newblock Nuovo Cimento, \textbf{33 C} (1) (2010), 71-78.

\bibitem{GP}
    \newblock V.I. Gerasimenko and D.O. Polishchuk,
    \newblock \emph{Dynamics of correlations of Bose and Fermi particles},
    \newblock Math. Meth. Appl. Sci., \textbf{34} (1) (2011), 76-93.

\bibitem{GP11}
    \newblock V.I. Gerasimenko and D.O. Polishchuk,
    \newblock \emph{On evolution equations for marginal correlation operators},
    \newblock arXiv:1105.5822, (2011), 28p.

\bibitem{G11}
    \newblock V.I. Gerasimenko,
    \newblock \emph{Heisenberg picture of quantum kinetic evolution in mean-field limit},
    \newblock Kinet. Relat. Models, \textbf{4} (1) (2011), 385-399.

\bibitem{Rul}
    \newblock D. Ruelle,
    \newblock \emph{Statistical Mechanics. Rigorous Results},
    \newblock World Sci. Publ. Co., 1999.

\bibitem{FG}
    \newblock Fedchun, Yu.Yu.; Gerasimenko, V. I.
    \newblock \emph{Evolution equations in functional derivatives of many-particle systems},
    \newblock arXiv:1107.0823, (2011), 21p.

\end{thebibliography}
\end{document}